\documentclass[aaspp4,11pt,preprint,apjfonts,natbib209,twocolappendix]{emulateapj}
\pdfoutput=1

\usepackage{graphicx}
\usepackage{times}
 \usepackage{xspace}
\usepackage{amsmath}
\usepackage{mathtools}

\usepackage[usenames,dvipsnames,svgnames,table]{xcolor}

\usepackage[colorlinks=True, 
	    linkcolor=blue,
            urlcolor=red,
            citecolor=blue]{hyperref}

\newcommand{\unit}[1]{\mathrm{#1}}
\newcommand{\deltPA}{\ensuremath{\Delta \mathrm{PA}}\xspace}

\newcommand{\axratio}{\ensuremath{b/a}\xspace}

\newcommand{\Halpha}{\ensuremath{\textrm{H}\alpha}\xspace}
\newcommand{\Hbeta}{\ensuremath{\textrm{H}\beta}\xspace}
\newcommand{\NII}{\ensuremath{\textsc{[Nii]}}\xspace}

\newcommand{\Mstar}{\ensuremath{M_*}\xspace}
\newcommand{\Mgas}{\ensuremath{M_{\mathrm{gas}}}\xspace}
\newcommand{\Mbar}{\ensuremath{M_{\mathrm{baryon}}}\xspace}
\newcommand{\Mdyn}{\ensuremath{M_{\mathrm{dyn}}}\xspace}
\newcommand{\krot}{\ensuremath{k_{\mathrm{rot}}}\xspace}
\newcommand{\kdisp}{\ensuremath{k_{\mathrm{disp}}}\xspace}
\newcommand{\keff}{\ensuremath{k_{\mathrm{eff}}}\xspace}

\newcommand{\aper}{\ensuremath{\mathrm{ap}}\xspace}
\newcommand{\re}{\ensuremath{R_E}\xspace}  
\newcommand{\intrm}{\mathrm{int}}

\newcommand{\Vrms}{\ensuremath{V_{\mathrm{RMS}}}\xspace}
\newcommand{\rms}{\ensuremath{{\mathrm{RMS}}}\xspace}
\newcommand{\sigmav}{\ensuremath{\sigma_{V}}\xspace}
\newcommand{\sigmavint}{\ensuremath{\sigma_{V, 0}}\xspace}
\newcommand{\sigmavobs}{\ensuremath{\sigma_{V, \, \mathrm{obs}}}\xspace}
\newcommand{\sigmavmodel}{\ensuremath{\sigma_{V, \, \mathrm{model}}}\xspace}

\newcommand{\vtosig}{\ensuremath{V/\sigma}\xspace}
\newcommand{\vtosigre}{\ensuremath{(V/\sigmavint)_{\re}}\xspace}
\newcommand{\vtosigtt}{\ensuremath{(V/\sigmavint)_{2.2}}\xspace}

\newcommand{\KMOSTD}{KMOS$^{\hbox{\scriptsize{3D}}}$\xspace}
\newcommand{\HST}{\textit{HST}}
\newcommand{\JWST}{\textit{JWST}}

\shorttitle{Masses and Kinematics of \lowercase{$z\sim2$} Star-Forming Galaxies}
\shortauthors{Price et al.}
\slugcomment{Accepted to ApJ}

\begin{document}

\title{The MOSDEF Survey: Dynamical and Baryonic Masses and Kinematic Structures of Star-Forming 
Galaxies at \lowercase{$1.4 \leq z \leq 2.6$}}

\author{Sedona H. Price\altaffilmark{1}}
\author{Mariska Kriek\altaffilmark{1}} 
\author{Alice E. Shapley\altaffilmark{2}} 
\author{Naveen A. Reddy\altaffilmark{3,5}} 
\author{William R. Freeman\altaffilmark{3}}
\author{Alison L. Coil\altaffilmark{4}} 
\author{Laura de Groot\altaffilmark{3}}
\author{Irene Shivaei\altaffilmark{3}}
\author{Brian Siana\altaffilmark{3}}
\author{Mojegan Azadi\altaffilmark{4}}
\author{Guillermo Barro\altaffilmark{1}}
\author{Bahram Mobasher\altaffilmark{3}}
\author{Ryan L. Sanders\altaffilmark{2}}
\author{Tom Zick\altaffilmark{1}} 

\altaffiltext{1}{Department of Astronomy, University of California,
  Berkeley, CA 94720, USA; sedona@berkeley.edu}
\altaffiltext{2}{Department of Physics \& Astronomy, University of
  California, Los Angeles, CA 90095, USA} 
\altaffiltext{3}{Department of Physics \& Astronomy, University of California, Riverside, CA
  92521, USA} 
\altaffiltext{4}{Center for Astrophysics and Space
  Sciences, University of California, San Diego, La Jolla, CA 92093,
  USA}
\altaffiltext{5}{Alfred P. Sloan Fellow}

% ++++++++++++++++++++++++++++

\begin{abstract}
We present \Halpha gas kinematics for 178 star-forming
galaxies at $z\sim2$ from  the MOSFIRE Deep Evolution Field survey. We
have developed models to interpret the kinematic measurements  from
fixed-angle multi-object spectroscopy, using structural parameters
derived from CANDELS \HST/F160W imaging. For 35 galaxies we
measure resolved rotation with a median $\vtosigre=2.1$. We derive
dynamical masses from the kinematics and sizes and compare them to
baryonic masses, with gas masses estimated from 
dust-corrected \Halpha  star formation rates (SFRs) and the 
Kennicutt-Schmidt relation. When assuming that galaxies with and
without observed rotation have the same median $\vtosigre$, we find
good agreement between the dynamical and baryonic masses, with a
scatter of $\sigma_{\rms}=0.34\,\mathrm{dex}$ and a median
offset of $\Delta\log_{10}M=0.04\,\mathrm{dex}$. 
This comparison implies a low dark matter fraction (8\% within an effective 
radius) for a Chabrier initial mass function (IMF), and disfavors a Salpeter IMF.
Moreover, the requirement that $\Mdyn/\Mbar$ should be independent of inclination 
yields a median value of $\vtosigre=2.1$ for galaxies without observed rotation. 
If instead we treat the galaxies without detected rotation as early-type galaxies,
the masses are also in reasonable agreement ($\Delta\log_{10}M=-0.07\,\mathrm{dex}$,
$\sigma_{\rms}=0.37\,\mathrm{dex}$). 
The inclusion of gas masses is critical in this comparison; if gas masses are excluded 
there is an increasing trend of $\Mdyn/\Mstar$ with higher specific SFR (SSFR).   
Furthermore, we find indications that $\vtosig$ decreases with increasing \Halpha SSFR 
for our full sample, which may reflect disk settling. 
We also study the Tully-Fisher relation and find that at fixed stellar mass 
$S_{0.5}=\left(0.5V_{2.2}^2+\sigmavint^2\right)^{1/2}$ was higher at earlier times. 
At fixed baryonic mass, we observe the opposite trend. 
Finally, the baryonic and dynamical masses of the active galactic nuclei in our sample are also in excellent agreement, 
suggesting that the kinematics trace the host galaxies.
\end{abstract}


\maketitle

\section{Introduction}

\setcounter{footnote}{0}

% ++++++++++++++++++++++++++++++++++++++++++++++++++++++++++++++++++++++++++++++++++++
% ++++++++++++++++++++++++++++++++++++++++++++++++++++++++++++++++++++++++++++++++++++

% Intro: local universe
In the local universe, most massive star-forming galaxies have
traditional Hubble-type morphologies and relatively smooth and thin
stellar disks (e.g., \citealt{Blanton09}). 
These disks are thought to form from the cooling of baryons within dark matter halos 
(\citealt{White78}, \citealt{Fall80}, \citealt{Blumenthal84}, \citealt{White91}).   
Galaxy formation models (both semi-analytic models, e.g. \citealt{Dalcanton97}; \citealt{Mo98}, and 
hydrodynamic simulations, e.g. \citealt{vandenBosch01}; \citealt{Governato07}; \citealt{Dutton09})  
are able to reproduce realistic local disk galaxies. 
However, testing these specific models requires direct observations of galaxies throughout cosmic time. 

%Details of technological advances 
Over the past two decades, technological advances have enabled observations that provide new
insights into the nature of star-forming galaxies at intermediate and
high redshifts, in particular due to the combination of high-resolution
multi-wavelength imaging with the {\em Hubble Space Telescope} ({\em HST}) 
and near-infrared integral-field spectroscopy with ground-based
telescopes. 
%% Look-back study results: z~2
The kinematics and structures of star-forming galaxies have been
measured out to $z\sim2$  (e.g., \citealt{Weiner06a},
\citealt{Kassin07, Kassin12}, \citealt{Noeske07}, \citealt{Miller11,Miller12, Miller13},  
\citealt{Szomoru11}, 
\citealt{Contini12}, 
\citealt{Nelson12,  Nelson13, Nelson15}, \citealt{vanDokkum13}, 
\citealt{Buitrago14}), 
the epoch during which the star formation rate (SFR) density in the universe is at its peak
value. At this time, massive star-forming galaxies generally look very
different from similar-mass star-forming galaxies today (e.g.,
\citealt{Fan01}, \citealt{Chapman05}, \citealt{Hopkins06},
\citealt{Reddy09}). 
They tend to be smaller (e.g., \citealt{Williams10}, \citealt{vanderWel14a}), 
morphologically clumpier  (e.g., \citealt{Elmegreen06}, \citealt{Elmegreen07, Elmegreen09},
\citealt{Law07, Law09, Law12a}, \citealt{Genzel08},
\citealt{ForsterSchreiber09, ForsterSchreiber14}), have thicker disks
\citep{Elmegreen06}, and higher gas fractions  (\citealt{Daddi08,
  Daddi10}, \citealt{Tacconi08, Tacconi10, Tacconi13},
\citealt{Swinbank11}). 

% Theory behind thicker disks at z~2
Many massive star-forming galaxies at $z\sim2$ do have rotating disks -- similar to their local 
counterparts -- but tend to have higher velocity dispersions (and therefore lower $\vtosig$, 
i.e. the ratio of rotation to velocity dispersion) than local star-forming galaxies 
(e.g., \citealt{Epinat08,Epinat10}, \citealt{Green14}). The higher velocity dispersions at higher redshifts 
are thought to reflect increased turbulence and thickened disks (e.g., \citealt{ForsterSchreiber06, ForsterSchreiber09},
\citealt{Wright07, Wright09}, \citealt{Genzel08, Genzel11},
\citealt{Law07a, Law09, Law12}, \citealt{Wisnioski12, Wisnioski15},
\citealt{Newman13}). Theoretical models suggest that the higher
turbulence and clumpier morphology of massive star-forming galaxies at
$z\sim2$ relative to their local counterparts are the result of the
higher gas fractions (e.g., \citealt{Dekel09}, 
\citealt{Bournaud11}, 
\citealt{Genel12}), and
that the gas-rich, thicker disks are built-up by smooth, cold-mode
gas accretion or minor mergers (e.g., \citealt{Keres05, Keres09},
\citealt{Dekel06},  \citealt{Dave08}, \citealt{Dekel09},
\citealt{Oser10},  \citealt{Cacciato12}, \citealt{Ceverino12}). 

% Limitations to previous studies: seeing 
However, the theoretical interpretation of the structures of distant
star-forming galaxies is complicated by observational limitations,
including low spatial resolution and small sample sizes. For
example, initial studies with SINFONI found that one third (14 of 47)
of star-forming galaxies at $z\sim2$ appeared to be small and
dispersion dominated (i.e., $V/\sigma < 1$;
\citealt{ForsterSchreiber09}). Nonetheless, \citet{Newman13} revealed that
objects that do not show evidence for rotation in the lower resolution
observations, especially objects with sizes close to the previous
resolution limit, do show evidence for rotation in follow-up
adaptive-optics assisted IFU observations. Thus, small
rotationally-supported galaxies may appear to be dispersion dominated
because of smearing caused by resolution limitations.

% Possible now because: KMOS and MOSFIRE, but challenges.
New near-infrared (NIR) spectrographs, including KMOS
\citep{Sharples04, Sharples13} and MOSFIRE \citep{McLean10, McLean12}
have multiplexing capabilities, and thus allow for extensive kinematic
studies of large, complete samples of galaxies at $z\sim2$.  However,
as both KMOS and MOSFIRE are seeing limited, kinematic measurements of
the majority of the star-forming galaxies at $z\sim2$ with $\log_{10}
(M_*/M_{\odot}) \lesssim 10$ will suffer from the same resolution
problem as the seeing-limited SINFONI studies. Additionally,
multi-slit spectrographs like MOSFIRE have no IFU and a constant
position angle for all slits in one mask. The mask orientation is
generally set by the algorithm to maximize the number of targets in a
mask, and thus the slit position angle is randomly oriented compared
to the galaxy major axes. The random slit orientations introduce
additional challenges to interpreting the observed kinematic
information.

% Why use multi-obj slit spectroscopy to study kinematics?
Despite these complications, we can take advantage of the large galaxy
surveys  afforded by multi-object NIR spectrographs by combining these
observations with high-resolution rest-frame optical imaging from 
\HST. Ancillary \HST/WFC3 data accurately show what portion of a
galaxy falls within  the slit, and can be used to interpret the
observed spectrum. Furthermore, by using large galaxy samples with
detailed ancillary  measurements, we can apply statistical approaches
to constrain the kinematic structures of galaxies. For example,
\citet{vanderWel14} use the distribution of observed axis ratios to
constrain the structures of star-forming galaxies. 

%Here we present...
In this paper we study the dynamical and baryonic masses and kinematic
structures of a sample of 178 star-forming galaxies using data from the MOSFIRE
Deep Evolution Field (MOSDEF) survey \citep{Kriek15}. The galaxies are
observed with random  orientations between the slit and kinematic
major axes, and rotation is detected in only 35 galaxies.  However,
for the galaxies without detected rotation we take advantage of our
large sample size, accurate \Halpha SFRs,  stellar masses and detailed
morphological information from imaging of the CANDELS survey
(\citealt{Koekemoer11}, \citealt{Grogin11}) to constrain their
kinematics. We derive dynamical masses for all galaxies, compare them
with baryonic masses, and discuss the implications for the structures
of the galaxies, the stellar initial mass function (IMF) and dark matter fraction, 
and the gas kinematics of active galactic nuclei (AGN) host galaxies.

% Organization
The paper is organized as follows. In Section~\ref{sec:data}, we
present our sample and the ancillary measurements. The methods of
extracting kinematic information from both 2D spectra and integrated
1D spectra are detailed in Section~\ref{sec:kinematic_measurements}.
In Section~\ref{sec:results}, we
present the baryonic and dynamical masses,  as well as $V/\sigma$, for
both the galaxies with and without detected rotation. 
The implications and caveats of our results  are presented in Section
\ref{sec:discussion}. We summarize our results in Section
\ref{sec:summary}.

% Info on assumed cosmology for distance calculations:
Throughout this paper we adopt a $\Lambda$CDM cosmology with $\Omega_m= 0.3$,  
$\Omega_{\Lambda} = 0.7$, and $H_0 = 70 \ \unit{km \, s^{-1} \, Mpc^{-1}}$.

% ++++++++++++++++++++++++++++++++++++++++++++++++++++++++++++++++++++++++++++++++++++
% ++++++++++++++++++++++++++++++++++++++++++++++++++++++++++++++++++++++++++++++++++++

\section{Data}
\label{sec:data}

\subsection{The MOSDEF Survey}
\label{sec:mosdef_survey}

% Survey introduction
We make use of data from the MOSDEF survey \citep{Kriek15}, conducted
using the MOSFIRE spectrograph \citep{McLean12} on the 10 m Keck I
telescope.  In this work, we use the spectra obtained during semesters
2012B, 2013A, and 2014A.  When complete, the MOSDEF survey will
contain moderate resolution ($R = 3000 - 3650$) rest-frame optical
spectra for  $\sim 1500$ $H$-band selected galaxies at $1.4 \leq z
\leq 3.8$ in several Cosmic Assembly Near-Infrared Deep Extragalactic
Legacy Survey (CANDELS; \citealt{Koekemoer11}, \citealt{Grogin11})
fields. 
A detailed overview of the survey, observations, data
reduction, line measurements and sensitivities, success rate,
redshift measurements, stellar population properties, and sample
characteristics are given in \citet{Kriek15}. 

% Structural parameter measurements
For all galaxies observed with MOSFIRE, we measure structural
parameters, including the S\'ersic index, $n$ \citep{Sersic68}, the effective radius,
$\re$ (assumed to be  the semi-major axis, unless explicitly stated
otherwise), the axis ratio, $b/a$, and the major axis position angle
from the  \HST/F160W images (released by the CANDELS team)
using \textsc{Galfit} \citep{Peng10a}.  See L. de Groot et al. (in
preparation) for more details on the structural parameter
measurements.

% Stellar masses
Stellar masses for all MOSDEF galaxies are derived by fitting the
$0.3-8.0 \, \mu\mathrm{m}$ photometry from the 3D-HST survey
(\citealt{Brammer12}, \citealt{Skelton14}, \citealt{Momcheva15})
with the flexible stellar population models
(\citealt{Conroy09}, \citealt{Conroy10a}) using FAST \citep{Kriek09}, while adopting
the MOSFIRE redshifts ($z_{\mathrm{MOS}}$). We assume a
\citet{Chabrier03} stellar IMF, along with a \citet{Calzetti00} dust
attenuation curve, a delayed exponentially-declining star formation
history, and solar metallicity. To account for template
mismatch, we assume the default FAST template error
function. Confidence intervals are calibrated using 500 Monte Carlo
simulations. Hence, the stellar mass uncertainties do not include
systematic uncertainties due to the choice of IMF, dust attenuation
curve, or other assumptions. See \citet{Kriek15} for more details on
the stellar population modeling. 

% Stellar mass galfit vs photometry correction:
Following \citet{Taylor10a}, we correct the stellar masses by the
difference between the \textsc{Galfit} ($m_{\textsc{Galfit}}$) and total
photometric F160W magnitudes ($m_{\mathrm{phot}}$), using
\begin{equation}
\log_{10} M_{*} = \log_{10} M_{*, \, \mathrm{FAST}} + 
0.4 (m_{\mathrm{phot}} - m_{\textsc{Galfit}}). 
\label{eq:mstel_corr}
\end{equation}
This correction makes the size and stellar mass measurements
self-consistent.

% Emission line measurements
Emission line fluxes are measured from the optimally extracted MOSFIRE
1D spectra by fitting adjacent lines simultaneously with Gaussians
plus a linear fit to account for the underlying continuum. The \Halpha
and \Hbeta lines are corrected for the underlying Balmer absorption,
as estimated from the best-fit stellar population models.  See
\citet{Kriek15} and \citet{Reddy15} for more details on the emission
line measurements. 

% Halpha SFR measurements: dust correction
We use the \Halpha emission lines to estimate SFRs and gas masses
(\Mgas) using the following method. For galaxies with detected \Hbeta,
the Balmer absorption-corrected \Halpha fluxes are corrected for dust
using the Balmer decrement, assuming a \citet{Cardelli89} extinction
curve \citep{Reddy15}.  When \Hbeta is undetected, we assume the
reddening of the nebular regions is related to that of the continuum
by $A_{\mathrm{V,neb,Calzetti}} = 1.86 \,
A_{\mathrm{V,cont,Calzetti}}$ \citep{Price14}. As this relation was
derived by assuming the \citet{Calzetti00} attenuation curve for both the
continuum and line emission, we convert the inferred nebular
attenuation to the \citet{Cardelli89} curve, and correct the \Halpha
fluxes accordingly. In order for \Hbeta to be used in the dust
correction, it must be detected with a signal-to-noise ratio (S/N)~$\ge$~3, and the spectrum
transmission at \Hbeta must be at least 50\% of the maximum
transmission.

% Halpha SFR measurements: corrected flux to SFR and gas measurements
The dust-corrected \Halpha fluxes are converted into \Halpha
luminosities, that are then used to calculate the \Halpha SFRs
following the relation of \citet{Kennicutt98} for a
\citet{Chabrier03} IMF \citep{Shivaei15}.  Finally, the relation
between $\Sigma_{\mathrm{gas}}$ and $\Sigma_{\mathrm{SFR}}$ by
\citet{Kennicutt98} is used to estimate the gas masses, using
$\Sigma_{\mathrm{gas}} = \Mgas/(\pi R_E^2)$ and $\Sigma_{\mathrm{SFR}}
= \mathrm{SFR}/(\pi R_E^2)$, where $R_E$ is the best-fit
\textsc{Galfit} major axis. In Section~\ref{sec:caveats} we discuss
the validity of this relation at high redshift. The gas mass
uncertainties include uncertainties on the \Halpha and \Hbeta fluxes
and on the slit-loss corrections. An uncertainty of $0.2 \, \mathrm{dex}$ on 
$A_{\mathrm{V,cont, Calzetti}}$ is assumed when \Hbeta is undetected.

% +++++++++++++++++++++++++++++++++++++++++++++++++++++++++++++++++++++++++++++++++++++++
\subsection{Sample selection}
\label{sec:sample_selection}

% Basic sample selection: redshift selection. Ha/Hb detection cut. F160W coverage. AGN (de)selection
For this work, we select objects in the redshift ranges $1.34 \leq z
\leq 1.75$ and $2.075 \leq z \leq 2.6$, to ensure coverage of the
\Halpha emission line. We also require that \Halpha is detected (i.e.,
S/N $\ge 3$), and that there is \HST/F160W coverage, to make use of the \textsc{Galfit} structural
parameter measurements.

% Quality cuts: spectra, stellar pop, structural parameters, interacting counterparts
We use additional selection criteria to ensure we include only
high-quality spectra. First, we consider only primary MOSDEF targets,
excluding any serendipitously detected galaxies that happened to fall
within the slit. Second, we exclude objects with non-negligible
contamination to the \Halpha flux from neighboring skylines, to provide clean
kinematic measurements. 
Third, we impose quality cuts for both the stellar population and
structural parameters to  ensure that the best fits adequately model
the data. For the stellar population fits, we exclude objects for
which the best-fit reduced chi-square $\chi_{\mathrm{red}}^2 >
10$. For the structural parameters, we flag and exclude objects for
which \textit{(a)} the \textsc{Galfit} runs did not converge, or
\textit{(b)} the \textsc{Galfit} and \HST/F160W total
magnitudes differ by more than 0.5 mag. 
Fourth, we exclude any objects that fall within the quiescent region 
in the UVJ diagram (\citealt{Wuyts07}, \citealt{Williams09}). 
Fifth, we exclude AGN with outflow signatures or with very broad emission lines (Freeman et al. in preparation).  
Sixth, we exclude objects that appear to have an interacting counterpart at a similar redshift, 
as the velocity signatures of these systems may not reflect the internal kinematics. 
We consider AGN, as identified by their X-ray luminosity, IRAC color, or rest-frame optical 
emission lines ratios \citep[][M. Azadi et al. in preparation]{Coil15} separately from our sample of
star-forming galaxies.

% Sample size and skyline contamination
Our final sample includes 178 unique galaxies, with \Hbeta detected
in 138. One object has been observed twice.  We also consider 21
unique AGN (14 with \Hbeta detected) that meet all selection criteria,
with 2 AGN having been observed twice.

%%%%%%%%%%%%%
% R_E vs M*
\begin{figure}
  \centering
  \hglue -4pt
  \includegraphics[width=0.5\textwidth]{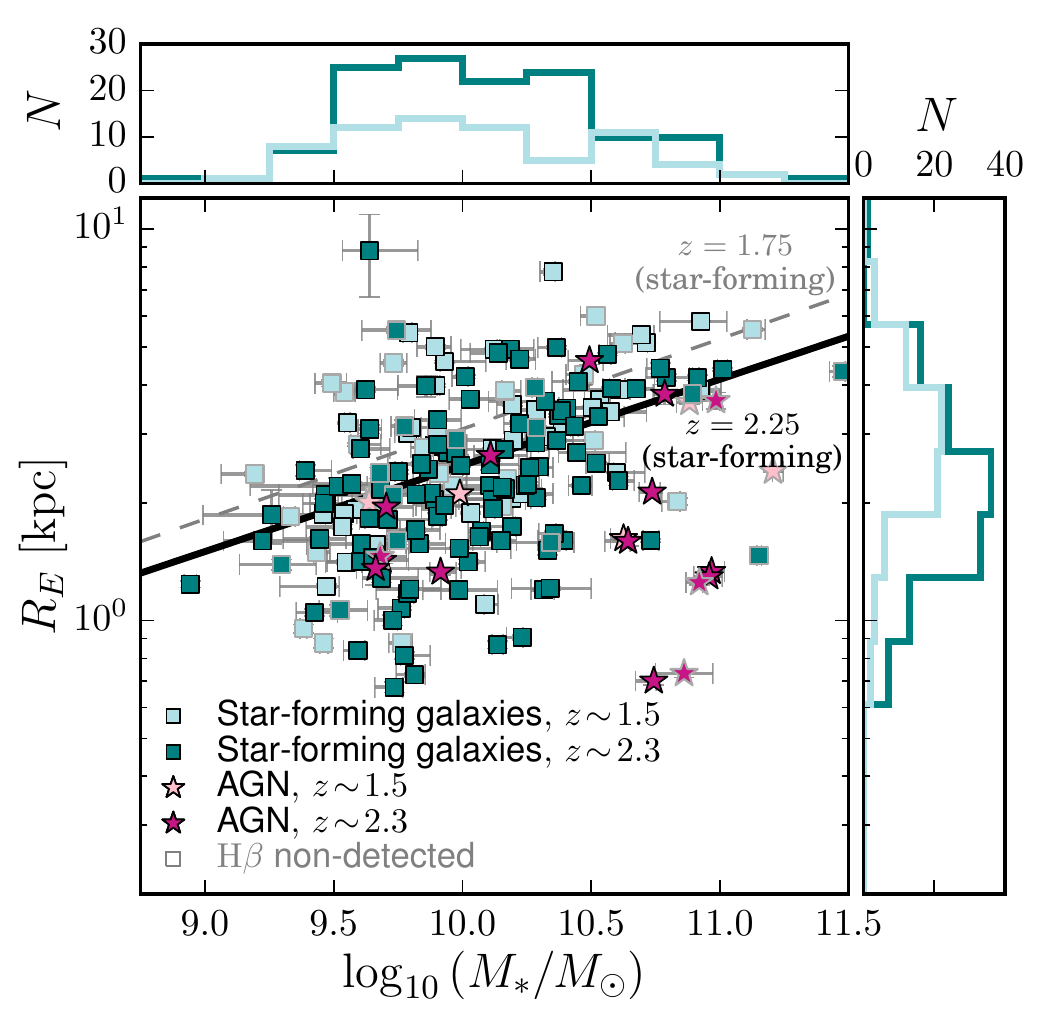}
  \vglue -4pt
  \caption{ Effective radius, $\re$, versus stellar mass, $\log_{10}
    (M_*/M_{\odot})$, for the galaxies and AGN in our sample, split by redshift
    range.  The galaxies at in the lower ($z \sim 1.5$) and higher
    ($z\sim 2.3$) redshift ranges are shown as light blue and teal squares,
    respectively.  The AGN in the same redshift ranges are shown as
    pink and purple stars, respectively.  
    Histograms of  $ \re $ and $\log_{10}(M_*/M_{\odot}) $ show the objects (galaxies and 
    AGN) in the lower (light blue) and higher (teal) redshift ranges. 
    Galaxies (and AGN) without \Hbeta 
    detections are marked with grey outlines. 
     The black solid and grey dashed lines
    represent the best-fit size-stellar mass relations for
    star-forming galaxies  from \citet{vanderWel14a} at $z=2.25$ and
    $z=1.75$, respectively.  Our sample of star-forming galaxies
    generally follow the best-fit size-stellar mass relations.  }

  \label{fig:mstar_re}
\end{figure}

% Stellar mass vs effective radius
We show the effective radii versus stellar masses for the galaxies and
AGN in our sample in Figure~\ref{fig:mstar_re}.  
For comparison, we also show the best-fit size-stellar mass relations found by 
\citet{vanderWel14a} for a complete sample of star-forming (late-type) galaxies at 
$z=2.25$ and $1.75$. 
The samples are complete down to $M_* \!
\sim \! 10^{9.5} M_{\odot}$ at $z=2.25$ and down to  $M_* \! \sim \!
10^{9.1} M_{\odot}$ at $z=1.75$, and are therefore a good
representation of the  star-forming galaxies at these redshifts.  
Our galaxies generally follow these best-fit size-stellar mass relations in both redshift ranges, 
though our galaxies at $z\sim 1.5$ may be somewhat smaller in size than the average as determined by \citet{vanderWel14a}.

% +++++++++++++++++++++++++++++++++++++++++++++++++++++++++++++++++++++++++++++++++++++++
\section{Kinematic Measurements}
\label{sec:kinematic_measurements}

\begin{figure*}	
  \centering
  \hglue -3pt
  \includegraphics[width=1.02\textwidth]{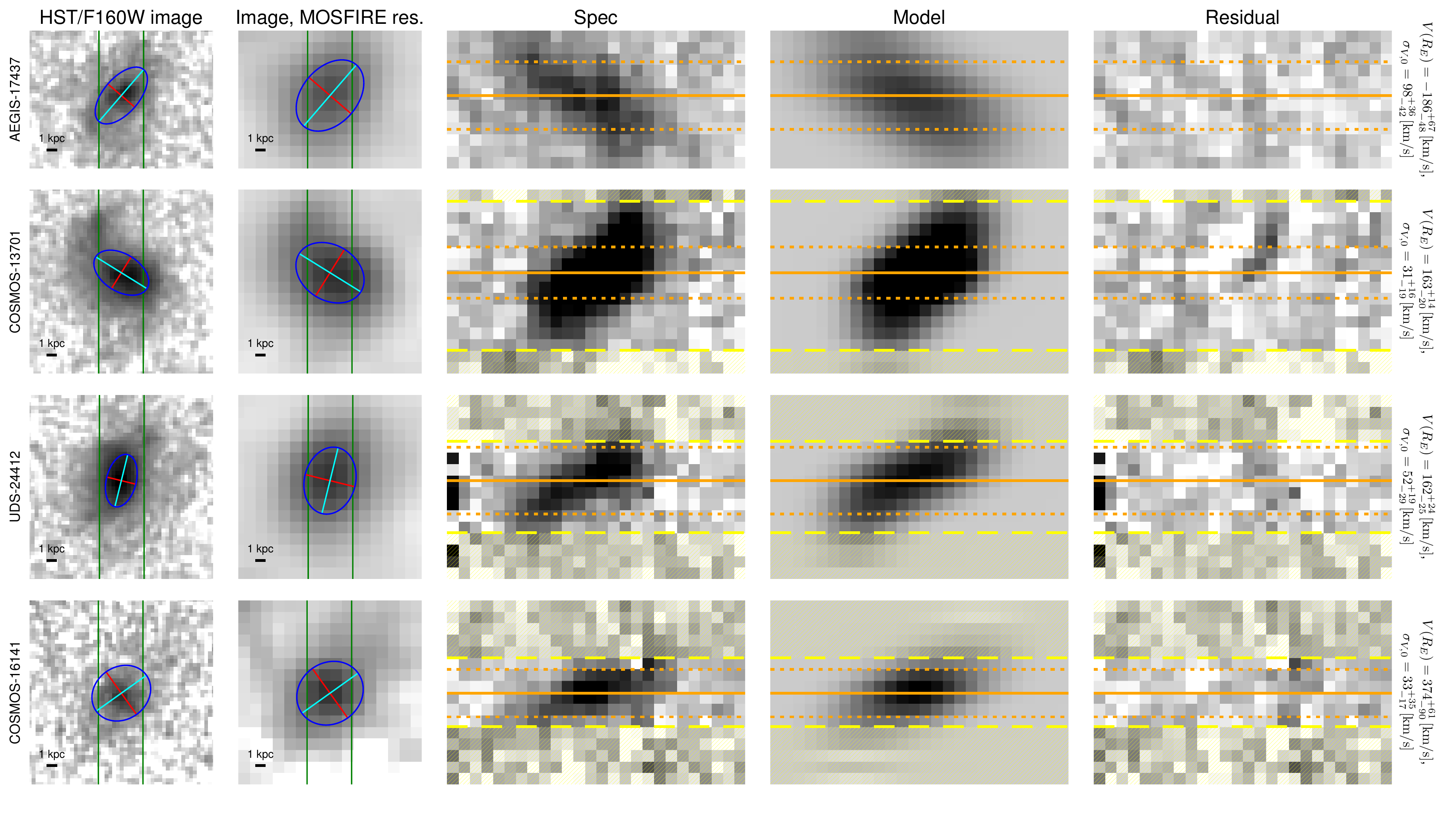}

  \caption{Example overviews of the spatially resolved kinematic
    modeling, as described in  Section~\ref{sec:rot_meas}.  The first
    column shows the \HST/F160W image,  the slit position
    (green lines) and the \textsc{Galfit} half-light ellipse (major
    axis in cyan, minor axis in red, ellipse in blue).  The second
    column shows the \HST/F160W image convolved to match the
    seeing resolution of the MOSFIRE spectra, with the \textsc{Galfit}
    parameters similarly convolved.  The third column shows the
    continuum-subtracted 2D spectrum centered at \Halpha.  The fourth
    column shows the best-fit kinematic model to the line emission.
    The fifth column shows the residual between the 2D spectrum and
    the best-fit model, on the same grey scale  as the 2D spectrum.
    In the third-to-fifth columns, the vertical and horizontal axes
    are the spatial position and wavelength, respectively.  The orange
    horizontal line shows the best-fit model center, $y_0$,  with the
    dotted lines showing the convolved and projected $\re$, and the
    yellow shading and yellow dashed lines indicate low
    signal-to-noise rows that are masked in the fitting procedure.
    For each object, the field and 3D-HST v4 ID number are shown at
    the left, and the best-fit $V(\re)$ and $\sigmavint$  are given on
    the right.  }

  \label{fig:example_bestfits}
\end{figure*}

% Introduction to gas kinematic measurements,
We measure the kinematic properties of our galaxy sample from the
\Halpha emission lines in combination with the \HST/F160W
structural parameters. For objects with spatially-resolved rotation
curves, we constrain the rotation and dispersion velocity components
by fitting models to the 2D \Halpha emission lines, as discussed in
Section~\ref{sec:rot_meas}. The kinematics for objects without
detected rotation are constrained from the 1D \Halpha
profile, by simultaneously fitting \Halpha and \NII lines following
the method listed in Section~\ref{sec:disp_meas}.  In Section
\ref{sec:resolved_unresolved}, we determine for which objects we may
reasonably expect to see rotation and for which  we do not expect to
see rotation at all. We compare these expectations  with our
observations and discuss what this may tell us about the kinematic
structures of galaxies. The method for calculating  the dynamical
masses is discussed in Section~\ref{sec:mdyn_eff}. Finally, in Section
\ref{sec:aper_corr_test}, the 2D and 1D  kinematic measurement methods
are compared using the galaxies with rotation.

% ###############################################################################

\subsection{Rotation velocity measurements}
\label{sec:rot_meas}

% Intro to method: models
In this section we constrain the rotational velocity and velocity dispersion simultaneously by modeling the 2D spectra 
in combination with the F160W structural parameters for each MOSDEF galaxy in our sample. Previous studies have 
presented methods for fitting 2D spectra, including \citet{Vogt96}, \citet{Simard99}, and \citet{Weiner06a}. 
However, the models of \citet{Simard99} and \citet{Weiner06a} do not account for misalignment between the slit and 
major axis, while \citet{Vogt96} and \citet{Simard99} exclude velocity dispersion. 

% Kinematic model definition
Instead, we define kinematic models that explicitly include the position angle misalignment and inclination, 
and simultaneously fit the rotation velocity and velocity dispersion. The kinematic models are discussed in detail 
in Appendix~\ref{sec:appendix2DsubA}. 
In summary, the kinematic models include both rotation and a constant velocity dispersion 
over the galaxy, and have a total of 3 free parameters:  the asymptotic
velocity ($V_a$) and turnover radius ($r_t$) of the arctan rotation
curve model, and the constant intrinsic velocity dispersion ($\sigmavint$).
We assume the best-fit \textsc{Galfit} parameters and the position
angles from the F160W observations. The model is collapsed along the
line-of-sight, and convolved to match the seeing conditions of each
spectrum. Using the position angle, inclination, brightness profile,
and seeing information, we determine what portions of the model fall
within the slit for each object. Finally, the model is collapsed in
the spatial direction perpendicular to the slit and convolved by the
instrumental resolution.

% Brief overview of how we extract the 2D Halpha emission line spectra "stamps"
To fit the emission lines, we start by subtracting the continuum from each \Halpha 2D
spectrum. We also trim the spectrum to exclude the \NII emission lines
and to include only the positive emission line image. We then
construct a mask for the emission line spectrum to exclude missing
data and low signal-to-noise rows from our fitting procedure. 
A detailed description of this procedure can be found in Appendix
\ref{sec:appendix2DsubB}.

% sigma to VRMS correction plot
\begin{figure*}
  \centering \hglue -10pt
  
  \includegraphics[width=0.9\textwidth]{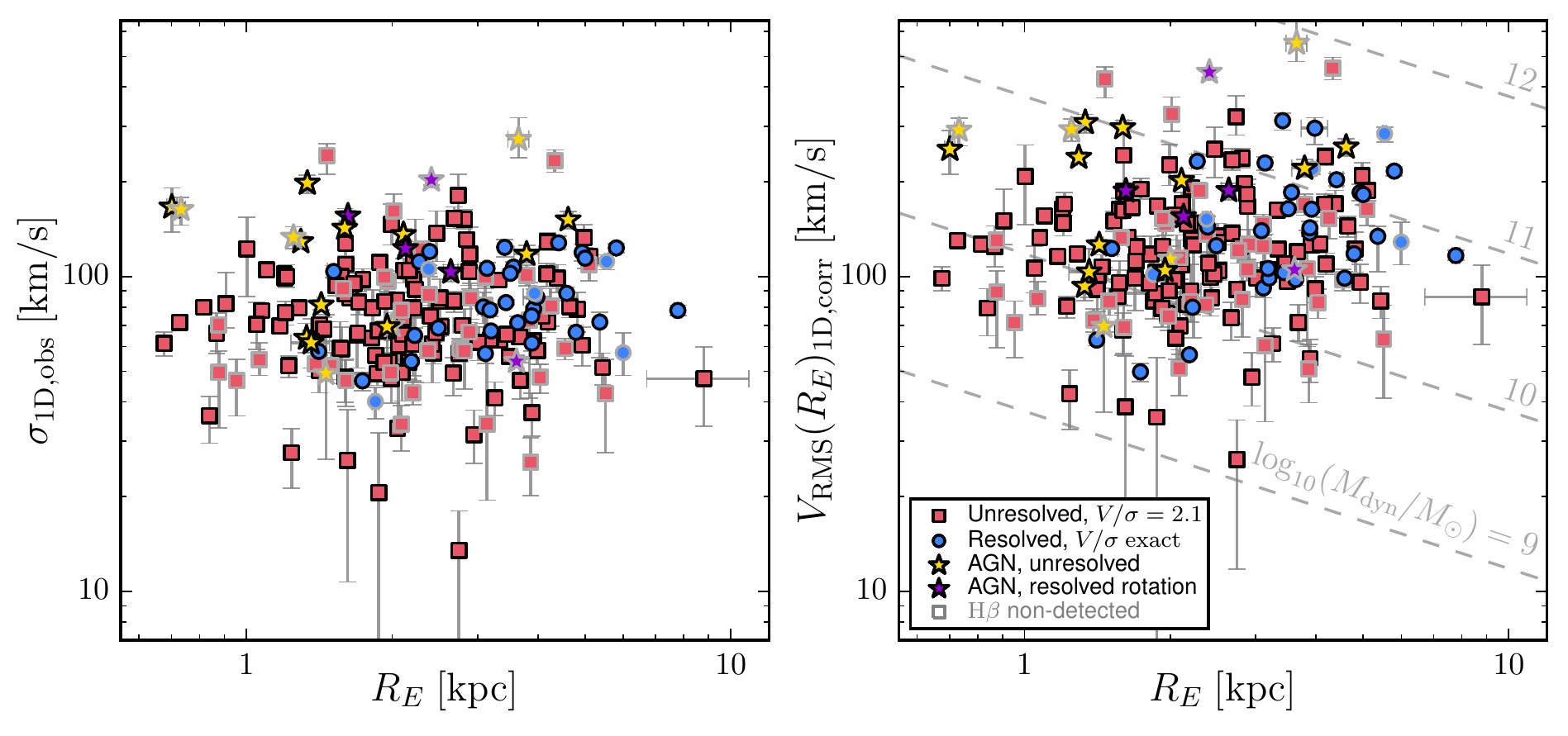}

  \caption{Observed velocity dispersion $\sigma_{\mathrm{1D,obs}}$
    (left) and  the corrected velocities
    $V_{\rms}(\re)_{\mathrm{1D,corr}}$ (right) vs. $\re$.  The
    corrections for the galaxies without resolved rotation (red squares) are derived 
    assuming $\left[\vtosigre\right]_{\mathrm{2D, median}} = 2.1$.  
    For comparison, the corrections for the
    galaxies with resolved rotation, derived using the measured
    $\vtosigre$ of each object, are shown as well (blue circles). We
    also show the AGN in our sample, and denote those with and without
    resolved rotation by purple and yellow stars,
    respectively. Galaxies (and AGN) without \Hbeta detections are
    marked with grey outlines. Lines of constant dynamical mass, 
    calculated using Equation~\ref{eq:mdyn_eff} and assuming $\vtosigre = 2.1$, 
    are shown in the right panel (dashed grey lines). The RMS
    velocities are on average a factor of $\sim2$ larger than the
    observed values. }

  \label{fig:sig_obs_corr_re} 
\end{figure*}

% MCMC fitting
We then find  the best-fit models to the trimmed 2D \Halpha spectra and the
corresponding confidence intervals by performing parameter space
exploration using the \texttt{python} Markov-Chain Monte Carlo (MCMC) package \texttt{emcee}
\citep{Foreman-Mackey13}, following the method detailed in Appendix
\ref{sec:appendix2DsubB}. As the rotation curve turnover is not well
constrained in our data, there is a degeneracy in the values of $V_a$
and $r_t$. Nonetheless, the values of $V(\re)$ and $V_{2.2} =
V(2.2r_s)$ (assuming the arctan rotation curve model, see Equation
\ref{eq:v_rot}) are well constrained. We note that we explicitly
include the structural parameters and projection effects in our model,
so we directly constrain the intrinsic galaxy parameters, without
projection or blending effects. 
% Example fits
Examples of the 2D \Halpha emission line fits are shown in Figure~\ref{fig:example_bestfits}.

%Sample splitting
We use the values of $V(\re)$ to determine which objects have
spatially resolved rotation. We take objects with $V(\re) \neq 0$
within the $95\%$ one-sided distribution to be our sample with resolved rotation, 
and the objects that fail this cut to be our dispersion-only sample. 
% PV appendix:
The position-velocity diagrams of the 35 galaxies with detected rotation are shown in Appendix~\ref{sec:appendix_PV}.

% ###############################################################################
\subsection{Integrated velocity dispersion measurements}
\label{sec:disp_meas}

% Intro
For all objects without resolved rotation (see Section
\ref{sec:rot_meas}), we measure the kinematics from the  1D
spectra. As our sample consists of star-forming galaxies, we may
expect that their intrinsic velocity support is at least partially
rotational. This assumption is reinforced by the work of
\citet{Newman13}, who find that galaxies that were initially
classified as dispersion-dominated in fact do show evidence for
rotation in observations with higher spatial resolution. Thus, we
model the composite unresolved kinematics by assuming a value for
$\vtosigre = V(\re)/\sigmavint$, while taking into account the \textsc{Galfit} parameters
and seeing conditions. We then use this model to convert the measured
velocity dispersion to intrinsic quantities.

% Non-resolved gaussian fitting, skyline width definition. 
We measure the velocity dispersion from the optimally extracted 1D
spectra by fitting  \Halpha, the \NII doublet, and the continuum
simultaneously with a triple Gaussian and a linear component.  We fit
the spectrum between $6480 \unit{\AA} \le \lambda/(1+z_{\mathrm{MOS}})
\le 6650 \unit{\AA}$,  and mask pixels with no coverage.  We vary the
coupled line centers, while constraining  $\lambda_{\Halpha,
  \mathrm{obs}}$ to within $\pm 20 \mathrm{\AA}$ of
$\lambda_{\Halpha}(1+z_{\mathrm{MOS}})$. The widths of the emission
lines (in \AA) are coupled in velocity space, with $\sigma_{\lambda, \NII
  \lambda\lambda 6584,48} = \sigma_{\lambda,
  \Halpha}\left(\lambda_{\NII \lambda \lambda
  6584,48}/\lambda_{\Halpha}\right)$.  We also assume $F(\NII\lambda
6548) = 1/3 \, F(\NII\lambda 6584)$ \citep{Osterbrock06}.  Finally, we
enforce $\sigma_{\lambda, \Halpha} \geq \sigma_{\lambda,
  \mathrm{sky}}$,  with the instrumental resolution measured from the
median skyline width.

% Instrumental resolution correction, velocity dispersion conversion
The \Halpha line widths are corrected for the line broadening due to
the instrumental resolution by  subtracting $\sigma_{\lambda,
  \mathrm{sky}}$ in quadrature from $\sigma_{\lambda, \Halpha}$. Each
corrected \Halpha line width $\sigma_{\lambda, \Halpha, \,
  \mathrm{corr}}$ is converted to an observed velocity dispersion,
$\sigmavobs$, using the best-fit redshift.

% Error estimation
The errors on the observed velocity dispersions are estimated by
creating 500 realizations where the  spectra are perturbed according
to the corresponding error spectra.  We then perform the same fitting
and correction procedure on the perturbed spectra, and  convert the
corrected line widths to velocity dispersions using the best-fit
redshift of each realization.

% Example pstamp plot
\begin{figure*}
  \centering
  \includegraphics[width=0.95\textwidth]{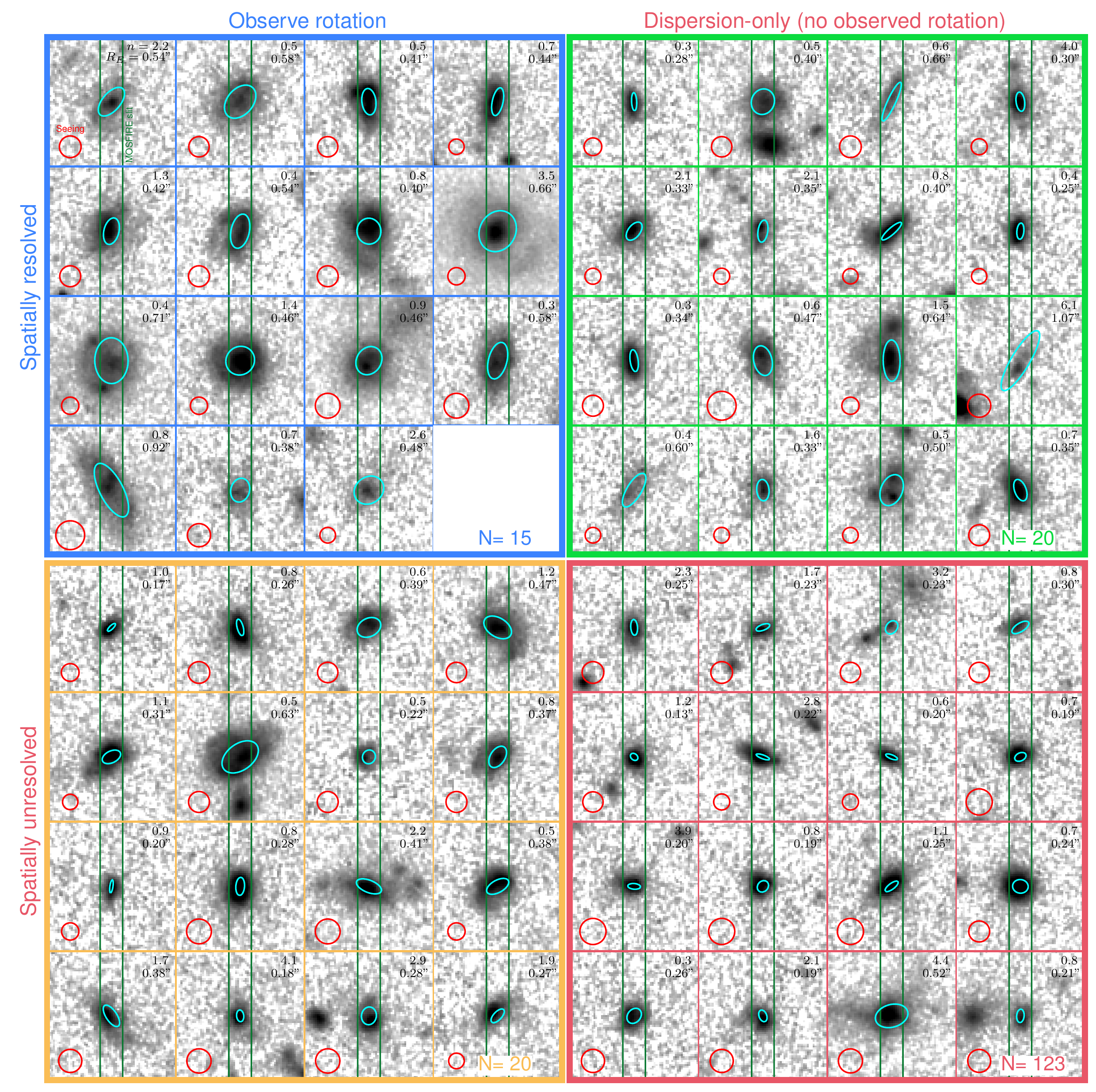}
  
  \caption{ \HST/F160W galaxy images ($4\arcsec \times
    4\arcsec$) for each of the 4 spatial/kinematic resolution
    categories: spatially resolved with observed rotation (upper left
    panel, blue), spatially unresolved with observed rotation (lower
    left, yellow),  spatially resolved with only dispersion observed
    (upper right, green), and spatially unresolved with only
    dispersion  observed (lower right, red). Each image is centered on
    the object center. The slit positions and orientations are shown
    with the green lines. We represent the \textsc{Galfit} effective
    radius ($\re$, measured from the major axis), axis ratio  ($b/a$),
    and position angle relative to the slit ($\deltPA$) for each
    object with the cyan ellipses. The values of $\re$ and  the
    S\'ersic index $n$ are annotated in the upper-right corners of the
    images. The atmospheric seeing FWHMs are  shown with the red
    circles. The total number of objects in each of the 4 categories
    is displayed in the lower right corner of each quadrant.  
    Spatially close companions of the target objects are located at different redshifts.
    }

  \label{fig:sample_pstamps} 
\end{figure*}

% Example spectra plot
\begin{figure*}
  \centering
  \includegraphics[width=0.95\textwidth]{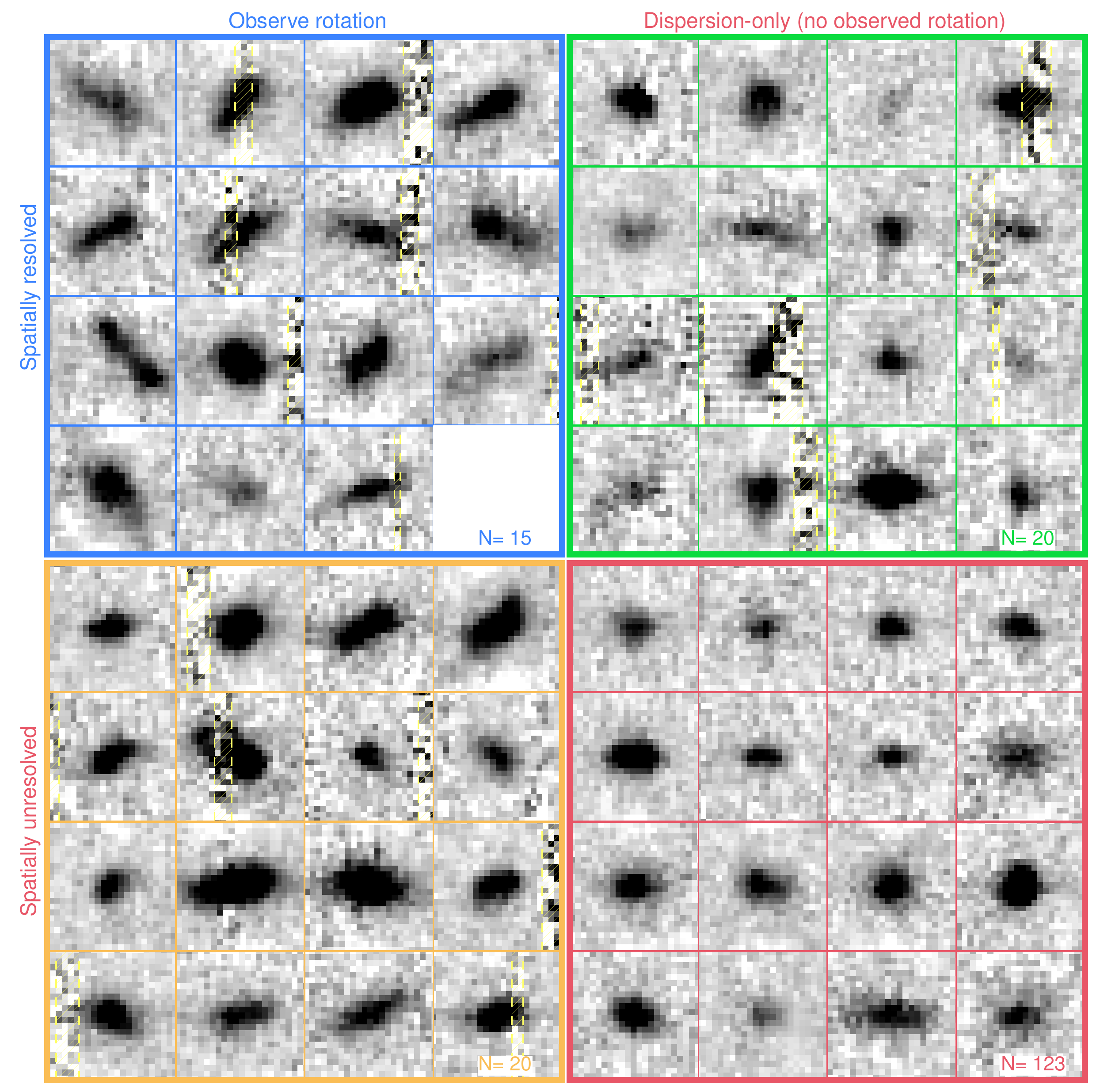}

  \caption{ MOSFIRE spectra centered on \Halpha for galaxies in the 4
    spatial/kinematic resolution categories. The displayed galaxies
    are the same as those shown in Figure~\ref{fig:sample_pstamps},
    with each object vertically centered on the same position and same
    spatial scale as the \HST/F160W
    images. The horizontal axis shows wavelength, where each stamp
    spans $\sim 36 \,$ \AA\ (H band) or $\sim 48 \,$\AA\ (K band) in the
    observed frame.}

  \label{fig:sample_spectra} 
\end{figure*}

% Aperture correction calculation
For each object, we convert the observed velocity dispersion into an intrinsic 
root mean square (RMS) velocity,  $V_{\rms} = \sqrt{V^2 + \sigmav^2}$, 
which explicitly includes both intrinsic
rotation and dispersion velocities. This method is discussed in detail
in Appendix~\ref{sec:appendix1D}. In summary, we model each object as
an inclined disk  (using the \textsc{Galfit} structural parameters
$\re$, $n$, $b/a$), with the major axis offset from the slit by
$\Delta \mathrm{PA}$. The rotation and velocity dispersion kinematics
are included by assuming a fixed ratio of $\vtosigre$, and then the
model is convolved to match the MOSFIRE seeing resolution.  
%%%%
%%%%
We determine which portions of the model fall within the slit width and
the extracted width in the spatial direction, and then apply the optimal-extraction weighting. 
For this model, we calculate the ratio of the luminosity-weighted second velocity moment 
($\sigmavmodel$) to the RMS velocity at \re 
($V_{\rms}(\re)_{\mathrm{model}}$), and use this ratio to
convert the observed, integrated velocity dispersion to the composite
RMS velocity at \re following
\begin{equation}
\Vrms (\re)_{\mathrm{1D, \, corr}}  = \sigmavobs
\left(\frac{\sigmavmodel}{
  V_{\rms}(\re)_{\mathrm{model}} } \right)^{-1}. 
\label{eq:vrms_1dcorr}
\end{equation}

%%%%%%%
\citet{vanDokkum15} use an $\alpha$ parameterization to infer a
rotational velocity from an observed velocity dispersion. This
$\alpha$ value is empirically derived and combined with an inclination
correction, with $\alpha = \sigmavobs/ \left(V \sin i
\right)$.  Hence, this correction does not take into account the exact
portion of the galaxy observed through the slit or partial support from random
motions. However, the galaxies by \citeauthor{vanDokkum15} are in general smaller than those in our sample, 
and will suffer less from slit losses.

% Observed sigma, VRMS vs R_E plot:
In Figure~\ref{fig:sig_obs_corr_re} we show $\sigmavobs$
and $\Vrms(\re)_{\mathrm{corr}}$ vs. $\re$ for galaxies and AGN without
detected rotation. 
For now, we assume $\vtosigre = 2.1$, the median
of the values measured for galaxies with detected rotation. For
reference, we also show the velocity dispersions measured from the
integrated 1D spectra of the galaxies with detected rotation, with
the corrections for these objects calculated using the exact
$\vtosigre$ measured for each object. 
The median observed 1D velocity dispersion for our sample with $M \geq 10^{9.5} M_{\odot}$ 
(the approximate completeness limit for star-forming galaxies in the MOSDEF survey, see 
\citealt{Shivaei15}) is 
$ \left(\sigmavobs\right)_{\mathrm{median}} = 78 \  \mathrm{km/s} $. 
The median observed velocity dispersion 
$\left(\sigmavobs \right)_{\mathrm{median}} = 70 \ \mathrm{km/s} $ 
of the galaxies at $ z\sim1.5 $ is slightly lower than the value of 
$ \left(\sigmavobs \right)_{\mathrm{median}} = 80 \ \mathrm{km/s}$ for the galaxies at 
$ z\sim2.3 $.

% ###############################################################################

% Delta PA vs R_E plot
\begin{figure}
  \begin{center}
    \hglue -4pt
    \includegraphics[width=0.48\textwidth]{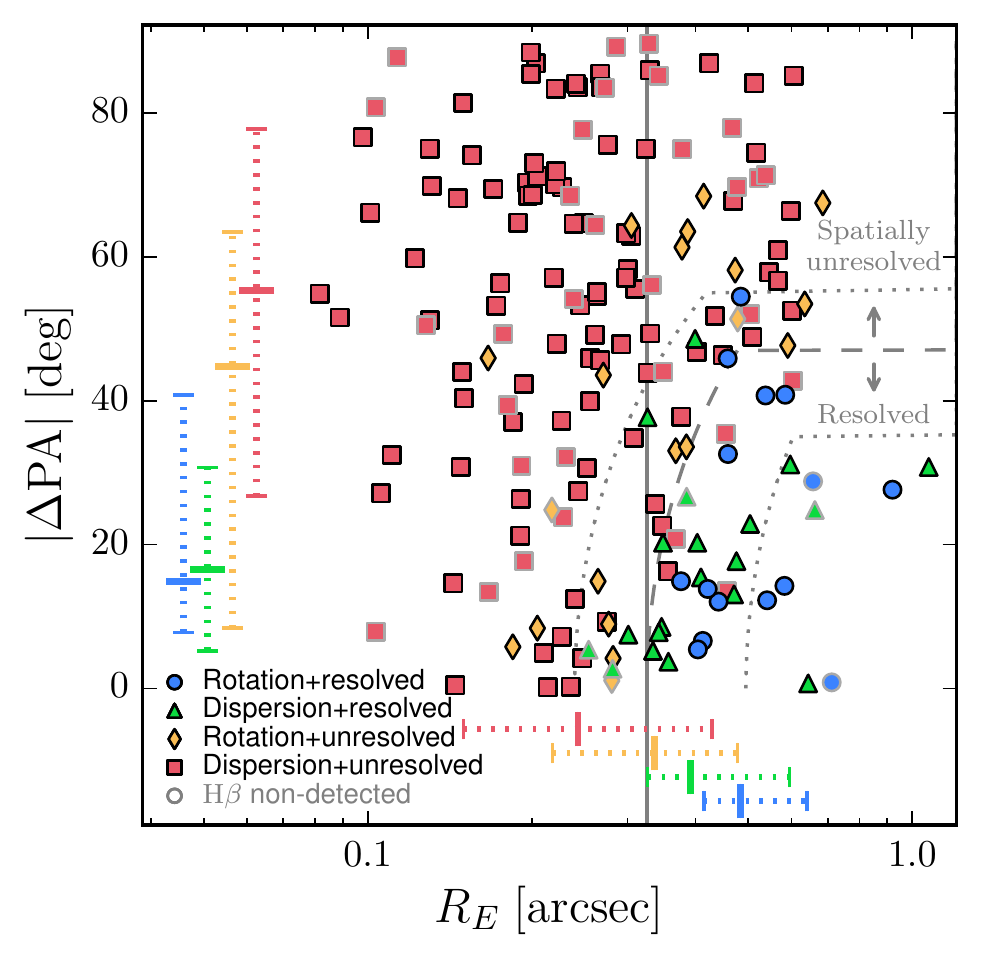}
    
    \caption{The position angle (PA) offset between the slit and
      photometric major axis versus the effective radius ($\re$) for
      galaxies in our sample. The color coding is based on whether we
      observe rotation or not (rotation vs. dispersion), and
      whether the projected major-axis is larger than the seeing size
      or not (spatially resolved vs. unresolved). The colored lines
      denote the median $\re$ and $|\deltPA|$ for each subsample, with
      the dotted lines showing the 68\%  value ranges. Galaxies
      without \Hbeta detections are marked with grey outlines.  The
      vertical grey line denotes 1/2 the median seeing FWHM
      ($0\farcs65$) for our sample. The dashed grey line shows the
      division between spatially resolved and unresolved objects
      described in Section~\ref{sec:resolved_unresolved}, assuming the
      median seeing. Objects to the right  and below this line would
      be classified as spatially ``resolved'' (assuming they were
      observed under the median  seeing conditions), while objects to
      the left and above would be classified as spatially
      ``unresolved''. 	The upper and lower dotted grey lines show the
      dividing lines if the seeing were equal to the minimum
      ($0\farcs48$) and maximum  ($0\farcs99$) effective seeing
      conditions, respectively (see Table 1, \citealt{Kriek15}).}
    \label{fig:galfit_params}
    
  \end{center}
\end{figure}

% ###############################################################################
\subsection{Resolved vs. unresolved kinematics}
\label{sec:resolved_unresolved}

% Intro: are our objects consistent with the assumption that they have rotation, but are just unresolved?
As our primary sample consists of star-forming galaxies, in Section
\ref{sec:disp_meas} we have treated  the objects for which we only
observe velocity dispersion  as being intrinsically disks, with some
amount of thickening.  Here we consider whether this is a reasonable
assumption by considering  the necessary conditions to observe
rotation in a galaxy.

% Reasons why objects may be unresolved
One reason why intrinsic rotation may not be observable is that the
galaxy is small with  respect to the seeing size. If there is only one
resolution element across the galaxy disk, then any rotation signature
will be  washed out and we would only observe velocity dispersion.
Additionally, the galaxy major axes may be misaligned with the slit axis.
A position angle (PA) offset reduces the ability to
detect rotation,  as for more misaligned objects, the rotational
information is projected into fewer resolution elements along the
slit.  At the most extreme, if a galaxy is completely misaligned with
the slit (i.e. $\deltPA = 90^{\circ}$), the rotational signature is
collapsed into the same resolution element, and again we would only
observe a velocity dispersion. 

% Which objects are can be spatially resolved? Projected size comparison
We consider the dual effects of object size and \deltPA, 
by calculating how much of the stellar light major axis falls
within the slit, projected along the slit direction. 
The projected size of the object falling within the slit ($2 \,
R_{E,\mathrm{proj}}$) should be larger than the seeing for the object
to be spatially resolved,  or $R_{E, \mathrm{proj}} \geq
\mathrm{FWHM}_{\mathrm{seeing}}/2$.

% Subsample sizes: 4 categories
We divide our sample into four categories, based on combination of the
projected spatial resolution criterion given above  (resolved
vs. unresolved) and whether we detect rotation or not (rotation
vs. dispersion,  see Section~\ref{sec:rot_meas}).  This classification
scheme gives 15 spatially resolved galaxies with observed rotation,
20 spatially unresolved galaxies with observed rotation,  20 spatially
resolved galaxies with only dispersion observed,  and 123 spatially
unresolved galaxies with only dispersion observed.  
In Figures \ref{fig:sample_pstamps} \& \ref{fig:sample_spectra} we
show example \HST{} images  and spectra, respectively, for
objects in each category.

% PA offset vs R_E plot and discussion 
We show PA offset versus object size ($\re$) as a function of seeing in Figure
\ref{fig:galfit_params}. 
% Trend with deltPA
At fixed $\re$, we tend to have more ``dispersion-only'' objects and
fewer objects with  detected rotation as the position angle offset
increases from aligned ($0^{\circ}$) to completely misaligned
($90^{\circ}$). 
% Trends in size:
We also note that on average, galaxies for which we observe 
rotation tend to be larger than those without observed rotation.
This finding supports the possibility that we may not observe rotation for
some objects simply because they are physically too small to resolve,
even if there is no position angle offset between the semi-major axis
and the slit.
% Comments based on example pstamps:
We can also see these trends in Figure~\ref{fig:sample_pstamps}, as
the objects with observed rotation (left panels)  tend to have better
alignment between their major axes and the slit,  and tend to have
larger angular sizes relative to the objects for which we only observe
dispersion (right panels).

% Issues with this classification: objects which don't follow our understanding!
However, not all objects follow the expected classifications.  For instance, we observe rotation for 
some spatially unresolved objects. These galaxies are shown as yellow diamonds in Figure
\ref{fig:galfit_params}, with example images and spectra shown in the
lower left panels of Figures \ref{fig:sample_pstamps} and
\ref{fig:sample_spectra}. Also, if all galaxies in our sample are
intrinsically disk galaxies and have at least partial rotational
support, we would expect to see rotation in all objects that are
spatially resolved. Yet we do not observe rotation in some of the
galaxies that meet the projected spatial resolution criterion, 
which are shown as green triangles in Figure~\ref{fig:galfit_params},
with examples in the upper right panels of Figures
\ref{fig:sample_pstamps} and \ref{fig:sample_spectra}.

% Issues with classification: possible sources of classification errors
Other effects may influence the classification of our sample into
these four sub-samples, which could explain why we see objects in the
unexpected classification categories. 
% delt PA estimation errors?
First, the position angle offset between the kinematic major axis and the slit could be incorrect. This error may be due to 
uncertainties in the photometric major axis position angle estimation or to a misalignment between the kinematic 
and photometric major axes. The latter effect has been observed by \citet{Wisnioski15} in galaxies at $z\sim2$, 
and possibly indicates disturbed kinematics due to mergers \citep{Epinat12}. 
If the position angle offset is incorrect, our projected size along the major
axis may not match the  true intrinsic projected size of the region
that we probe with the kinematics. Thus objects may scatter from the
``spatially resolved'' category into the ``spatially unresolved''
category or vice versa.

% Halpha size variations
Second, we use the rest-frame optical \re in this classification, but we
measure the kinematics from \Halpha emission. 
We show in Section~\ref{sec:light_profiles} that the
rest-frame optical and \Halpha sizes of the galaxies with detected
rotation are very similar, so using \re to determine spatial
resolution is a reasonable approximation. Still, if an object is
close to the detection limit, small differences between \re and
\Halpha size could change the  spatial resolution classification. 

% Inclination angle not included
Third, we do not incorporate inclination angle in our classification
procedure. For face-on galaxies, we do not expect to detect
rotation. If we consider galaxies with 
$ b/a \geq 0.9 $ (i.e., $ i \lesssim 26^{\circ} $, 
assuming $ (b/a)_0 = 0.19 $)
to be face-on, only 5 of our galaxies satisfy this criterion, three of which have detected
rotation, and two of which had been classified as ``spatially resolved''.  

% Also general problems with non-ideal Sersic fits.
Fourth, we rely on single-component \textsc{Galfit} fits to determine
the photometric position angle, axis ratio $b/a$,  and stellar
effective radius \re. Our galaxies often exhibit clumps, so they are
not perfectly fit by a smooth S\'ersic profile.  
Furthermore, \textsc{Galfit} is unable to recover extreme inclination angles 
(i.e., close to edge-on or face-on; \citealt{Epinat12}).
These limitations could further influence the accuracy of the position angle, 
effective radius, and axis ratio, and could influence whether an
individual galaxy is categorized as ``spatially resolved'' or not.

% S/N of spectra affecting kinematic classification
Fifth, the S/N of the observed spectra will influence the object
kinematic categorization.  Our $95\%$ one-sided $V(\re)$ detection
requirement may result in classifying objects with intrinsic 
rotation but low S/N spectra as ``dispersion-only.'' 
The example spectra in Figure~\ref{fig:sample_spectra} in
the spatially-resolved, dispersion-only quadrant  (green upper-right
panel) do appear to have either similar or lower signal-to-noise ratios 
relative to the  spectra of the objects with detected rotation (shown
in the left panels). 

% Skyline confusion in spectra
Sixth, neighboring skylines may overlap portions of a rotation curve,
which may also cause an object to fail the  $V(\re)$ detection
criterion. In Figure~\ref{fig:sample_spectra},  we see some objects in
the green quadrant with signifiant overlap with skyline contaminated
columns (i.e., the fourth object,  top row, and second object, third row, of the green, 
upper-right quadrant of Figure \ref{fig:sample_spectra}).

% Consistent with assumption of rotation?
Between the four categories, 89\% of the objects are consistent with
having rotation. This includes the 69\% of the galaxies that are unresolved and have
no detected rotation, for which the kinematic structures of the individual galaxies are unknown. 
In Section~\ref{sec:vtosig}, we find that galaxies without detected rotation 
are consistent with having kinematic support from both rotation and random motions. 
The only galaxies inconsistent with the assumption of
intrinsic rotational support are those that are spatially resolved
without observed rotation (green objects, Figure
\ref{fig:galfit_params}).  However, as mentioned previously, there are
sources of uncertainty in our categorizations that may imply these
objects may still have intrinsic rotation.

\subsection{Dynamical mass measurements}
\label{sec:mdyn_eff}

% RMS velocity
We combine the kinematic and structural information to calculate the 
dynamical masses of our galaxies.  The rotation and dispersion
velocities are combined by taking the RMS velocity, $\Vrms (\re) =
\sqrt{\sigmavint ^2 + V(\re)^2}$,  and we calculate the total
dynamical mass as
\begin{equation}
\label{eq:mdyn_eff}
\Mdyn = \keff \frac{\Vrms (\re)^2 \re}{G}, 
\end{equation}
where $G$ is the gravitational constant. 
% k_eff definition
Here we define an ``effective'' virial coefficient to account for the
relative contribution to the RMS velocity  from the rotation and
dispersion velocities (i.e., including an ``asymmetric drift''
correction from the velocity dispersion,  as in \citealt{Meurer96},
\citealt{Epinat09}, \citealt{Daddi10}, \citealt{Newman13}): 
\begin{equation}
\label{eq:k_eff}
\keff = \frac{\kdisp + \krot (\vtosigre)^2}{1+(\vtosigre)^2}, 
\end{equation}
where $\vtosigre = V(\re)/\sigmavint$.\footnote{Note that when
  $\vtosigre \rightarrow \infty$  (i.e. only rotational support), we
  have  $\Vrms (\re) = V(\re)$ and  $\keff \rightarrow \krot$, and
  Equation~\ref{eq:mdyn_eff} is equivalent to  the dynamical mass
  assuming only rotational support, $\Mdyn = \krot V(\re)^2 \re/G$ .
  In the opposite limit, when $\vtosigre = 0$ (i.e. only pressure
  support), we have  $\Vrms (\re)  = \sigmavint $ and $\keff =
  \kdisp$, and we recover the case for pressure-only support:  $\Mdyn
  = \kdisp \sigmavint^2 \re/G$.}  
We assume $\kdisp = 5$ as the virial coefficient corresponding to the  
dispersion kinematic component,  from the simple case of a sphere of uniform density
\citep{Pettini01}.  We estimate the virial coefficient for the
rotational kinematic component following \citet{Miller11},  who find
$k = 1.33$ for the dynamical mass within $r = 2.2 r_s =1.3 \re$.  To
convert to the total dynamical mass, we approximate $\krot \approx 2k
= 2.66$.

% Calculating Mdyn for 1D, 2D galaxies
To calculate $\keff$ and the dynamical masses,  we use the best-fit
values of $V(\re)$ and $\sigmavint$ measured in Section
\ref{sec:rot_meas}  for the galaxies with detected rotation.  For the
galaxies without observed rotation, we have to assume a value of
$\vtosigre$ to  calculate $\keff$,  $\Vrms (\re) = \Vrms
(\re)_{\mathrm{1D, \, corr}}$, and $\Mdyn$.  We will discuss the
assumption of $\vtosigre$ in Section~\ref{sec:results}.

%%%%%%%%%%%%%%%%%%%%%%%%%

% Mass calibration: rotation objects
\begin{figure}
  \begin{center}
    \includegraphics[width=0.48\textwidth]{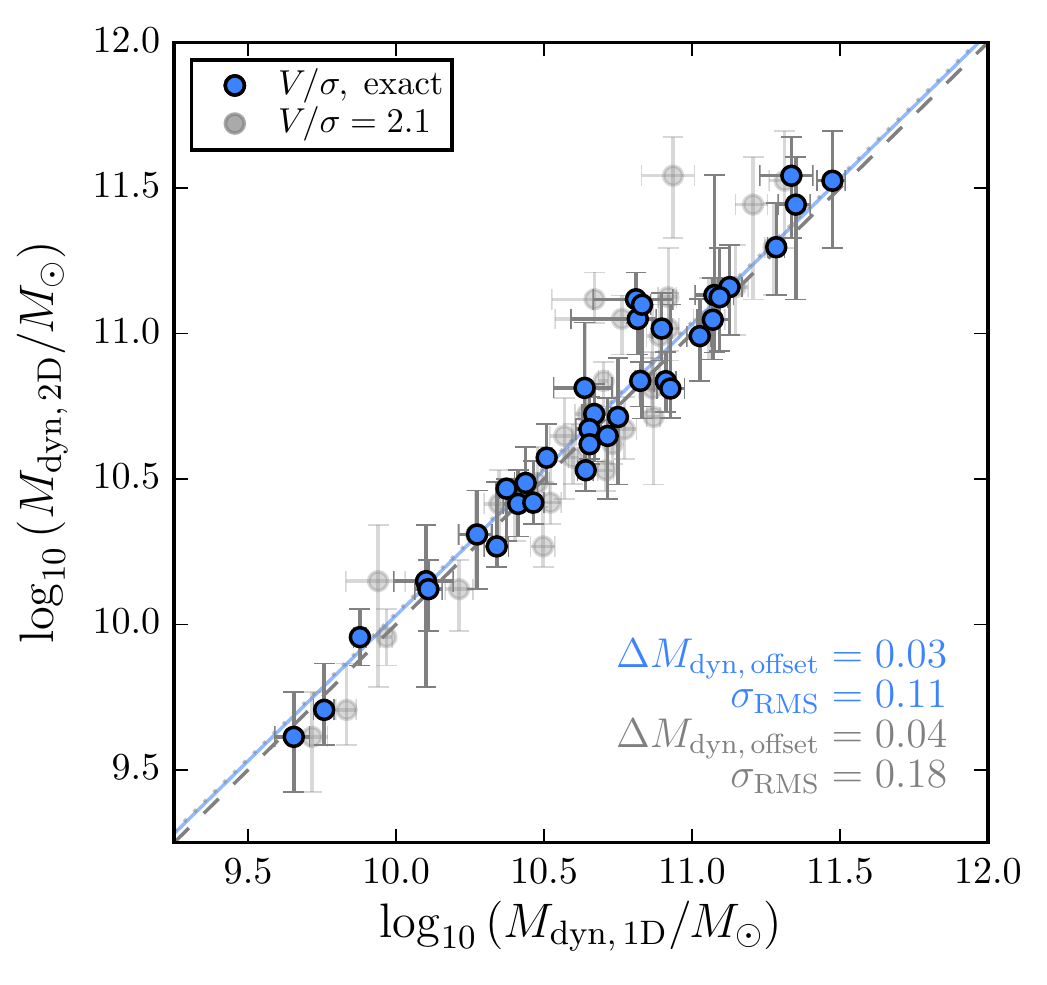}

    \caption{  Comparison of the dynamical mass measurement
      methods. The masses of the galaxies with observed 
      rotation are measured with both the resolved kinematic
      information ($M_{\mathrm{dyn, \, 2D}}$) and from the
      aperture-corrected optimally-extracted 1D-spectra
      ($M_{\mathrm{dyn, \, 1D}}$) assuming the exact value of
      $\vtosigre$ for each objects (blue circles).  There is little
      scatter ($\sigma_{\rms} = 0.11  \, \mathrm{dex}$)
      between the two measurements, and a small median offset $\Delta
      M_{\mathrm{dyn}} = 0.03 \, \mathrm{dex}$  (blue line), with
      $M_{\mathrm{dyn, \, 2D}}$ being slightly higher than
      $M_{\mathrm{dyn, \, 1D}}$.  If instead a constant $\vtosigre = 2.1$ 
      were assumed in calculating the 1D velocity dispersion
      corrections (grey circles) the median offset (dotted grey line) and scatter are slightly larger 
      ($\Delta M_{\mathrm{dyn}} = 0.04 \, \mathrm{dex}$, $\sigma_{\rms} = 0.18 \, \mathrm{dex}$). 
      Nonetheless, there is still excellent
      agreement between the measurements.  }
    \label{fig:mass_method_comparison}
	
  \end{center}
\end{figure}

% ###############################################################################

\subsection{Dynamical mass method comparison:\\spatially resolved galaxies}
\label{sec:aper_corr_test}

% Test correction method
We test the method for correcting the kinematics of disk galaxies
without detected rotation using the spatially-resolved rotation
sample of galaxies, for which we have more detailed kinematic
information.

% Calculation of Mdyn 1D, Mdyn 2D
First, we measure the dynamical masses $M_{\mathrm{dyn, \, 2D}}$ using
the exact values of $V(\re)$ and  $\sigmavint$ from the 2D spectral
fitting method in Section~\ref{sec:rot_meas}.  We then measure the
velocity dispersions from the optimally-extracted 1D spectra for the
same sample of galaxies.  We assume the exact $\vtosigre =
V(\re)/\sigmavint$ measured from the 2D spectral fitting for each
object  to calculate the corrected 1D velocity dispersions (Equation
\ref{eq:vrms_1dcorr}),  $\keff$ (Equation~\ref{eq:k_eff}), and the
resulting dynamical masses  $M_{\mathrm{dyn, \, 1D}}$.

% Figure comments
We compare $M_{\mathrm{dyn, \, 1D}}$ with $M_{\mathrm{dyn, \, 2D}}$ in
Figure~\ref{fig:mass_method_comparison}.  The corrected
$M_{\mathrm{dyn, \, 1D}}$ values are in excellent agreement with the
$M_{\mathrm{dyn, \, 2D}}$ values,  with little scatter between them
($\sigma_{\rms} = 0.11  \, \mathrm{dex}$).  We find a median
offset of $\Delta \log_{10} M_{\mathrm{dyn, offset}} = 0.03 \,
\mathrm{dex}$ between the two measurements,  such that the
2D-kinematic derived  values of \Mdyn are slightly larger than the
values derived from the aperture-corrected 1D spectra.

% Figure: if assuming const V/sigma
However, if these galaxies would not have been resolved, we would not
have known their intrinsic $\vtosigre$, to be used in the dynamical
mass estimate. Thus, we  also calculate $M_{\mathrm{dyn, \, 1D}}$
using the median $\left[\vtosigre \right]_{\mathrm{2D, median}} = 2.1$ 
for each object (shown as the grey points in Figure~\ref{fig:mass_method_comparison}).  
We find a slightly larger offset ($\Delta
\log_{10} M_{\mathrm{dyn, offset}} = 0.04  \, \mathrm{dex}$) and scatter  
($\sigma_{\rms} = 0.18  \, \mathrm{dex}$). 
Hence, for the galaxies with detected rotation,
assuming the average value of $\vtosigre$ for each object yields
dynamical masses that are nearly as accurate as the dynamical masses
derived from the rotation curves.  Based on this test, we conclude
that the 1D velocity dispersion correction method works well, and
should produce reasonable dynamical masses for the remainder of the galaxies 
without observed rotation if the average $\vtosigre$ is known.

%%%%%%%%%%%%%%%%%%%%%%%%%%%%%%%%%%%%%%%%%%%%%%%%%%%%%%%%%%
%%%%%%%%%%%%%%%%%%%%%%%%%%%%%%%%%%%%%%%%%%%%%%%%%%%%%%%%%%
%%%%%%%%%%%%%%%%%%%%%%%%%%%%%%%%%%%%%%%%%%%%%%%%%%%%%%%%%%
%%%%%%%%%%%%%%%%%%%%%%%%%%%%%%%%%%%%%%%%%%%%%%%%%%%%%%%%%%

\section{Results}
\label{sec:results}

% Intro
We now consider the total sample, combining the samples with and without observed rotation, 
and compare the dynamical and the baryonic masses, and assess the kinematic structures of star-forming
galaxies at $z\sim2$.

%%%%%%%%%%%%%%%%%%%%%%%%%%%%%%%%%%%%%%%

\subsection{Comparison of dynamical and baryonic masses}
\label{sec:mbar_mdyn}

% Dynamical masses: assumed V/sig for unresolved objects
In order to measure the dynamical masses for all galaxies in our
sample, we  need a $\vtosigre$ ratio for the galaxies without resolved
kinematics. We assume  that the kinematically resolved objects have a
similar structure as the unresolved objects, and adopt the median
$\vtosigre = 2.1$ as measured from the rotation objects  (see Section
\ref{sec:rot_meas} and Appendix~\ref{sec:appendix2D}).

\begin{figure}
  \centering
  \includegraphics[width=0.48\textwidth]{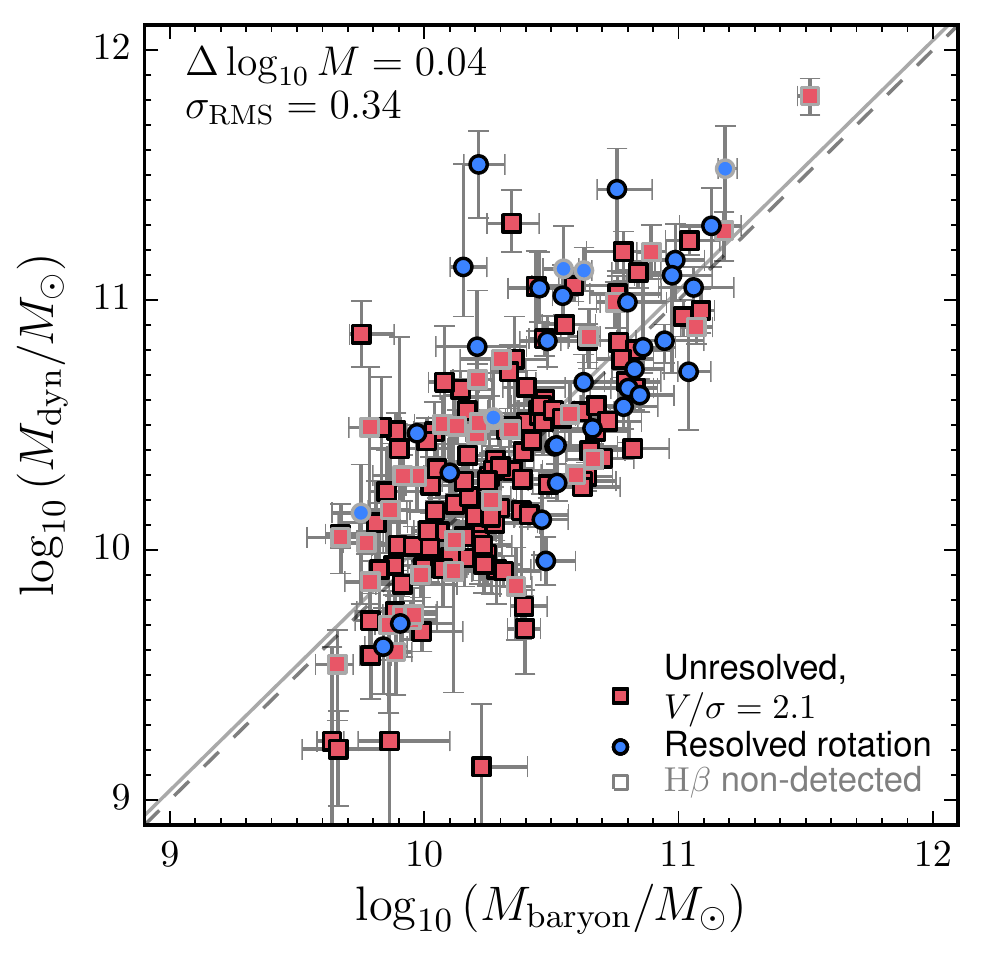}
  
  \caption{ Comparison of dynamical and baryonic (stellar + gas)
    masses for all galaxies in our sample. Symbols are similar as in
    Figure~\ref{fig:sig_obs_corr_re}. To calculate the velocity
    corrections for the sample without observed rotation, we assume
    the median $\vtosigre = 2.1$ from the sample with observed
    rotation.  The grey dashed line indicates equal \Mbar and \Mdyn,
    and the solid grey line indicates  the median offset of $\Delta
    \log_{10} M = \log_{10}\Mdyn- \log_{10}\Mbar$.  The scatter of the
    data around the median offset is $\sigma_{\rms} = 0.34 
    \, \mathrm{dex}$.  The error bars do not include the systematic
    uncertainties discussed in Section~\ref{sec:caveats}.  }

  \label{fig:mdyn_mbar}
\end{figure}

% Mbar vs Mdyn: Comment on trends in the figure.
In Figure~\ref{fig:mdyn_mbar} we compare the dynamical masses to the baryonic masses, $\Mbar =
\Mstar + \Mgas$ (as given in Section~\ref{sec:mosdef_survey}). 
They show a remarkable
agreement, with a median offset of  $\Delta \log_{10} M = 0.04  \,
\mathrm{dex}$, where \Mdyn is slightly larger than \Mbar at a given
\Mbar.  The scatter about the median $\Delta \log_{10} M$ is low, with
$\sigma_{\rms} = 0.34 \, \mathrm{dex}$. Additionally, objects
with and without detected rotation follow the same $\Mbar-\Mdyn$ relation, 
which may support the assumption that the galaxies are all intrinsically rotating disks. 
We will investigate this in more detail in the next section.

% Dark matter fraction discussion
The offset between the masses
implies a dark matter fraction within $\re$ of 8\%. 
This fraction is lower than the $\sim 30-50\%$ dark matter fractions
within 2.2 $r_s$ for disk galaxies at $z \sim 0$ (which increase with
decreasing stellar mass; \citealt{Pizagno05}, \citealt{Dutton11a}),
and the $\sim 20-30\%$ within $r<10\, \mathrm{kpc}$ at $z\sim 2$
\citep{ForsterSchreiber09}. This is expected, as these studies consider larger radii, 
and the dark matter fraction increases with increasing radius. Additionally, our measurement 
is dependent on several systematic uncertainties that we discuss in Section~\ref{sec:caveats}.

% Mdyn vs Mbar for all objects

%%%%%%%%%%%%%%%%%%%%%%%%%%%%%%%%%%%%%%%

%%%%%%%%%%%%%%%%%%%%%%%%%%%%%%%%%%%%%%%
% Mdyn vs Mbar for different physical condition assumptions
\begin{figure*}
  \centering
  \includegraphics[width=0.95\textwidth]{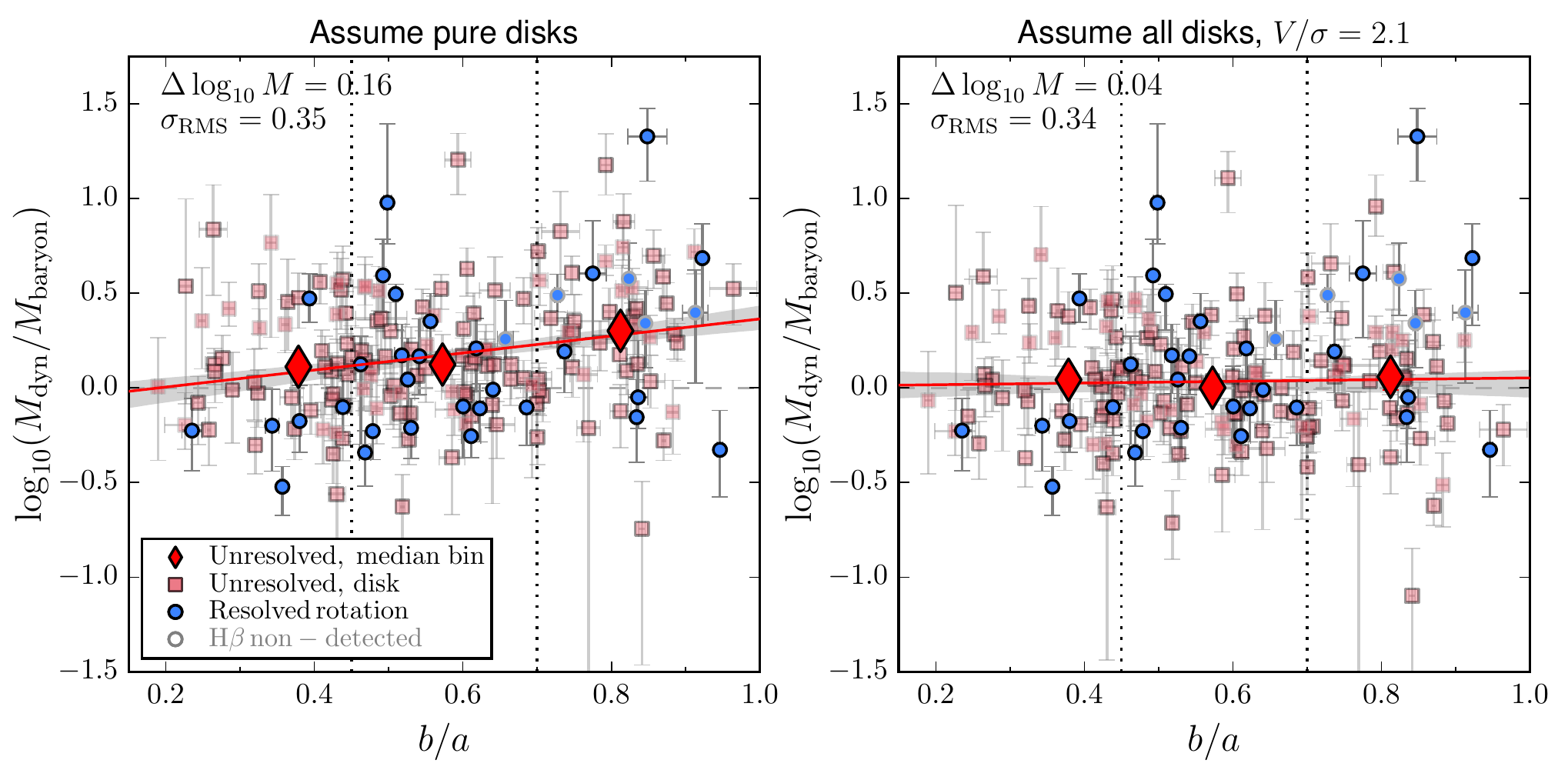}
  
  \caption{ The difference between the dynamical and baryonic masses
    $\Delta \log_{10} M$ vs. axis ratio $b/a$ under different
    assumptions of $\vtosigre$.  The galaxies with and without detected rotation
    are shown as blue circles and red squares, respectively. Galaxies
    without \Hbeta detections are marked with grey outlines. In the
    left and right panel, we assume no intrinsic velocity dispersion
    (i.e., $\vtosigre \rightarrow \infty$) and $\vtosigre = 2.1$,
    respectively, for the objects without observed rotation. 
    We show the median values of $\Delta
    \log_{10} M$ and $b/a$  for the objects without observed rotation 
    within bins of $b/a$ ($[0,0.45), [0.45, 0.7), [0.7,1]$) as red
    diamonds, and show the linear best-fit to the median points as
    the red line. The bin boundaries are shown as the black dotted
    lines.  }

	\label{fig:m_diff_splits}
\end{figure*}
%%%%%%%%%%%%%%%%%%%%%%%%%%%%%%%%%%%%%%%

%%%%%%%%%%%%%%%%%%%%%%%%%%%%%%%%%%%%%%%%%

\subsection{Rotational versus pressure support for unresolved galaxies}
\label{sec:vtosig}

% V/sigma: have resolved value, want independent value for unresolved sample.
In the previous section we simply assumed that all kinematically
resolved and unresolved objects have a similar median
$\vtosigre$. However, from Figure~\ref{fig:galfit_params} we know that
on average  the unresolved objects are smaller, and  thus they may be
structurally different. In this section we use the average properties
of the sample without resolved rotation to independently estimate the
average $\vtosigre$ for these objects.

% Demonstration of varying V/sigma for unresolved object: Figure
The effects of varying the $\vtosigre$ for all objects without observed rotation 
are demonstrated in Figure~\ref{fig:m_diff_splits}.  In the left panel, 
we show the assumption of $\vtosigre \rightarrow \infty$,
or no intrinsic velocity dispersion, for the objects without detected rotation. 
For comparison, we show the galaxies with detected rotation, using the the dynamical masses calculated from the 
rotation velocities and velocity dispersions measured from the 2D fitting procedure 
(see Section~\ref{sec:rot_meas}). 
There is a positive correlation between the mass offset
$\Delta \log_{10} M$ and  $b/a$ with Spearman correlation coefficient 
$\rho= 0.18$ at $2.2 \sigma$.  We quantify this trend by fitting a
line to median binned $\Delta \log_{10} M$ and $b/a$ in bins of $b/a$.
This trend of increasing \Mdyn relative to \Mbar as $b/a$ increases
indicates that we have over corrected for inclination, and that our
assumption of $\vtosigre \rightarrow \infty$ is too extreme.

% V/sigma constraint method for unresolved sample
If we instead assume lower values of $\vtosigre$, the inclination
correction will be reduced at higher $b/a$, resulting in a reduced
mass offset. Hence, we can constrain $\vtosigre$ for the galaxies without detected rotation 
by examining the offset  $\Delta \log_{10} M =
\log_{10} (\Mdyn/M_{\odot}) - \log_{10} (\Mbar/M_{\odot})$ versus the
axis ratio, $b/a$ over a range of  assumed $\vtosigre$ values.  For
each assumed $\vtosigre$, we calculate the dynamical masses for the
galaxies without detected rotation. We then measure  the $\chi^2$ between
the data and the value if there were no trend with $b/a$, or a
constant $\Delta \log_{10} M$ equal to the median of the $\Delta \log_{10} M$ values.
We determine the best-fit $\vtosigre$ by minimizing the $\chi^2$ statistic.

% Compare V/sigma for resolved objects and the avg value from unresolved objects.
We find a best-fit $\vtosigre = 2.1 _{-0.3}^{+0.2}$  for the objects without detected rotation. 
We show the effects of assuming this $\vtosigre$ value in the right panel of Figure
\ref{fig:m_diff_splits}. When we adopt  $\vtosigre = 2.1$, we notice
very little offset in $\Delta \log_{10} M$ as a function of $b/a$, and
the total scatter in  \Mdyn-\Mbar is also slightly lower. We note that this
measurement reflects an estimate of the typical $\vtosigre$ of this
sample; the scatter in the \Mdyn-\Mbar relation also includes 
variations introduced by a range of intrinsic $\vtosigre$ values for
the sample without observed rotation. The typical $\vtosigre$ of the
galaxies with and without observed rotation are identical, suggesting
that on average all galaxies have similar support from random
motions.

\subsection{Trends between $V/\sigma$ and other galaxy properties}
\label{sec:structure_interp}

% General intro to V/sigma
The ratio $\vtosig$ is a measure of the amount rotational support
versus support from random motions, and thus provides information
about the structure of a galaxy. Higher $\vtosig$ are indicative of
thin disks, while lower $\vtosig$ values may indicate thicker disks or high gas turbulence. 
To understand the physical processes setting the internal structure of star-forming galaxies 
at $z\sim2$, we investigate the relation between $\vtosig$ and other galaxy properties.

%%%%%%%%%%%%%%%%%%%%%%%%%%%%%%%%%%%%%%%

% V/sigma vs lmass, lSSFR_Ha plot
\begin{figure*}
  \centering
  \hglue -10pt
  \includegraphics[width=0.95\textwidth]{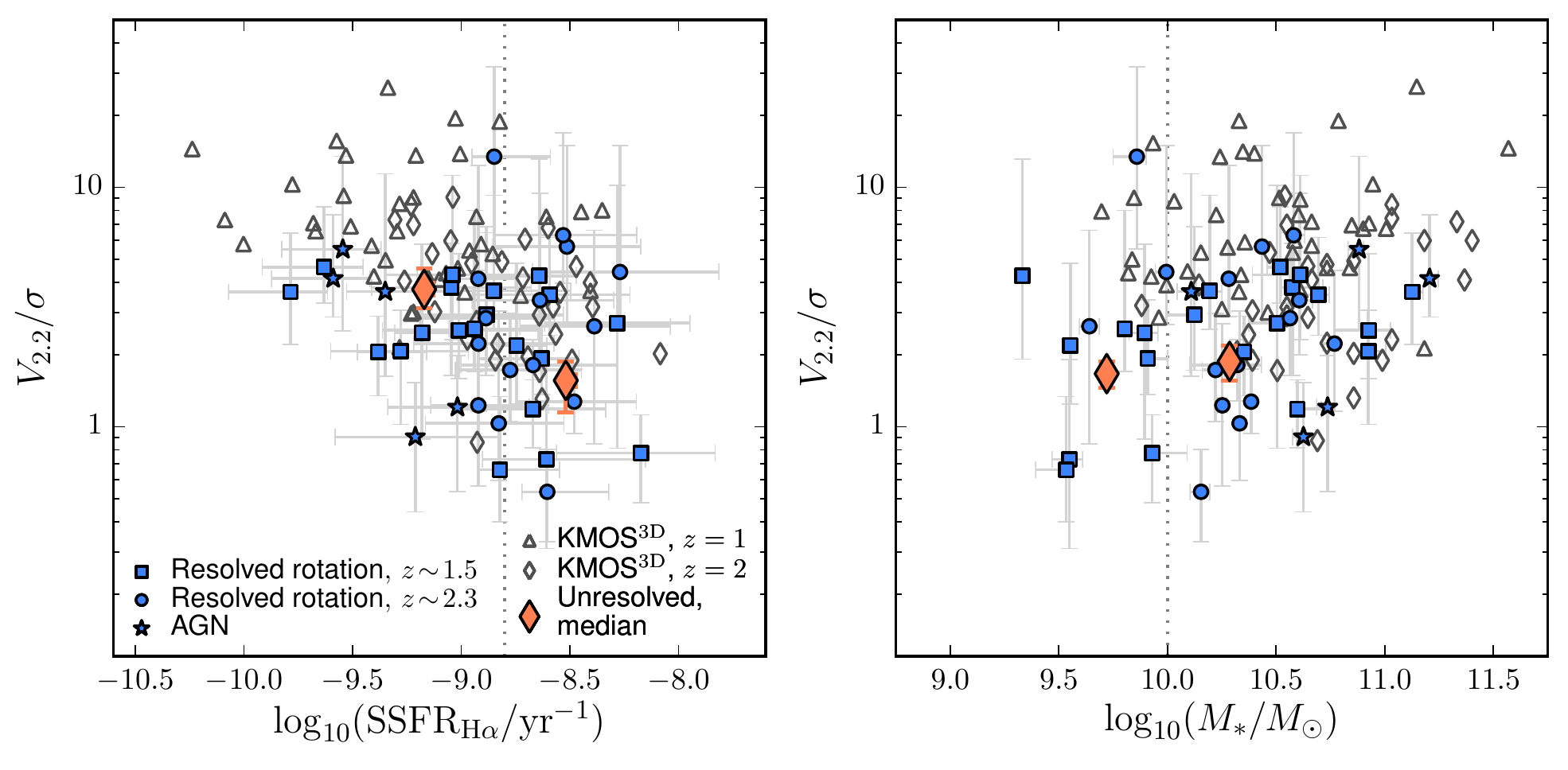}

  \caption{The ratio of support from rotation and random
      motions $\vtosigtt$ vs.  \Halpha SSFR (left panel) and stellar mass (right panel) 
      for our sample of galaxies with detected rotation. 
      Galaxies at $z\sim1.5$ and $z\sim 2.3$ are shown with blue squares and circles, respectively. 
      We also show the AGN with detected rotation as blue stars.  
      For the galaxies without observed rotation, we show the median $ \vtosigtt $ in 
      bins of \Halpha SSFR and stellar mass (orange diamonds). 
      The bin boundaries are shown by the vertical grey dotted lines.  
      For comparison, we also show the \KMOSTD values \citep{Wisnioski15}
      at $z= 1$ and $z = 2$ as the  grey open triangles and diamonds,
      respectively. When considering both our data and the
      \KMOSTD data, there is a trend of decreasing $V/\sigma$ with increasing SSFR, 
      consistent with the trend found by  \citet{Wisnioski15}. 
      This trend may reflect disk settling, with the velocity
      dispersions decreasing as the gas fractions decrease. }

	\label{fig:vtosig_lmass_lssfr}
\end{figure*}

% V/sigma vs lmass, SSFR
We show the measured $\vtosigtt = V(2.2 r_s)/\sigmavint$ values for our sample of galaxies as a
function of \Halpha specific star formation rate
(SSFR) and stellar mass in Figure~\ref{fig:vtosig_lmass_lssfr}.  
For the galaxies without observed rotation, we show the median $\vtosigtt$ values in bins of 
\Halpha SSFR and stellar mass, calculated using the method described 
in Section~\ref{sec:vtosig} and assuming $ r_t = 0.4 r_s $. 
% Comparison with other studies,  possible systematic difference with definition of Sigma_vint
For comparison, we also show the \KMOSTD results of \citet{Wisnioski15} at 
$z=1$ and $z=2$ as the grey open triangles and diamonds, respectively. 
We note that \citeauthor{Wisnioski15} measure the rotation velocity from the difference between the observed 
minimum and maximum velocities along the major kinematic axis. 
However, we do not directly constrain the flat portions of the rotation curves, so instead we consider 
$V_{2.2} = V(r=2.2 r_s)$ -- the radius at which an exponential rotation curve peaks -- to attempt to provide a 
reasonable comparison with existing measurements.
We also note that we assume a constant $\sigmavint$, while \citeauthor{Wisnioski15} measure 
$\sigma_0$ in the outer portions of the galaxies.

% Interpretation:
We see a trend of decreasing $\vtosigtt$ with increasing \Halpha
SSFR, especially when considering the binned measurement for galaxies without observed rotation 
and when adding the results by \citet{Wisnioski15}. As suggested by \citet{Wisnioski15}, this trend 
may reflect disk settling, where galaxies with lower gas fractions (and lower SSFRs)
have lower velocity dispersions.  For the galaxies without observed rotation, this trend may
also reflect a higher fraction of dispersion dominated galaxies at high SSFRs. 
Our measured $\vtosigtt$ values do not show a trend with stellar mass,
but when considering our data together with the results of
\citet{Wisnioski15}, we may see a weak trend of increasing $\vtosig$ with
increasing stellar mass.  

% Caveat
When measuring the kinematics, we assume that the velocity dispersion is
constant. If  the true velocity dispersion profile rises towards the
center of a galaxy (as discussed in Section~\ref{sec:caveats}),  our
measured $\sigmavint$ may be systematically higher than what would be
measured in the outer portions of our galaxies.  This would
systematically shift our measured $\vtosigtt$ to lower values than those
found by \citet{Wisnioski15}.  Thus the trends of $\vtosig$ with
\Halpha SSFR and stellar mass found from the combined samples may be
partially  due to a systematic difference between the adopted
$\vtosig$ values.

%%%%%%%%%%%%%%%%%%%%%%%%%%%%%%%%%%%%%%%%%%%%%%%%%%%%%%%%%%
%%%%%%%%%%%%%%%%%%%%%%%%%%%%%%%%%%%%%%%%%%%%%%%%%%%%%%%%%%
%%%%%%%%%%%%%%%%%%%%%%%%%%%%%%%%%%%%%%%%%%%%%%%%%%%%%%%%%%

% Mstel-Mbar-Mdyn plot
\begin{figure*}
  \centering
  \includegraphics[width=0.95\textwidth]{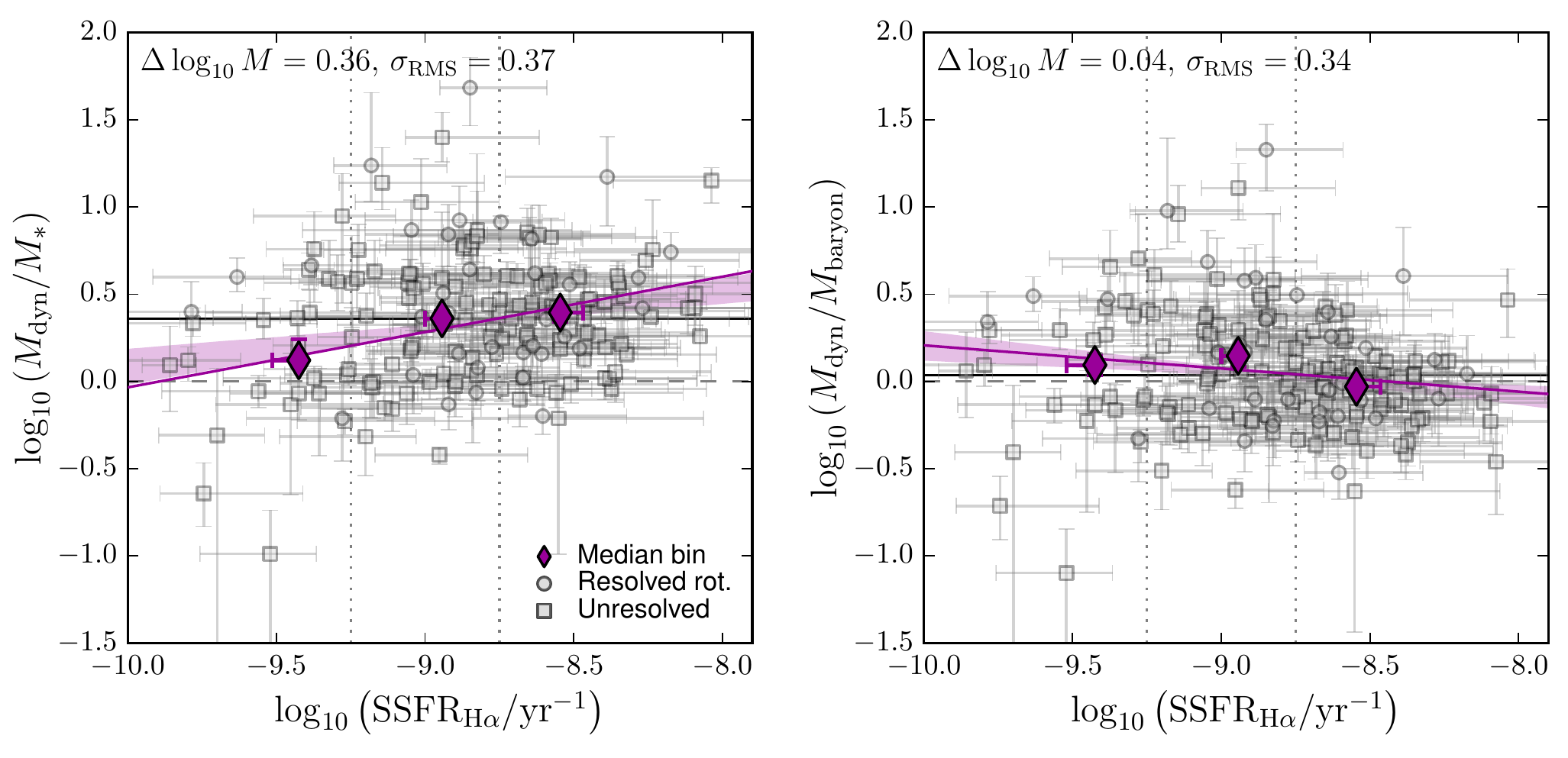}
  
  \caption{ Difference between the dynamical and the stellar (left)
    and baryonic masses (right),  versus the \Halpha SSFR.  The
    objects with observed rotation are shown as circles, and the objects
    without observed rotation as squares. The median $\Delta
    (\log_{10} M)$ for the entire sample is shown with the black line.
    The data are binned in SSFR$_{\Halpha}$, with the median values
    shown as the purple diamonds, and  a linear fit to the median
    points is shown with the purple line.  In the left panel, we see a
    larger offset between the dynamical and stellar masses for
    galaxies with higher SSFR than those with lower SSFR.  This
    offset is reduced when instead we compare the dynamical and
    baryonic masses.  Additionally, the scatter between the mass
    values is larger when comparing the dynamical and the stellar
    ($\sigma_{\rms} = 0.37 \, \mathrm{dex}$) rather than the
    baryonic ($\sigma_{\rms} = 0.34 \, \mathrm{dex}$)
    masses. This figure illustrates the importance of including gas
    masses in the mass comparison.}
    
  \label{fig:mstel_mbar_mdyn}
\end{figure*}

\section{Discussion}
\label{sec:discussion}

% Layout of discussion
In this section we analyze how different assumptions and caveats may
influence our results. In Section~\ref{sec:mstel_vs_mdyn}, we
consider the results if instead of baryonic masses  we had compared
just stellar and dynamical masses. Section~\ref{sec:assume_ellip}
presents the implications of treating our galaxies as 
dispersion-dominated galaxies. 
In Section~\ref{sec:IMF}, we discuss constraints on the IMF based on
our measured \Mbar and \Mdyn values. In Section~\ref{sec:tullyfisher},
we use our data to investigate the modified $S_{0.5}$ Tully Fisher relation (e.g., \citealt{Kassin07}). 
We compare the $\Mbar-\Mdyn$ relation of the selected AGN 
to the relation measured for our star-forming galaxy sample in Section~\ref{sec:mbar_mdyn_agn}.  
Section~\ref{sec:light_profiles}
examines differences between the stellar continuum and \Halpha
intensity profiles.  Finally, we discuss caveats to assumptions we
have made in Section~\ref{sec:caveats}.

%%%%%%%%%%%%%%%%%%%%%%%%%%%%%%%%%%%%%%%%%%%%%%%%%%%%%%%%%%
\subsection{Importance of including gas in comparisons to dynamical masses}
\label{sec:mstel_vs_mdyn}

% Why should we include gas masses?
Previous studies have compared stellar masses to dynamical masses,
especially  using the stellar kinematics of quiescent galaxies (e.g.,
\citealt{ANewman10}, \citealt{Taylor10a}, \citealt{Bezanson13},
\citealt{vandeSande13},  \citealt{Belli14a}). For quiescent galaxies
this may be a fair comparison, but for high redshift star-forming galaxies the gas mass
is found to be a non-negligible fraction of the total baryonic mass (e.g., \citealt{Tacconi13}). 

% Comparison of Mstar-Mdyn vs Mbar-Mdyn: Mdyn-Mstar vs SSFR
To assess this finding, we consider the difference between the dynamical masses and the
stellar masses alone versus \Halpha SSFR in the left panel of  Figure
\ref{fig:mstel_mbar_mdyn}.  There is an offset between the masses,
with the dynamical masses generally exceeding the stellar masses.
When we split the sample into bins of \Halpha SSFR ($\log_{10}
(\mathrm{SSFR_{\Halpha}}) < -9.25$,  $-9.25 \leq \log_{10}
(\mathrm{SSFR_{\Halpha}}) < -8.75$, $\log_{10}
(\mathrm{SSFR_{\Halpha}}) \geq -8.75$),  we find that the higher SSFR
bins have larger offsets between the dynamical and stellar masses than
the lower SSFR bins.  This result is not surprising, as the median
inferred ratios of  $f_{\mathrm{gas}} = \Mgas/(\Mstar+\Mgas) $ are
$f_{\mathrm{gas}} = 0.22$, $f_{\mathrm{gas}} = 0.44$, and
$f_{\mathrm{gas}} = 0.58$  for the lowest to highest SSFR bins,
respectively.

% Mdyn-Mbar vs SSFR
In contrast, the baryonic masses (right panel of Figure
\ref{fig:mstel_mbar_mdyn})  show a much weaker SSFR-dependent offset
with respect to the dynamical masses. The $\Mdyn/\Mbar$ -- SSFR relation also
show a smaller observed scatter ($\sigma_{\rms} = 0.34 \,
\mathrm{dex}$) than the $\Mdyn /\Mstar$ -- SSFR relation ($\sigma_{\rms} =
0.37 \, \mathrm{dex}$). 
The agreement between the baryonic and dynamical masses, as well as the larger $\Mdyn /\Mstar$ 
offset for higher SSFR galaxies, suggests that our \Halpha SFR-based gas masses
are reasonable estimates of the true values.

% Summary of why Mgas is important, and why one should use Mbar in comparisons.
Thus, at least for star-forming galaxies, it is important to include
the gas masses in the total baryonic mass when comparing it against the dynamical
masses.

%%%%%%%%%%%%%%%%%%%%%%%%%%%%%%%%%%%%%%%%%%%%%%%%%%%%%%%%%%
\subsection{What if we assume that all galaxies without detected rotation are early-type galaxies? }
\label{sec:assume_ellip}

% What if the unresolved galaxies were all dispersion-dominated?
As our sample consists of star-forming galaxies, we have assumed that
there is some amount of rotational support for all  galaxies, even for
those that are not kinematically resolved.  However, it might be the
case that some, if not all, of these objects without observed rotation are early-type galaxies. 
To assess this possibility, we
derive dynamical masses assuming instead that all the kinematically
unresolved galaxies are dispersion-dominated ellipticals or lenticulars (S0s).

% Mass plot assuming all dispersion-dominated, beta(n), CIRCULARIZED effective radii
\begin{figure}
  \centering
  \includegraphics[width=0.48\textwidth]{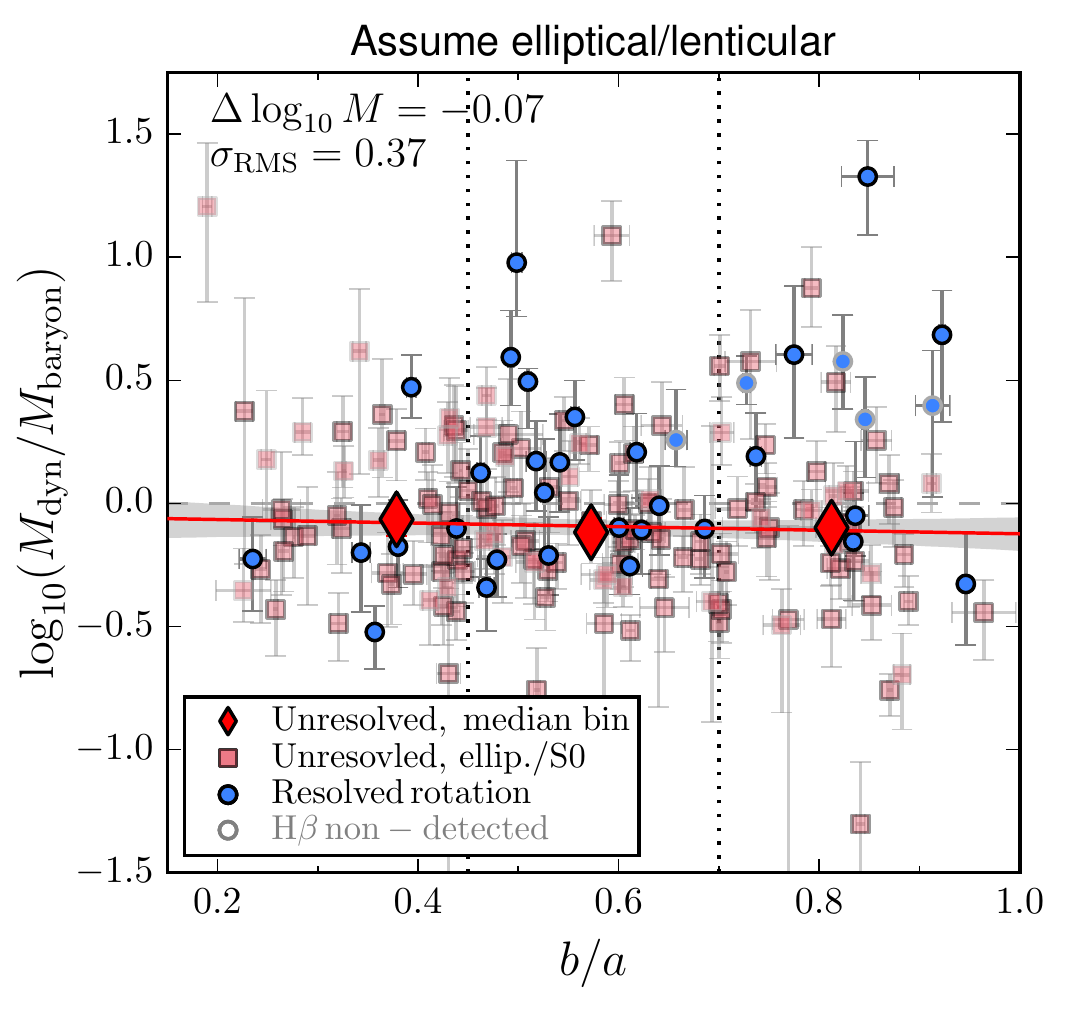}

  \caption{ 
    The difference between dynamical and baryonic masses
    vs. axis ratio $b/a$  assuming that the galaxies without observed 
    rotation are ellipticals or lenticulars (S0s).  The observed velocity dispersions for
    the galaxies without apparent rotation are corrected using a
    modified aperture correction based on the method by
    \citet{vandeSande13}.  The dynamical masses are then calculated
    using virial coefficient $\beta(n)$ and the circularized effective radii
    $R_{E,\mathrm{circ}}$. The plot point and line definitions are the
    same as in Figure~\ref{fig:m_diff_splits}. 
    The S\'ersic dependent virial coefficient combined with the circularized radius 
    results in no trend with axis ratio, similar to the effect of applying inclination corrections 
    when assuming the galaxies have partial rotational support. 
    }

  \label{fig:mdyn_elliptical}
\end{figure}

% Calculation of Mdyn for dispersion-dominated galaxies.
We calculate the dynamical masses  as 
\begin{equation}
\Mdyn = \beta(n) \frac{ \sigma_{e, \mathrm{corr}}^2
  R_{E,\mathrm{circ}} }{ G } .
\label{eq:mdyn_disp}
\end{equation} 
Here we use the S\'ersic index dependent virial coefficient, $\beta (n)$, given in
\citet{Cappellari06}, where $\beta(n)$ is a quadratic function of
$n$ (e.g., $\beta(n=1) \sim 8$ and $\beta (n=4) \sim 6$). 
The observed integrated velocity dispersions ($\sigmav$) are corrected 
(to $\sigma_{e,\mathrm{corr}}$) using an aperture correction similar to
that presented in \citet{vandeSande13}, modified to include the axis
ratio and position angle offset relative to the slit. When calculating the dynamical masses, we use the
circularized effective radii,  $R_{E,\mathrm{circ}} = R_E \sqrt{b/a}$,
with $\re$ the \textsc{Galfit} effective radius (which is equal to the semi-major axis) and $b/a$ the axis ratio.
It is necessary to use $R_{E,\mathrm{circ}}$ in this case, as $\beta (n)$ is derived using
circularized radii.  The resulting mass difference $\Delta \log_{10} M$ as a function of axis ratio, $b/a$, 
is shown in Figure~\ref{fig:mdyn_elliptical}.

%%%%%%%%%%%%%%%%%%%%%
% IMF offset
\begin{figure*}
  \centering
  \includegraphics[width=0.95\textwidth]{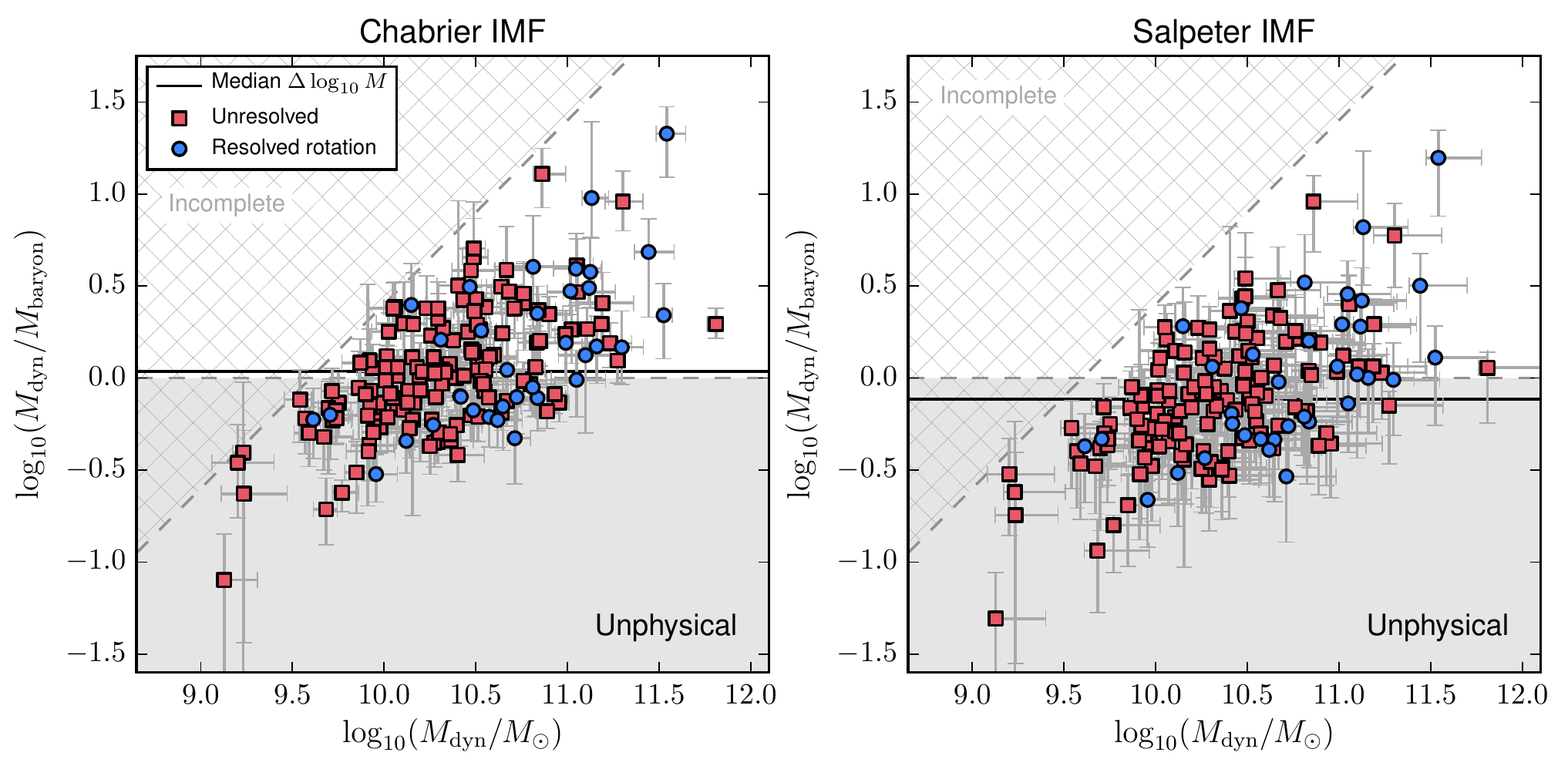}

  \caption{ $\log_{10}(\Mdyn/\Mbar)$ vs. $\log_{10} \Mdyn$ for a
    Chabrier (left panel) and a Salpeter IMF (right panel).  The
    galaxies with and without detected rotation are shown
    with the blue and red points, respectively.   
    In each panel, the zero-point $\log_{10} (\Mdyn/\Mbar) = 0$ is presented by the
    dotted grey line, and the median $\Delta \log_{10} M = \log_{10}
    (\Mdyn/\Mbar)$ of the whole galaxy sample by the solid black
    line. The unphysical region where $\Mdyn < \Mbar$ is shaded grey.
    The mass incomplete region ($\Mbar \lesssim 10^{9.6} M_{\odot}$) is shown  with the grey hatched
    region. A Chabrier IMF is consistent with our measurements, while
    a Salpeter IMF is not; this IMF shifts the median to the region
    where $\Mdyn < \Mbar$, which is unphysical.  }

  \label{fig:IMF}

\end{figure*}

%%%%%%%%%%%%%%%%%%%%%

% Figure analysis: good agreement, no trend with b/a
Under these assumptions, the dynamical masses are also in reasonable
agreement with the baryonic masses  ($\Delta \log_{10} M = -0.07 \,
\mathrm{dex}$, $\sigma_{\rms} = 0.37 \, \mathrm{dex}$),
though the mass offset is somewhat lower than measured for the galaxies
with detected rotation.  Remarkably, there is no trend of $\Delta
\log_{10} M$ with $b/a$, which is due to the circularization of the
effective radii.  For objects with smaller axis ratios $b/a$,  the
circularized radii will be smaller, and thus the dynamical masses will
be lower than if they were calculated using  non-circularized radii.  
% Implications: 
Thus, the S\'ersic dependent virial coefficient, combined with the
circularized radius has a similar effect as the inclination correction
applied when assuming that the galaxies are primarily supported by rotation.

% Why "one plane for all" may work while treating all galaxies as dispersion-dominated
The agreement between the dynamical masses for the two separate
methods may explain why both star-forming and quiescent galaxies,
when treated as early-type galaxies,  
share ``one mass fundamental plane,''  as found by \citet{Bezanson15}.
In this work both star-forming and quiescent galaxies fall on the same
surface in the 3D parameter space defined by  effective radius \re,
velocity dispersion $\sigmav$, and stellar mass surface density
$\Sigma_*$.  We note that when assuming all unresolved galaxies are
early-type galaxies, the median offset $\Delta \log_{10} M$ is negative,  
with the typical galaxy lying in the unphysical regime where $\Mdyn < \Mbar$.  
This result could suggest that more accurate dynamical masses are
obtained when assuming the galaxies are rotationally supported. However, both the baryonic and
dynamical masses are subject to systematic uncertainties, and thus we
cannot definitively state whether the unresolved galaxies are rotationally supported or not.

%%%%%%%%%%%%%%%%%%%%%%%%%%%%%%%%%%%%%%%%%%%%%%%%%%%%%%%%%%

\subsection{Stellar initial mass function constraints}
\label{sec:IMF}

% Implications of IMF on stellar, gas masses
In this section we consider the implications of the measured \Mbar and
\Mdyn values of our galaxy sample for the stellar IMF.  In estimating
the baryonic masses, we have assumed a \citet{Chabrier03} IMF. An
alternative choice is a \citet{Salpeter55} IMF,  which would result in higher stellar
masses, by a factor of 1.8 (e.g., \citealt{Erb06c}). 
The relations between $L(\Halpha)$ and SFR, and $\Sigma_{\mathrm{SFR}}$ 
and $\Sigma_{\mathrm{gas}}$  given by \citet{Kennicutt98} both 
contain an IMF dependence, but the relation between the intrinsic 
$L(\Halpha)$ and $\Sigma_{\mathrm{gas}}$ does not depend on the IMF, 
so the gas masses do not change whether using a Salpeter or a 
Chabrier IMF.

% Plot overview
We show the implications of assuming a Salpeter rather than a Chabrier IMF in Figure~\ref{fig:IMF}. 
The median $\log_{10}(\Mdyn/\Mbar)$ for the whole sample is shown as 
the solid black line.  If we had instead adopted a Salpeter IMF 
(right panel), the zero-point of $\log_{10} (\Mdyn/\Mbar)$ is lower. 

% Interpretation of plot: Salpeter is inconsistent with our measurements
Physically, we expect that the dynamical mass, which traces all mass
in a galaxy, must at least be as large  as the observed baryonic mass,
depending on the dark matter fraction within an effective radius.  The
median $\log_{10} (\Mdyn/\Mbar)$ we measure assuming a Chabrier IMF is
consistent with this expectation, with 46\% of the 
galaxies fall within the unphysical regime. However, for a
Salpeter IMF 63\% of the galaxies fall within the unphysical
regime where $\Mdyn < \Mbar$. Thus a Salpeter IMF is inconsistent
with our measured values of \Mbar and \Mdyn.   This inconsistency with
a Salpeter IMF is in agreement with the findings of other studies of
star-forming, disk galaxies  (\citealt{Bell01}, \citealt{Tacconi08},
\citealt{Dutton11a}, \citealt{Brewer12}). 
% Systematic uncertainties
Nonetheless, our result is subject to potential systematic
uncertainties that might decrease the measured dynamical masses, such
that our measurements would be inconsistent with both a \citet{Salpeter55}
and a \citet{Chabrier03} IMF. We discuss these systematic
uncertainties in detail in Sections \ref{sec:light_profiles} and
\ref{sec:caveats}.

% M*-Mbar-S0.5
\begin{figure*}
	\centering
	\includegraphics[width=0.95\textwidth]{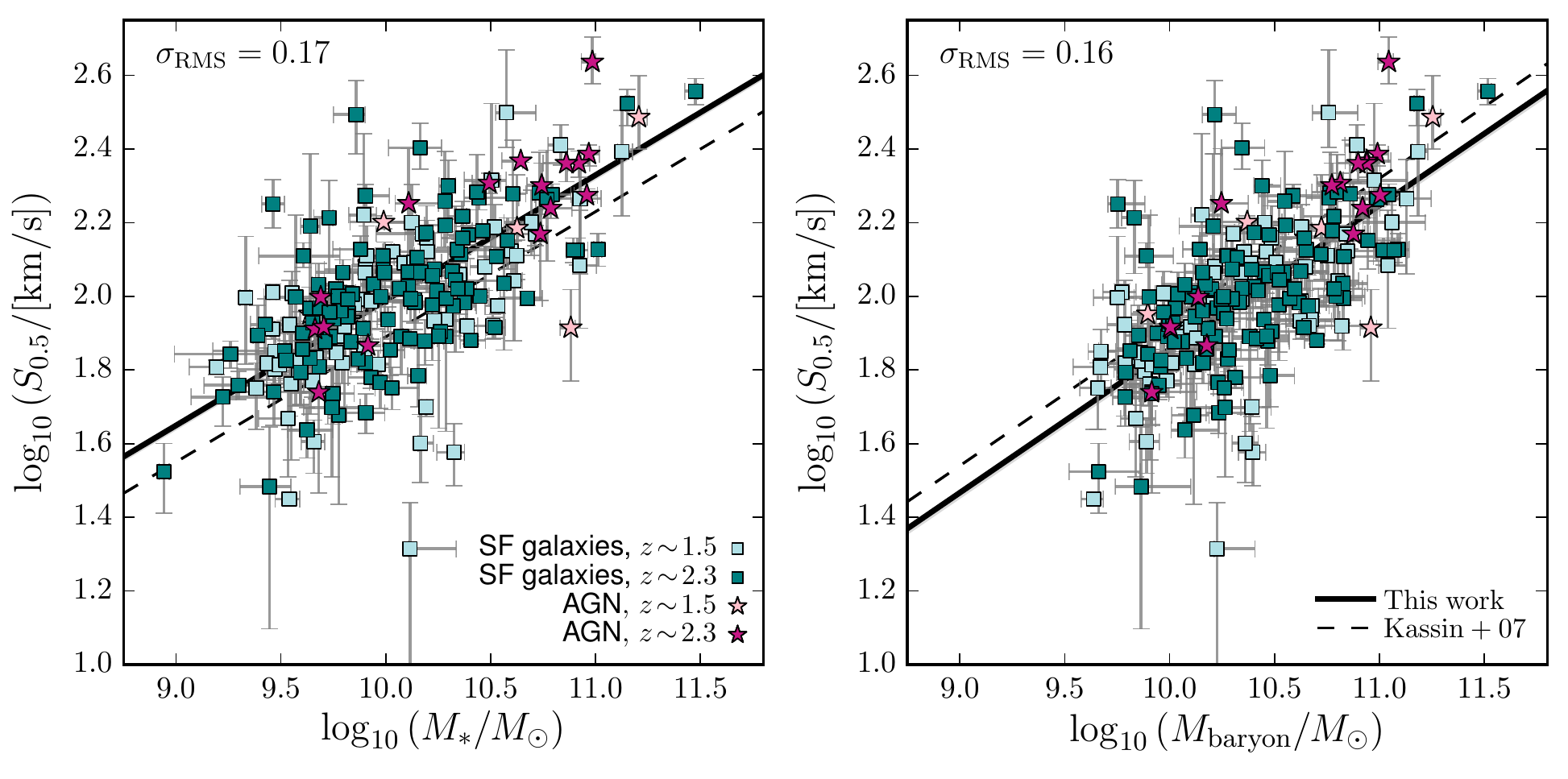}
	\caption{
		The modified $S_{0.5}$-$M$ Tully-Fisher (TF) relation \citep{Kassin07} for our sample galaxies 
		with and without detected rotation. 
		The left and right panels show the stellar and baryonic $S_{0.5}$-$M$ TF relations, respectively, 
		with $S_{0.5} = \sqrt{0.5 V_{2.2}^2 + \sigmavint^2}$ for our measurements. 
		In each panel, the star-forming galaxies at $z\sim 1.5$ and $\sim 2.3$ are shown as light blue and teal squares, 
		respectively, and the AGN at $z\sim 1.5$ and $\sim 2.3$ are shown as pink and purple stars, respectively. 
		% S0.5-Mstar
		For the same slope of the $S_{0.5}-\Mstar$ relation, our sample has higher $S_{0.5}$ at fixed $\Mstar$ 
		than the sample of \citeauthor{Kassin07}, which may reflect the trend of a decreasing average gas fraction. 
		% S0.5-Mbaryon
		For the $S_{0.5}-\Mbar$ relation we find that, for the same slope, our sample has lower $S_{0.5}$ at 
		fixed $\Mbar$ than the sample of \citeauthor{Kassin07}
		An increase in the average dark matter fraction over time could explain the higher $S_{0.5}$ 
		values observed by \citeauthor{Kassin07}
		}
	\label{fig:mstar_mbar_s05}
\end{figure*}

% Incompleteness with Mbaryon -- upper envelope
Finally, there is a suggestive trend between $\Mdyn/\Mbar$ and
\Mdyn, such that more massive galaxies may have a steeper IMF (or a
larger dark matter fraction). However, this trend primarily reflects the cutoff in observed baryonic masses
(upper envelope). Thus, at fixed dynamical mass, we miss galaxies with the lower baryonic masses. 
These missed galaxies could increase the median
baryonic-dynamical mass offset, leaving room for more dark matter, or
bringing the Salpeter IMF into agreement with our data.  

%%%%%%%%%%%%%%%%%%%%%%%%%%%%%%%%%%%%%%%%%%%%%%%%%%%%%%%%%%

\subsection{Modified stellar mass Tully-Fisher relation}
\label{sec:tullyfisher}

%%%%%%%%%%%%%%%%%%%%%%%%%%%%%%%%%%%%
% Intro to M*, Mbar TFR
The Tully-Fisher relation (TFR, \citealt{Tully77}) - 
which relates the luminosity of disk galaxies to their rotation
velocity -  captures information about the interplay between the
build-up of disk galaxies and their dark matter halos. As the
luminosity-based TFR is subject to luminosity evolution (due to aging
populations)  and a possible evolution in the gas mass fraction, more
recent works have focused on measuring the stellar mass or baryonic
mass TFRs, as mass allows for more straight-forward comparisons between redshifts
(e.g., \citealt{Dutton11}, \citealt{Gnerucci11}, \citealt{Miller12}, \citealt{Vergani12}).

% Intro to S0.5 TFRs
Furthermore, \citet{Kassin07} argue that the stellar and baryonic TFRs may evolve due to the increase of non-rotational 
support in higher redshift galaxies. To account for the increased non-rotational support, they argue for the adoption 
of the $S_{0.5}$ kinematic indicator, with $S_{0.5} = \sqrt{0.5 V_{\mathrm{rot}} ^2 + \sigmav^2}$. 
This study shows a reduction of scatter in the $S_{0.5}$-\Mstar TFR relative to the standard \Mstar-TFR at 
$z \sim 0.2 - 1$. Furthermore, they find that there is barely any evolution in the $S_{0.5}$-\Mstar TFR out to $z \sim 1$.

%  S0.5 measurements : Fit S0.5-Mstar (masses as independent variable), figure interpretation
We use our kinematic measurements to examine the $S_{0.5}$-\Mstar and $S_{0.5}$-\Mbar TFRs 
for our sample of star-forming galaxies at $z\sim1.5-2.6$, shown in Figure~\ref{fig:mstar_mbar_s05}. 
We perform a weighted linear fit of $S_{0.5}$ vs \Mstar (left panel, Figure~\ref{fig:mstar_mbar_s05}) to our sample of star-forming 
galaxies by fixing the slope to the average value \citet{Kassin07} find at $z \sim 0.1-1.2$ (i.e., $\mathrm{slope} = 0.34$). 
We find a moderate correlation between $S_{0.5}$ and \Mstar, with scatter in $S_{0.5}$ of 
$\sigma_{\rms}=0.17 \, \mathrm{dex}$ for our star-forming galaxies. 
The scatter is similar to the average scatter \citet{Kassin07} find (0.16 dex). 
Our best-fit relation is offset to higher $S_{0.5}$ compared to the average relation found by \citet{Kassin07} 
(black dashed line, left panel of Figure~\ref{fig:mstar_mbar_s05}), which may be explained by lower average gas 
fractions of star-forming galaxies at lower redshifts. 
We find a similar intercept if we fit the $S_{0.5}-\Mstar$ relation using only the galaxies with detected rotation.

% S0.5-Mbar, figure interpretation
We follow the same general method to fit $S_{0.5}$ vs \Mbar (right panel, Figure~\ref{fig:mstar_mbar_s05}), adopting the 
slope of the $S_{0.5}$-\Mbar TFR at $z\sim 0.2$ measured by \citet{Kassin07} (i.e., $\mathrm{slope} = 0.39$), 
and find a correlation between $S_{0.5}$ and \Mbar with a scatter in $S_{0.5}$ of 0.16 dex. 
Our $S_{0.5}$ intercept is somewhat lower than found by \citet{Kassin07} at $z\sim0.2$ (dashed black line, right panel), 
suggesting an increase in $S_{0.5}$ at fixed \Mbar over time. 
We find the same result if we exclude the galaxies without detected rotation. 
The offset between $S_{0.5}$ intercepts may be explained by a higher dark matter fraction at later times. 
This trend may reflect the increasing radii of similar-mass star-forming galaxies with decreasing redshift 
(e.g., \citealt{Williams10}, \citealt{vanderWel14a}); as dark matter halo profiles are less concentrated than 
stellar mass profiles, a larger radius results in a higher dark matter fraction.

% Caveats to comparison: different velocity definitions.
Nonetheless, systematic differences may affect the comparison of the $S_{0.5}-M$ TFRs. 
In particular, the rotation velocities are not measured uniformly, which could introduce systematic offsets. 
\citet{Kassin07} use the maximum rotation velocity $V_{\mathrm{max}}$ (i.e., the velocity at the flat portion of an arctan 
rotation curve or at $2.2 r_s$  for an exponential disk). 
Our data do not sample the flat portion of the arctan rotation curve, so we instead adopt $V_{2.2}$ 
as our velocity measurement, as we have reasonable velocity constraints at this radius. 
Future work is necessary to quantify the differences between the literature measurements, in order to study the 
$M$-TFR evolution over time in a consistent manner.

%%%%%%%%%%%%%%%%%%%%%%%%%%%%%%%%%%%%%%%%%%%%%%%%%%%%%%%%%%
\subsection{Comparison of baryonic and dynamical masses for AGN}
\label{sec:mbar_mdyn_agn}

% Mbar Mdyn including AGN
\begin{figure}
  \centering 
  \includegraphics[width=0.48\textwidth]{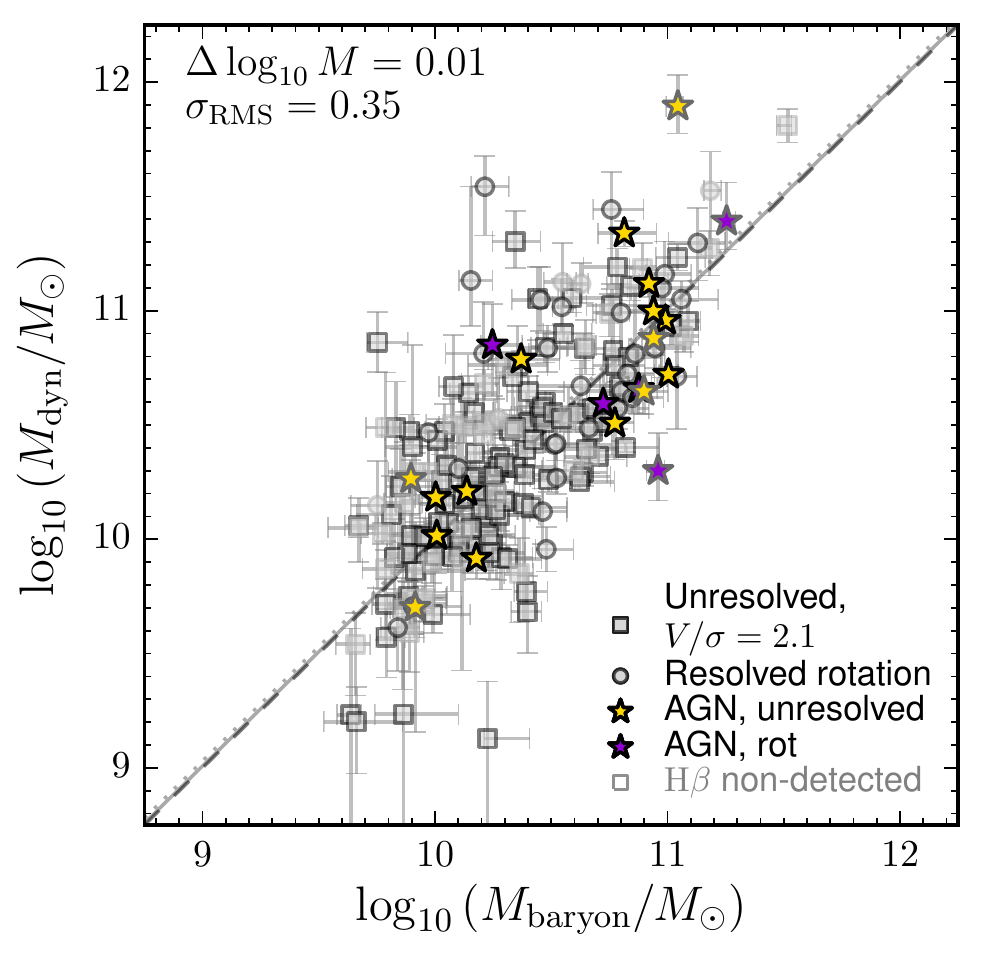}

  \caption{ \Mdyn vs. \Mbar for the AGN in comparison with the galaxy
    sample. Galaxies with and without detected rotation galaxies are
    represented by the grey circles and squares, respectively. The AGN
    with and without detected rotation are shown with the purple and
    yellow stars, respectively. The grey line shows the median offset
    of $\Delta (\log_{10} M) = 0.01 \, \mathrm{dex}$ for the AGN, and
    the dashed grey line shows $\Mdyn=\Mbar$. For reference, the grey
    dotted line shows the median offset for the star-forming galaxies.
    The relation between baryonic and dynamical masses for the AGN is
    similar to the relation for primary galaxy sample, and thus the gas
    kinematics likely trace the dynamics of the host galaxies. }

  \label{fig:mbar_mdyn_agn}
\end{figure}

%%%%%%%%%%%%%%%%%%%%%%
% Example light profiles
\begin{figure*}
  \centering
  \includegraphics[width=0.95\textwidth]{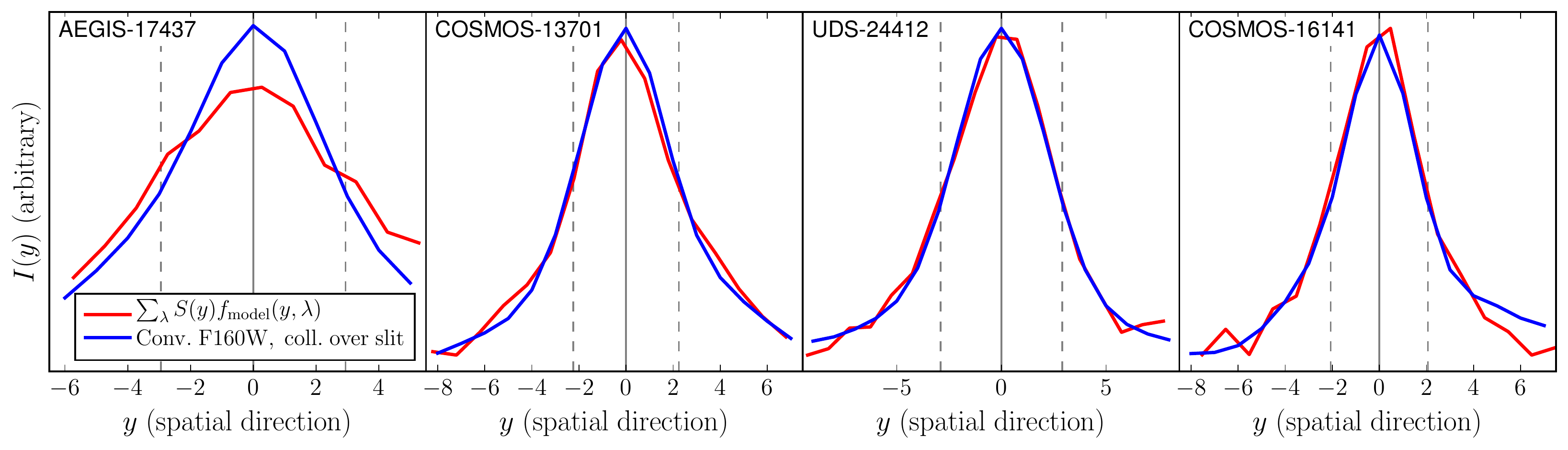}

  \caption{ Intensity profiles for the stellar continuum and \Halpha
    for the four objects shown in Figure~\ref{fig:example_bestfits},
    with the field and 3D-HST v4 ID noted in the upper left corner of
    each panel.  The normalized stellar light and \Halpha profiles are shown as
    the blue and red lines, respectively.  The solid grey line shows
    the emission line center measured during the 1D spectra
    extraction.  The dashed grey lines show the convolved, projected
    effective radius for each object.  The profiles are generally in
    good agreement, with only the first object having a noticeably
    larger  \Halpha profile width.  }

	\label{fig:light_profiles}
\end{figure*}
%%%%%%%%%%%%%%%%%%%%%%

% Intro: compare Mbar-Mdyn for AGN
In this section we consider the kinematics of the AGN that fall within our galaxy sample. Interpreting 
the kinematics of AGN can be challenging, as the line profiles may have contributions from nuclear 
emission tracing non-virial motions. Therefore, we did not include the AGN in our analysis so far. 
Here we assess whether the kinematics may still provide a probe of the host-galaxy structure. 

% Method, intro to figure
We calculate the stellar, gas, and baryonic masses following the 
procedure of Section~\ref{sec:mosdef_survey}.  We measure the \Halpha 
kinematics from the resolved 2D or integrated 1D spectra, and derive
dynamical masses following the procedures of Sections 
\ref{sec:rot_meas}, \ref{sec:disp_meas}, and \ref{sec:mdyn_eff}.  The
resulting baryonic and dynamical masses for the AGN, along with those
of the galaxy sample, are shown in Figure~\ref{fig:mbar_mdyn_agn}.

% Interpretation of figure: AGN seem to follow same trend
The AGN that meet our sample selection criteria generally follow the 
same relation of $\Mdyn-\Mbar$ as the primary galaxy sample. For the 
AGN we find a median offset $\Delta \log_{10} M = 0.01 \, \mathrm{dex}$,  
which is slightly lower than the median offset for the star-forming galaxies 
($\Delta \log_{10} M = 0.04 \, \mathrm{dex}$), and a scatter of 
$\sigma_{\rms }= 0.35 \, \mathrm{dex}$. 
We note that the effective radii may be underestimated for some 
objects, due to the influence of a nuclear source.

% What the similar trend implies: probing AGN host galaxies
The good agreement between the AGN and galaxies in the $\Mdyn-\Mbar$ plane suggests that 
the rest-frame optical lines of most AGN in our sample are not dominated by line emission from the nuclear
regions, and that we are probing the kinematics of the host galaxies. 
% Interpretation:
Our findings support the results of \citet{Gabor14}, 
who show that in high-resolution hydrodynamic simulations of disk galaxies at $z\sim2$, AGN have 
relatively little impact on the gas in galaxy disks.

%%%%%%%%%%%%%%%%%%%%%%%%%%%%%%%%%%%%%%%%%%%%%%%%%%%%%%%%%%
\subsection{Mass-to-(H$\alpha$)-luminosity variations}
\label{sec:light_profiles}

% Halpha sizes and distribution relative to stellar light and distribution
When constructing kinematic models, we rely on structural parameters and radii measured 
from the stellar light distribution, but measure the kinematics from \Halpha emission. 
Ideally, we would measure the kinematics from stellar absorption features, but our galaxies 
are too faint for these measurements. Instead, we assume that 
the ionized gas and stellar mass have the same distribution and that the gas follows the 
gravitational potential of the galaxy.

% Does Halpha follow the stellar light profile? Test assumption! 
When using kinematic models to measure rotation and interpret the
velocity dispersion of unresolved objects, we weight the model
velocity field by a luminosity distribution. For the galaxies with
detected rotation, this weighting determines the composite velocity
profile within  a spatial slice, as there is a mix of line-of-sight
velocity components, as well as components that fall within the slit,
parallel to the  spatial direction. The weighting also predicts a
light profile along the spatial direction, but our method of fitting
the 2D spectra with  the models removes this variation in the spatial
direction through the scaling factor $S(y)$.  For the kinematically
unresolved galaxies, the  luminosity weighting affects all
directions.

% What we should have done to be fully consistent.  
To be fully consistent, we should weight the model velocities by the 
\Halpha light profile, so there are no mis-matches between the model 
and observed luminosity profiles. However, the \Halpha profiles are 
not well constrained by the MOSDEF data, and most galaxies in our 
sample lie at redshifts where \Halpha falls outside  of the wavelength coverage  
of the WFC3 grism \citep{Brammer12}. 
We therefore assume that the stellar light profiles are a reasonable 
approximation for the luminosity weighting.  We assess this assumption 
by comparing the stellar light profiles 
with the \Halpha profiles for the galaxies with detected rotation.

% Method of measuring Halpha and stellar light profiles. Figure interpretation
To measure the stellar light profile, we convolve the \HST/F160W image
to match the seeing resolution of the corresponding  MOSFIRE
spectrum. We then collapse all light falling within the slit along the
slit direction. We approximate the \Halpha profile  by collapsing the
scaled emission line model, $S(y) f_{\mathrm{model}}(y, \lambda)$,
over the wavelength direction.  We show example F160W and \Halpha
profiles in Figure~\ref{fig:light_profiles} for the same galaxies
shown in Figure~\ref{fig:example_bestfits}. 
For three of the four examples, the stellar and \Halpha profiles are
very similar.  Only the first object, AEGIS-17437, has noticeable
deviation between the \Halpha and stellar profiles, where the \Halpha
profile is wider than the stellar light profile.

%%%%%%%%%%%%%%%%%%%%%%
% FWHM comparison for all galaxies with resolved rotation
\begin{figure}
  \centering \hglue -5pt
  \includegraphics[width=0.48\textwidth]{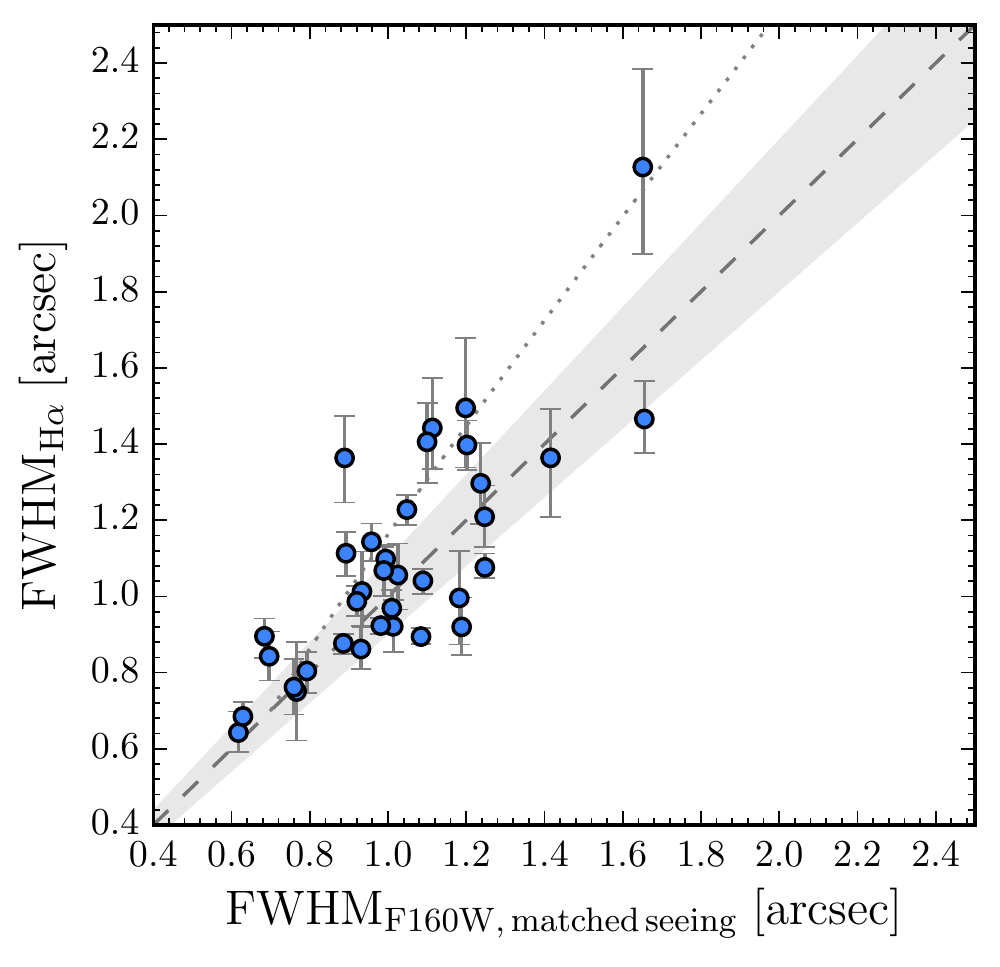}
  \caption{ $\mathrm{FWHM}_{\Halpha}$ vs. $\mathrm{FWHM_{F160W}}$ for
    the galaxies with observed rotation.  The \Halpha FWHM is measured
    from the scaled 2D emission line model, and  the F160W FWHM is
    measured from the F160W image within the slit,  convolved to match
    the MOSFIRE seeing conditions.  The dashed grey line shows the
    line of equality. The grey shaded region shows the  range of
    values when $\mathrm{FWHM}_{\Halpha}$ is 10\% smaller or larger
    than $\mathrm{FWHM_{F160W}}$.  The FWHMs are similar for most of
    the galaxies with detected rotation.  Also, all but a few of the
    galaxies lie below the relation $R_{E,\Halpha} = 1.3 \re$ found by
    \citet{Nelson12},  with the converted FWHMs convolved with a
    median seeing FWHM of $\sim 0\farcs7$.  }
  \label{fig:fwhm_comparison}
\end{figure}
%%%%%%%%%%%%%%%%%%%%%%

% FWHM Ha, F160W comparison for all galaxies with resolved rotation: Similar
We quantify the profile differences for all galaxies with detected
rotation by fitting the  \Halpha and F160W profiles with Gaussians. We
note that the profiles of a number of objects  are not well-described
by a Gaussian profile, but the FWHM measurements should provide a
reasonable, albeit rough, comparison.  We compare the widths of the
seeing-matched stellar light and \Halpha profiles in Figure~\ref{fig:fwhm_comparison}.  
Generally, the FWHMs are in reasonable agreement, and only a few objects have FWHMs that differ by more than
10\% (objects lying outside the grey shaded region). Additionally, we
show the relation $R_{E,\Halpha} = 1.3 \re$  found by \citet{Nelson12}
as the dotted grey line, converted to FWHMs and convolved with a
median seeing FWHM of  $\sim 0\farcs7$. All but a few of the
galaxies lie below this line, suggesting that the \Halpha sizes for
our sample are closer  to the stellar light sizes than for the
\citeauthor{Nelson12} sample.  
Thus, for most of our objects the
stellar light profile is a reasonable substitute for the \Halpha
profile, and hence  the measured kinematics will not be biased.

%%%%%%
% Changes for the 2D model
However, for galaxies with more extended \Halpha, such as AEGIS-17437,
we may overestimate the velocity dispersion  and possibly the rotation
velocity, when assuming the stellar light profile in the model
construction.  This velocity difference can be explained by the fact
that the high velocities at large radii have been down-weighted  when
using the less extended stellar light instead of the \Halpha profile.
The exact changes in the measured velocity and dispersion velocity
from the 2D models depend on the misalignment between  the major axis
and the slit.  If $\deltPA = 0$, there is symmetric mixing of
velocities at different radii within a spatial slice,  and the
narrower stellar light profile therefore results in less broadening in
the wavelength direction.  Additionally, by weighting with the
narrower stellar light profile, the composite of the velocity
components along  the line-of-sight direction also results in less
broadening in the wavelength direction.  Thus, when $\deltPA = 0$ and
we weight the velocity fields with the stellar light profile,  the
measured $V(\re)$ is the same, and $\sigmavint$ is larger than we
would measure when using the  \Halpha light profile. 
If a galaxy is misaligned with the slit, the measured $V(\re)$ using
the stellar light profile may also be  larger than if we weighted with
the \Halpha profile.

%%%%
% 1D model changes for this object
The 1D model for AEGIS-17437 would suffer a greater discrepancy if we 
would have weighted the velocities of the model with the  \Halpha 
instead of the stellar light profile. The increased weight at large 
radii would increase the weighted integrated velocity  dispersion of 
the model within the aperture. Thus the ratio $\sigmavmodel/\Vrms(\re)_{\mathrm{model}}$ 
is higher for the \Halpha profile than for the stellar light profile, when using the 
same underlying rotation curve  and assumed $\vtosigre$. Therefore, 
the corrected RMS velocity values measured using the stellar light 
profile  overestimate the true values.

% Mass sizes: 
An additional question is whether the stellar light sizes correctly probe the characteristic size of the 
underlying matter density profile.  In particular, our current calculations have assumed that half of 
the total mass is enclosed within the half-light radius. However, the  half-mass sizes are on average 
$\sim 25 \%$ smaller than the half-light sizes \citep{Szomoru13}. If we assume the same intrinsic 
rotation velocity curve, $V(\re) > V(r_{1/2,\mathrm{mass}})$, and thus the measured $V(\re)$ 
(for the resolved models) and  $\Vrms(\re)$ (for the unresolved models) are larger than the 
velocities at $r_{1/2,\mathrm{mass}}$.  If we assume a constant $\sigmavint$, this difference 
implies a lower $\vtosig$. In combination with a smaller $R_e$, this results in a considerably lower
dynamical mass. For example, if $r_{1/2, \mathrm{mass}}$ is 25\% lower than $\re$ (as in \citealt{Szomoru13}), 
then a galaxy with $\vtosigre = 2.1$ at $\re$ and  $r_t = 0.4 r_s = 0.4 \re /1.676$ (following \citealt{Miller11},
see Appendix~\ref{sec:appendix1D})  has $\Mdyn(r_{1/2,  \mathrm{mass}})$  which is  31\% lower than 
$\Mdyn(\re)$. 
When applying the correction found by \cite{Szomoru13} the dynamical masses
decline by 0.16 dex  (to $\Delta \log_{10} M = -0.12 \, \mathrm{dex}$)
and are inconsistent with both a \citet{Chabrier03} and a \citet{Salpeter55} IMF.

% Rough systematics plot
\begin{figure*}
  \centering
  \hglue -5pt
  \includegraphics[width=0.95\textwidth]{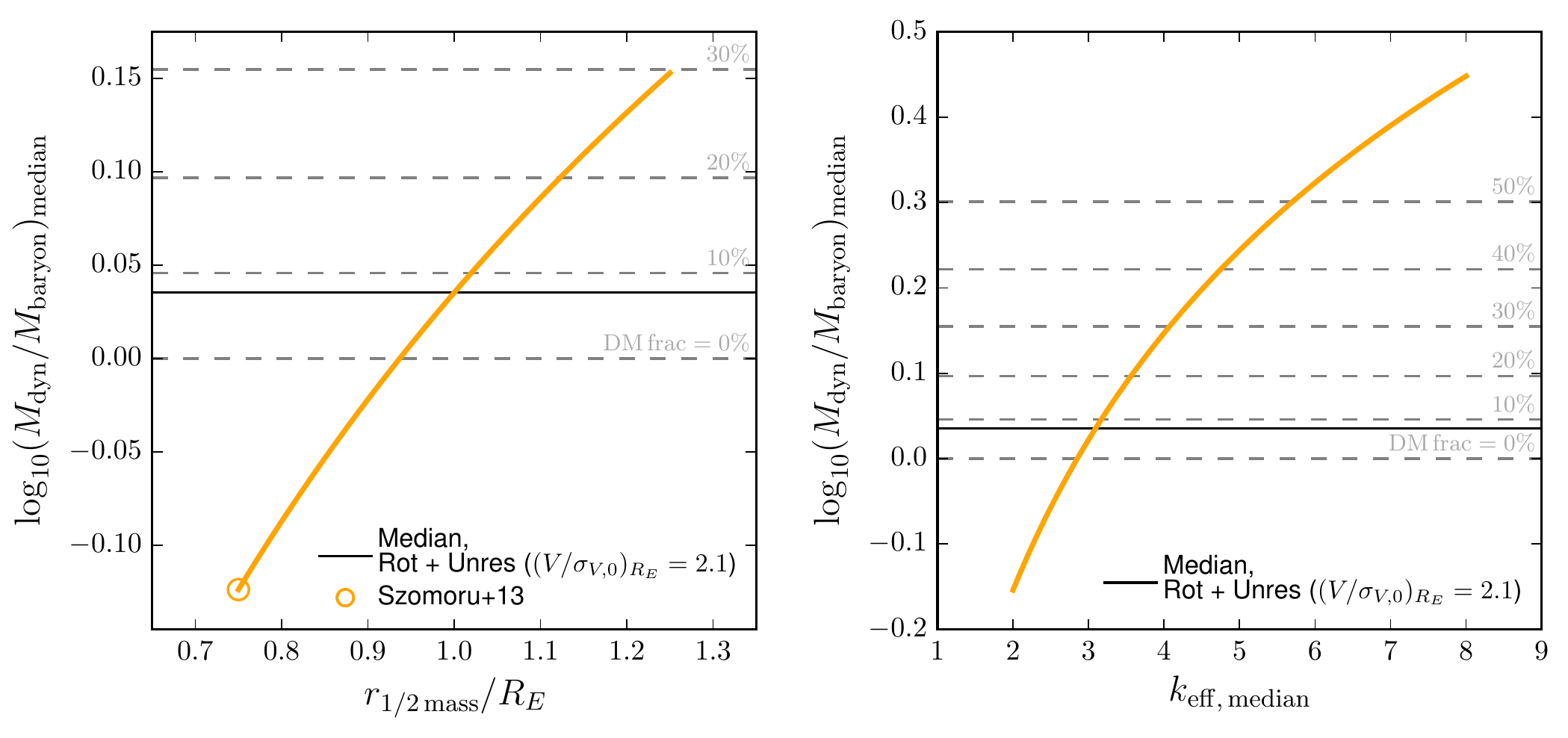}

  \caption{ 
  	Systematic changes in the median $\log_{10}(\Mdyn/\Mbar) = \Delta \log_{10} M$ with 
  	varying half-mass to  half-light radius ratio ($r_{1/2 \, \mathrm{mass}}/R_{E,*}$, left panel) and 
	effective virial coefficient ($k_{\mathrm{eff, \, median}}$, right panel).  
	In both panels the black horizontal line shows the median $ \Delta \log_{10} M$ for the whole sample 
	when adopting the default assumptions. Lines of constant dark matter fraction (assuming  
	$f_{\mathrm{DM}} = M_{\mathrm{DM}}/\Mdyn = (\Mdyn - \Mbar)/\Mdyn$) are shown with grey dashed lines.  
	% Left panel:
	In the left panel, the systematic changes with half-mass to half-light radius ratio include 
	variations of the RMS velocity and virial coefficient assuming an arctan rotation curve 
	with $\vtosigre = 2.1$ and $r_t = 0.4 \re /1.676$.  \citet{Szomoru13} find that $r_{1/2 \, \mathrm{mass}}$
    	is on average 25\% lower than $\re$ (orange circle),  which corresponds to the median offset 
	$\Delta \log_{10} M$ shifted to the unphysical region, or $\Delta \log_{10} M \sim -0.12 \, \mathrm{dex}$.
	% Right panel
    	In the right panel, the systematic changes with the effective virial coefficient are approximated by applying the
    	same $k_{\mathrm{eff, \, median}}$ to all galaxies and calculating the resulting median offset $\Delta \log_{10} M$.  
	Assuming $k_{\mathrm{eff, \, median}} \sim \krot = 2.66$ for all galaxies (without changing the assumed $\vtosigre$)  
	results in an unphysical $\Delta \log_{10} M < 0$, while assuming $k_{\mathrm{eff, \, median}} \sim \kdisp= 5$ 
	results in a much higher inferred dark matter fraction of $f_{\mathrm{DM}} \sim 45\%$.  
	}
		
	\label{fig:systematics}
\end{figure*}

% Example systematic variation, possible caveats
We illustrate a lower limit of this systematic variation in the left panel of Figure~\ref{fig:systematics}, 
by approximating the changes of $\Delta \log_{10} M$ caused by varying
$r_{1/2,\mathrm{mass}}/\re$ and using  $\Mdyn = \Mdyn(r_{1/2,\mathrm{mass}})$ (i.e., including
 variations to the  RMS velocity and virial coefficient with the assumption of an arctan curve with 
 $\vtosigre = 2.1$ and $r_t = 0.4 r_s = 0.4 \re /1.676$). 
Even $\sim 10\%$ changes in $r_{1/2,\mathrm{mass}}/\re$ result in a $\sim 10\%$ change to 
the inferred dark matter fraction.  
However, if we instead assume a decreasing velocity dispersion with
increasing radius, as we discuss in the next section, the
$\Vrms(r_{1/2, \mathrm{mass}})$ values would be larger. This moves in
the opposite direction as the above trend, and may change the masses
so they remain consistent with a \citet{Chabrier03} IMF.

%%%%%%%%%%%%%%%%%%%%%%%%%%%%%%%%%%%%%%%%%%%%%%%%%%%%%%%%%%
\subsection{Other caveats}
\label{sec:caveats}

% Intro. What other caveats have we not mentioned previously?
In this section we consider caveats to assumptions made in the preceding analysis.  
Specifically, we focus on possible variations due to assumptions about the accuracy of the 
structural parameters, misalignment of the photometric and kinematic major axes,
the intrinsic thickness of galaxy disks, the accuracy of our derived gas masses, the shape of the rotation curve, 
and the velocity dispersion profile being constant.

% Galfit parameter accuracy
First, we have not fully accounted for the accuracy of the \textsc{Galfit}-derived structural parameters. 
We depend on the structural parameters to model the kinematics of the detected rotation curves, 
to correct the kinematics from integrated 1D spectra, and to calculate the dynamical masses. 
We include estimated errors on the effective radii when calculating the dynamical masses, 
but do not include any errors when fitting the kinematic models. Thus, uncertainties in the structural 
parameters introduce scatter in our derived dynamical masses.

% Kin/phot major axis misalignment
Second, for objects where the photometric and kinematic major axes are misaligned, the inferred velocities and 
dispersion velocities will be incorrect. If the true $\deltPA$ is closer to aligned, the corrected RMS velocities will 
be over-estimated, while if it is more misaligned, the velocities will be under-estimated (as seen in the 
first panel in Figure~\ref{fig:aper_corr}).  We expect similar over- and under-estimates in the measured velocities 
and dispersions for the galaxies with detected rotation. Additional misalignment uncertainties are introduced by slit 
alignment issues, which introduce the same trends as stated above.

% Disk thickness assumption
Third, we assume an intrinsic disk thickness of $(b/a)_0=0.19$ to estimate inclination angles. 
If a galaxy is intrinsically thicker than the assumed value, the inferred inclination angle underestimates 
the true value. In this case, the inferred intrinsic rotation velocity and $\vtosigre$ would be overestimated.  
If a galaxy is thinner, the inclination angle will be overestimated, producing an underestimate of both the rotation 
and $\vtosigre$. Thus, variations in disk thickness within our sample will add scatter and a possible 
systematic offset in our dynamical masses.

% Gas mass accuracy
Fourth, we assume that the galaxies in our sample follow the SFR-gas mass relation of \citet{Kennicutt98} 
for star-forming galaxies in the local universe. 
Based on a local sample of normal and starburst galaxies, \citet{Kennicutt98} measure $N=1.4$ for 
$\Sigma_{\mathrm{SFR}} \propto \Sigma_{\mathrm{gas}}^N$, where $\Sigma_{\mathrm{gas}}$ 
includes both  atomic and molecular gas.  Analysis of galaxies at $z\sim 1-3$ find slopes that vary between 
$N=1.28$ \citep{Genzel10} and $N=1.7$ \citep{Bouche07}.  These values bracket the local slope, 
so our gas masses may be reasonable. However, if the  true slope is lower than the local relation, 
our gas masses underestimate the true value, while a  higher slope implies our gas masses overestimate 
the true value. An alternate method would be to adopt the gas mass scaling relations presented in \citet{Genzel15}, 
which relate the gas mass to the stellar mass, the offset from the star-forming main sequence, and the 
redshift. If we adopt this scaling relation, we see an offset $\Delta \log_{10} M = -0.12 \, \mathrm{dex}$ and a 
scatter of $\sigma_{\rms} = 0.368 \, \mathrm{dex}$ between the dynamical and baryonic masses, 
with the median mass difference lying in the unphysical region where $\Mbar > \Mdyn$. However, the gas mass 
scaling relations of \citet{Genzel15} were calibrated for UV+IR SFRs, while we use \Halpha SFRs in this work. 
Mismatches between these SFR indicators could be causing the large ($\sim 0.2 \, \mathrm{dex}$) difference 
between the inferred baryonic masses when using the scaling relation method and the inverted Kennicutt-Schmidt relation. 
As \citet{Kennicutt98} calibrated the  $\Sigma_{\mathrm{SFR}} - \Sigma_{\mathrm{gas}}$ relation using \Halpha SFRs, 
we opt to estimate the gas masses following this prescription.

% V(r) assumed to be arctan
Fifth, we have assumed that the rotation velocity profiles of our disk galaxies are well described by arctan models, 
as shown by \citet{Courteau97}, \citet{Weiner06a}, and \citet{Miller11}. 
Some distant star-forming galaxies exhibit rotation following a Freeman exponential
disk model \citep{Freeman70}, as found by \citet{Wisnioski15}, while \citet{vanDokkum15} 
find indications of Keplerian rotation in compact star-forming galaxies. However, preliminary 
analysis suggests that using an exponential disk rotation curve with our modeling produces 
poor fits to some of our galaxies. More detailed modeling is required to determine in detail whether
an alternative rotation profile provides better  agreement with our 
data and to assess the uncertainties introduced by this assumption.

% Constant sigma profile
Sixth, we have assumed that the intrinsic velocity dispersion is constant over all radii. However, the true velocity 
distribution may decrease with increasing radius, as seen in \citet{Genzel14} and \citet{Wisnioski15}.
For the unresolved objects, a decreasing velocity dispersion distribution would produce higher model RMS 
velocities for a given $\vtosigre$ measured at $\re$,\footnote{Other studies (e.g., \citealt{Newman13}, \citealt{Wisnioski15}) 
measure $\sigmavint$ in the outer portions of a galaxy, so taking $\sigmav(\re) \approx \sigmavint$ in 
the case of a non-constant velocity profile should be a reasonable comparison, though may be higher than the true 
value depending on  the exact form of the additional velocity dispersion term. This notation is in contrast to 
velocity dispersions of elliptical galaxies, where the \textit{central} velocity dispersions are often denoted as 
$\sigma_0$ or $\sigmavint$. }  
as matching the same $\sigmav(\re) \approx \sigmavint$ implies a higher central velocity dispersion, 
$\sigmav(r=0)$.  A velocity dispersion profile which rises towards the center would increase the integrated 
model velocity dispersions  but not the model RMS velocity at \re, leading to lower corrected RMS velocities and 
lower dynamical masses.  The trend of decreasing integrated RMS velocity with increasing inclination will also be 
less strong for a fixed $\vtosigre$  than with a constant \sigmavint. The median $\vtosigre$ required to remove the 
$\Delta \log_{10}M$ trend for the kinematically  unresolved galaxies will therefore be higher, increasing the implied 
amount of rotational support relative to the random motions, which would indicate more settled or thinner disks. 
Preliminary calculations assuming an additional dispersion term that rises towards the center of a galaxy confirms these 
general trends, but more careful analysis is necessary to determine the correct form of a rising velocity dispersion profile. 

% Virial coefficient
Seventh, in our derivation of the dynamical masses we have not included the systematic uncertainties 
arising from the choice of virial coefficients, $\kdisp$ and $\krot$. The matter distributions assumed when deriving
the virial coefficients may not match  the underlying profiles of the star-forming galaxies in our sample, 
but more detailed analysis is required to quantify the  uncertainties introduced by the adopted virial coefficients. 
We note that the systematic shifts from a different choice of  virial coefficient can be non-negligible, and
have implications especially for the IMF constraints.  For instance, the combination of $\kdisp = \beta(n)$ 
(from \citet{Cappellari06}), $n=1$, and $\vtosigre = 2.1$ would have resulted in dynamical masses that 
are larger by $\sim 0.07 \, \mathrm{dex}$ and an inferred dark matter fraction of $f_{\mathrm{DM}} \sim 22\%$. 
We approximate the systematic changes due to changing only the combined $k_{\mathrm{eff, \, median}}$ 
in the right panel of Figure~\ref{fig:systematics}.  In this plot, we see that if we assume 
$k_{\mathrm{eff, \, median}} \sim \krot = 2.66$ for all galaxies (without changing the conversion from the 
integrated velocity dispersions to the RMS velocities for the unresolved galaxies), then the median 
$\Delta \log_{10} M < 0$, which lies in the unphysical region where $\Mdyn < \Mbar$.  If we instead assume 
$k_{\mathrm{eff, \, median}} \sim \kdisp= 5$ (again with no other changes),  we would instead infer a much higher 
dark matter fraction of $f_{\mathrm{DM}} \sim 45\%$ rather than the $8\%$ measured from the default assumptions.

%%%%%%%%%%%%%%%%%%%%%%%%%%%%%%%%%%%%%%%%%%%%%%%%%%%%%%%%%%
%%%%%%%%%%%%%%%%%%%%%%%%%%%%%%%%%%%%%%%%%%%%%%%%%%%%%%%%%%
%%%%%%%%%%%%%%%%%%%%%%%%%%%%%%%%%%%%%%%%%%%%%%%%%%%%%%%%%%

\section{Summary}
\label{sec:summary}

% What we aim to do. Basic measurements
In this paper, we use spectra from the MOSDEF survey to study the masses and kinematic structures of a sample of 
178 star-forming galaxies at $1.4 \leq z \leq 2.6$. For all galaxies, structural parameters from CANDELS 
\HST/F160W imaging, stellar masses from multi-wavelength photometry, and gas masses 
from dust-corrected \Halpha SFRs and the relation by \citet{Kennicutt98} are available. The gas kinematics have been 
measured from the $\Halpha$ emission lines: for 35 of the galaxies we detect resolved rotation, 
while in the remaining 143 galaxies we only measure the velocity dispersion.

% Which galaxies are consistent with rotation?
As our galaxies are observed with random orientations compared to the slit angle, we may not see rotation 
for some objects that are intrinsically rotating. Additionally, we may not resolve rotation due to seeing limitations, 
as found by \citet{Newman13}. To estimate how many of our galaxies are consistent with rotation, we compare 
the projected \Halpha major axis size within the slit to the seeing and use this to estimate whether a galaxy is 
spatially resolved or not. The majority of our sample (80\%) is too small relative to our seeing, and thus these galaxies 
may indeed be unresolved rotating disks. 

% 2D, 1D models and kinematic measurements
We have developed models to convert the observed kinematic measurements into intrinsic rotation and dispersion 
velocities. These models use the sizes, S\'ersic indices, axis ratios, and position angles measured from the F160W 
imaging to simultaneously account for the inclination of the galaxy, the misalignment of photometric major axis and 
the slit, and determine which portions of the galaxy fall within the slit. In the case of galaxies with detected rotation, 
we directly constrain $V(\re)$ and $\sigmavint$, and find a median $\left[\vtosigre\right]_{\mathrm{2D,median}} =2.1$. 
For the galaxies without observed rotation, the models allow us to convert the observed velocity 
dispersion into an RMS velocity for an assumed ratio of  $\vtosigre$.

% Baryonic and dynamical mass comparison, V/sigma measurement
When assuming that the galaxies with and without detected rotation have a similar $\vtosig$, 
we find that the baryonic ($\Mbar = \Mstar + \Mgas$) and dynamical masses of the total sample are in good agreement, 
with a median offset of $\Delta (\log_{10} M) = 0.04 \, \mathrm{dex}$ and a scatter of 
$\sigma_{\rms} = 0.34 \, \mathrm{dex}$. Moreover, we directly constrain the mean $\vtosigre$ for 
the galaxies without detected rotation by removing any trend of $\log_{10}(\Mdyn/\Mbar)$ with axis ratio $b/a$ 
and find $\left[\vtosigre\right]_{\mathrm{1D,median}}=2.1_{-0.3}^{+0.2}$.  
The offset between the dynamical and baryonic masses implies a dark matter fraction within \re of 8\% for a \citet{Chabrier03} IMF, 
which is lower than the value measured within $2.2 r_s$ for local star-forming galaxies \citep{Pizagno05, Dutton11a} or within 
$r < 10\, \mathrm{kpc}$ for galaxies at $z\sim2$ \citep{ForsterSchreiber09}.

% Mbar, not Mstar!
The consistency between the dynamical and baryonic masses relies on the inclusion of gas masses. 
When comparing the dynamical masses with only stellar masses, we find a larger scatter ($\sigma_{\rms} = 0.37 \, \mathrm{dex}$). 
Furthermore, the median offset between the stellar and dynamical mass increases with increasing \Halpha SSFR, 
which is suggestive of a larger gas fraction at higher SSFRs.

% V/sigma trends with SSFR, stellar mass
We examine trends in the ratio of support from rotation and random motions, $\vtosig$, as a function of \Halpha SSFR 
and stellar mass. For galaxies without detected rotation, we bin by \Halpha SSFR and stellar mass  
and estimate $\vtosig$ by removing any variation of $\log_{10}(\Mdyn/\Mbar)$ with axis ratio. 
We see a trend of decreasing $\vtosig$ with increasing \Halpha SSFR  
and a possible weak trend of increasing $\vtosig$ with increasing stellar mass when combining our 
measurements with the sample by \citet{Wisnioski15}. 
The trend of decreasing $\vtosig$ with increasing \Halpha SSFR may reflect disk settling,  
such that galaxies with lower SSFRs have lower gas fractions and therefore lower velocity dispersions. 

% Why the plane-for all works
While our assumption that all galaxies without detected rotation are disks results in highly consistent dynamical and 
baryonic masses, we also find a strong correspondence between the two masses if we had instead assumed that all 
unresolved galaxies are dispersion dominated. Differences in the methods of calculating the dynamical masses 
(i.e., using circularized radii for early-type galaxies vs. inclination corrections for disks, different virial coefficients) may explain 
why the dynamical masses are so similar, and why there is no observed trend of $\log_{10}(\Mdyn/\Mbar)$ with axis ratio.

% IMF constraint
The measured masses also provide insight into the stellar IMF in $z\sim2$ star-forming galaxies. 
The baryonic and dynamical masses of our sample are consistent with a \citet{Chabrier03} IMF.  
A  \citet{Salpeter55} IMF is disfavored by our data, as it would lead to 
baryonic masses that exceed the dynamical masses by $ \sim 0.1 \ \mathrm{dex}$ on average. 
However, when assuming that the half-mass sizes are 25\% smaller than the 
half-light sizes \citep{Szomoru13}, the inferred masses are also inconsistent with a \citet{Chabrier03} IMF. 
Nonetheless, other systematic uncertainties, as discussed in detail in the discussion section, may reduce this mass difference.

% M-S0.5 TFRs:
We examine the modified $S_{0.5}$-\Mstar and $S_{0.5}$-\Mbar Tully Fisher relations (TFRs) for our sample of galaxies, 
with $S_{0.5} = \sqrt{0.5 V_{2.2}^2 + \sigmavint^2}$. 
We find a higher intercept of $S_{0.5}$ than \citet{Kassin07} measure for the average $S_{0.5}-\Mstar$ TFR at 
$z\sim0.1-1.2$, which may be caused by a decrease in the average gas fraction of star-forming galaxies with time. 
For the $S_{0.5}-\Mbar$ TFR, we measure a lower intercept than \citet{Kassin07} find at $z\sim0.2$. 
The change in the $\Mbar$-TFR may reflect an increase in the average dark matter fraction with time, 
possibly caused by the increase in average galaxy size at fixed mass with decreasing redshift.

% AGN
Our sample also contains 21 AGN, selected by either X-ray luminosity, IRAC colors, or optical line ratios.  
As the line emission may be associated with nuclear accretion activity, the broadening may not only probe the 
kinematics of the host galaxy. We measure the baryonic and dynamical masses for the AGN in our sample, 
and find that they follow the same trend as the star-forming galaxies. 
This finding suggests that on average the line profiles do indeed reflect the host galaxy kinematics.

% Future
This paper demonstrates the power of using large samples of galaxies observed with multi-object near-infrared spectrographs 
under seeing-limited conditions to study the average kinematic properties of high redshift galaxies. In particular, combining such 
observations with high-resolution imaging makes it possible to model the effects of axis misalignment, seeing, and velocity rotation 
and dispersion on the observed spectra. This technique will prove useful in future studies of galaxy kinematics with 
\JWST/NIRSPEC, as this multi-object spectrograph will also suffer from random orientation of galaxies within the slits. 
Measurements from seeing-limited multi-object spectrographs are not sufficient to constrain kinematic properties of 
individual high-redshift galaxies, and need to be complimented by adaptive-optics assisted IFU observations. 
Together, these approaches will provide a powerful probe of the nature of galaxies at this key period of structure formation.

%%%%%%%%%%%%%%%%%%%%%%%%%%%%%%%%%%%%%%%%%%%%%%%%%%%%%%%%%%
%%%%%%%%%%%%%%%%%%%%%%%%%%%%%%%%%%%%%%%%%%%%%%%%%%%%%%%%%%
%%%%%%%%%%%%%%%%%%%%%%%%%%%%%%%%%%%%%%%%%%%%%%%%%%%%%%%%%%

\acknowledgements
We acknowledge valuable discussions with J. van de Sande, R. Feldmann, P. van Dokkum, M. Franx, R. Genzel, 
N. F\"orster-Schreiber, E. Toloba, M. George, and I. Shivvers. 
We thank the anonymous referee for constructive comments, which have improved this paper. 
We thank the MOSFIRE instrument team for building this powerful instrument, and for taking data for us during 
their commissioning runs. We are grateful to M. Kassis at the Keck Observatory for his many valuable contributions 
to the execution of this survey, and to I. McLean,  K. Kulas, and G. Mace for taking observations for us in May and June 2013. 
This work would not have been possible without the 3D-HST collaboration, who provided us with the spectroscopic and 
photometric catalogs used to select the MOSDEF targets and derive stellar population parameters. 
We acknowledge support for the MOSDEF survey from NSF AAG collaborative grants AST-1312780, 1312547, 
1312764, and 1313171, and archival grant AR-13907, provided by NASA through a grant from the Space Telescope 
Science Institute. 
SHP acknowledges support from the National Science Foundation Graduate Research Fellowship under grant DGE 1106400. 
MK acknowledges support from the Hellman Fellows fund. 
ALC acknowledges funding from NSF CAREER grant AST-1055081. 
NAR is supported by an Alfred P. Sloan Research Fellowship. 
The data presented in this paper were obtained at the W. M. Keck Observatory, which is operated as a scientific partnership 
among the California Institute of Technology, the University of California, and the National Aeronautics and Space Administration. 
The Observatory was made possible by the generous financial support of the W. M. Keck Foundation. 
The authors wish to recognize and acknowledge the very significant cultural role and reverence that the summit of Mauna Kea 
has always had within the indigenous Hawaiian community. We are most fortunate to have the opportunity to conduct 
observations from this mountain. 
This work is also based on observations made with the NASA/ESA Hubble Space Telescope, which is operated by the 
Association of Universities for Research in Astronomy, Inc., under NASA contract NAS 5-26555. 
Observations associated with the following GO and GTO programs were used: 12063, 12440, 12442, 12443, 12444, 12445, 
12060, 12061, 12062, 12064 (PI: Faber); 12177 and 12328 (PI: van Dokkum);  12461 and 12099 (PI: Riess); 
11600 (PI: Weiner); 9425 and 9583 (PI: Giavalisco); 12190 (PI: Koekemoer); 11359 and 11360 (PI: O'Connell); 
11563 (PI: Illingworth).

% ###############################################################################

\appendix

\section{Appendix A: Modeling of rotation in resolved disk galaxies}
\label{sec:appendix2D}

\subsection{Kinematic model definition}
\label{sec:appendix2DsubA}

% ############################################################

% Geometry for misaligned disk for ``aperture'' correction
\begin{figure*}
	\centering
	\includegraphics[width=0.88\textwidth]{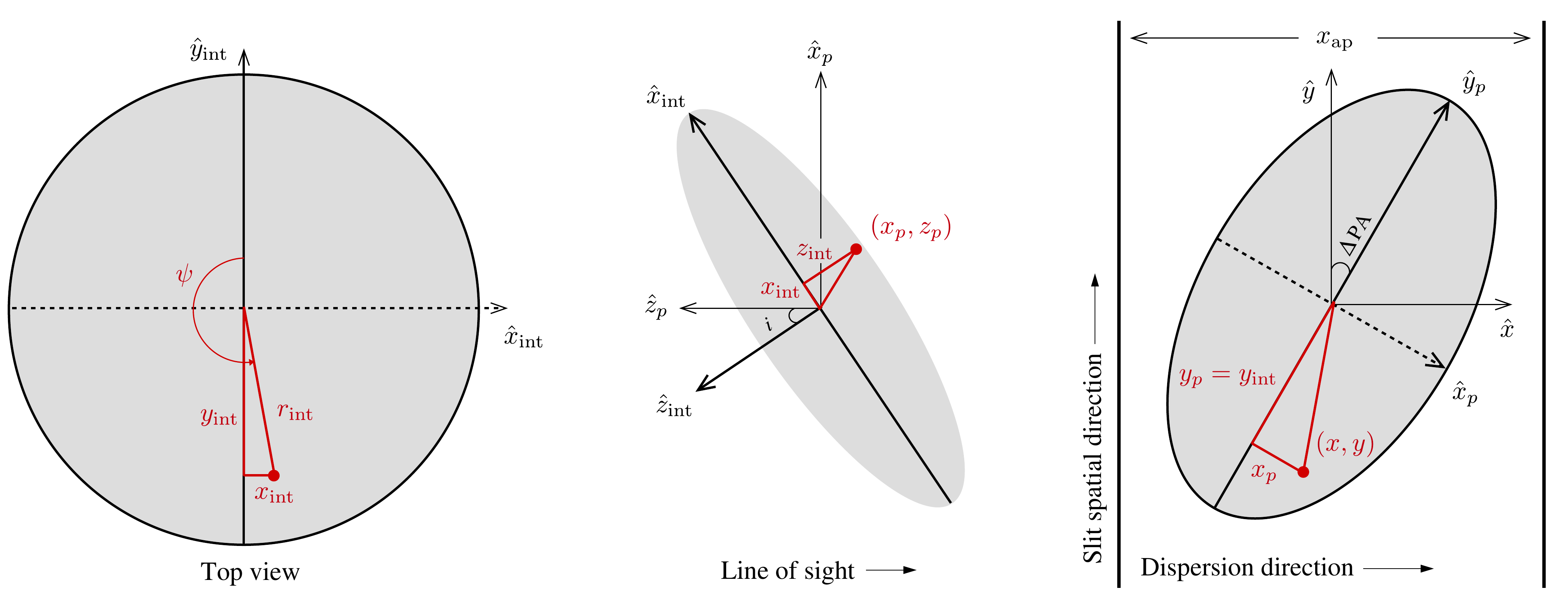}
	\caption{ 
		Coordinate definition for an inclined disk galaxy misaligned with the slit axis. The left panel shows a top down 
		view of the disk galaxy, depicting the $\hat{x}_{\intrm}-\hat{y}_{\intrm}$ plane, and the definition of 
		$r_{\intrm}$ and the angle $\psi$. The center panel shows a side view of the inclined disk galaxy, 
		with the line of sight extending to the right. Here we show the coordinate transformation due to inclination 
		from the intrinsic $(x_{\intrm}, z_{\intrm})$ coordinates to the projected $(x_p, z_p)$. 
		The right panel shows the disk galaxy relative to the slit, including the position angle misalignment ($\deltPA$). 
		The projected coordinates $(x_p, y_p)$ are shown relative to the slit coordinates $(x,y)$.
	}
	\label{fig:appendix_coords} 
\end{figure*}

% V_los plot for misaligned disk
\begin{figure}
	\centering
	\hglue -5pt
	\includegraphics[width=0.53\textwidth]{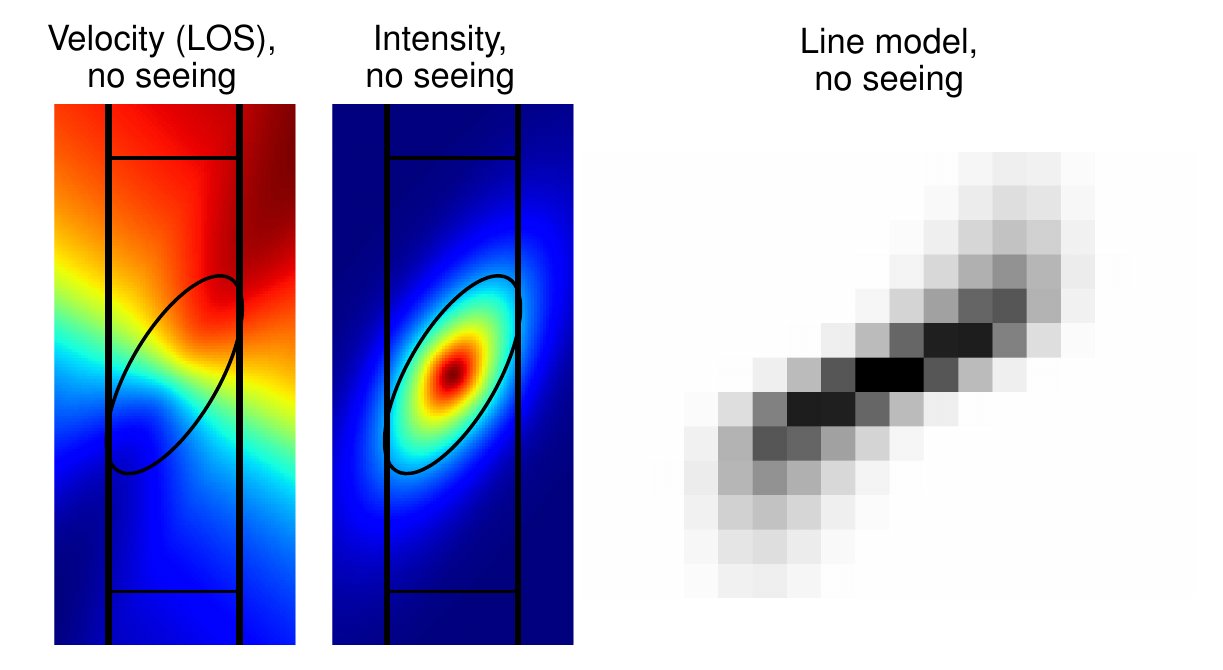}
	\caption{
		Example of the line of sight (LOS) velocity field $V_{\mathrm{obs}}(x,y)$ collapsed along the LOS, $z$ (left panel), 
		assumed model intensity collapsed along $z$, $I(x,y) = \sum_z I(x,y,z) \Delta z$ (center panel), and 
		final composite \Halpha emission line model (right panel) for a disk.  
		The model shown has $\sigmavint = 0$, $R_E = 0\farcs 6$, $b/a = 0.4$, $n=1$, 
		$\Delta \mathrm{PA} = 30^{\circ}$, 
		$V_a = 200 \ \mathrm{km/s}$, and $r_t = 0.4 r_s = 0.4 (\re/1.676)$. 
		Here we ignore any seeing effects, assuming $\mathrm{FWHM}_{\mathrm{seeing}} = 0\arcsec$, and 
		use a typical instrumental resolution width in calculating the \Halpha line model. 
		The wide black lines show the slit width, $0\farcs7$, 
		and the horizontal lines show the spatial aperture extent. 
	}
	\label{fig:v_los_example}
\end{figure}

% Problem of kinematics of disk misaligned with slit axis:
The multiplexing capabilities of MOSFIRE, which allow us to study many galaxies simultaneously, 
come at the price of not observing the galaxies along the kinematic major axis. 
Misalignment of the kinematic major and slit axes poses problems for the interpretation of kinematic 
measurements even for resolved disk galaxies. Issues to address include: 
How much kinematic information is lost because portions of the galaxy fall outside the slit? 
How much of the line broadening in a spatial row is caused by intrinsic velocity dispersion, 
and how much is caused by the  inclusion of multiple line-of-sight velocities in that slice of the galaxy?

% Model intro.
In this appendix, we describe how we model the internal kinematics of a disk galaxy, apply the appropriate 
inclination and position angle offset to the model, and then collapse the model along the line-of-sight and slit 
direction to calculate the observed kinematic signature of the object as a function of position along the slit.

% Coordinate system definition for misaligned disk
To model an ideal disk galaxy with an arbitrary position angle offset with respect to the slit, we define coordinates as 
shown in Figure~\ref{fig:appendix_coords}.  
% Intrinsic coordinates first
First, we consider a point on the galaxy at $(x_{\intrm}, y_{\intrm}, z_{\intrm})$, with distance in the plane of the disk of 
\begin{equation}
r_{\intrm}= \sqrt{x_{\intrm}^2 + y_{\intrm}^2}
\label{eq:r_int}
\end{equation}
from the axis of rotation, and define the angle $\psi$ as
\begin{equation}
\cos \psi = y_{\intrm}/r_{\intrm}, 
\label{eq:cos_psi}
\end{equation}
which is the counterclockwise angle between the  major axis $\hat{y}_{\intrm}$ and $(x_{\intrm}, y_{\intrm}, z_{\intrm})$ 
with respect to the rotation axis (see  the left panel of Figure~\ref{fig:appendix_coords}). 
% Apply inclination coordinate transformation: projected coordinates
We incline our galaxy model by rotating around the major axis $\hat{y}_{\intrm}$.  The inclination angle $i$ is estimated as 
\begin{equation}
\label{eq:sini}
\sin i = \sqrt{\frac{1- (\axratio)^2}{1-(\axratio)_0^2}}, 
\end{equation}
where $a$ and $b$ are semi-major and semi-minor axes lengths, respectively, from the \textsc{Galfit} parameterization. 
We assume an intrinsic disk axis ratio of $(b/a)_0 = 0.19$ \citep{Miller11}.  By inclining the model 
(see the middle panel of Figure~\ref{fig:appendix_coords}), the intrinsic coordinates are mapped to projected coordinates by 
\begin{align}
\label{eq:xyz_int}
x_p &= \ \  \, x_{\intrm}  \cos i + z_{\intrm} \sin i \notag \\ y_p &=
\ \  \, y_{\intrm} \notag \\ z_p &= -x_{\intrm} \sin i + z_{\intrm}
\cos i.
\end{align}

%% Apply PA offset between major axis and slit: transform to slit coordinates
Finally, we apply the position angle offset, \deltPA, between the galaxy major axis and the slit. 
We rotate the projected model by \deltPA in the $\hat{x}_p -\hat{y}_p$ plane, mapping the projected coordinates 
into observed coordinates relative to the slit layout (see the right panel of Figure~\ref{fig:appendix_coords}) by
\begin{align}
\label{eq:xyz_slit}
x &= \ \  \, x_p \cos \deltPA + y_p \sin \deltPA \notag \\ y &= -x_p
\sin \deltPA + y_p \cos \deltPA + y_0 \notag \\ z &= \ \  \, z_p, 
\end{align}
where we also allow for an offset of the object center relative to the slit center in the $\hat{y}$ direction, 
through the parameter $y_0$. If we invert this set of coordinate transformations, we may calculate the 
intrinsic position $(x_{\intrm}, y_{\intrm}, z_{\intrm})$ and intrinsic radius $r_{\intrm}$ within the galaxy for any given point 
$(x,y,z)$ in the slit coordinate system (see Figure~\ref{fig:appendix_coords}, right panel).

% Arctan model for disk: kinematic profile assumption
To model the kinematics of a disk galaxy, we adopt the arctan model for rotation in exponential disks 
(\citealt{Courteau97}, \citealt{Miller11}),  
\begin{equation}
\label{eq:v_rot}
V\left(r, r_t, V_a\right) = \frac{2}{\pi} \, V_a \arctan
\left(\frac{r}{r_t}\right),
\end{equation}
where $V_a$ is the asymptotic velocity and $r_t$ is the turnover radius, which encodes a transition between 
the rising and flat parts of the rotation curve \citep{Courteau97}.

% Intrinsic rotation profile to projected, LOS kinematics
We must account for LOS velocity reductions due to projection effects.  
First, the inclination reduces the LOS velocity by a factor of $\sin i$. The LOS velocity is also reduced 
by $\cos \psi$, which accounts for the position of every point around the rotational axis of the galaxy. 
Together, the LOS velocity at each point $(x,y,z)$ is 
\begin{align}
\label{eq:v_los}
V_{\mathrm{los}} &(x,y,z) = V \left( r , r_t, V_a \right) \cdot \sin i
\cdot \cos \psi, 
\end{align}
where $r=r_{\intrm}$ and $\cos \psi$ are evaluated given the slit coordinates $(x,y,z)$. 
We assume the rotation is independent of $z_{\intrm}$, 
so the model  consists of nested cylindrical shells of varying radii, with each shell rotating at the appropriate velocity. 
We show an example LOS velocity field of a galaxy, integrated along the line-of-sight, in the left panel of Figure~\ref{fig:v_los_example}.

% Case of non-ideal disk
However, our galaxies may not be ideal disks. Thus the galaxies may also have an intrinsic velocity dispersion component, 
as is the case with thickened disks.  We assume the intrinsic dispersion component $\sigmavint$ is 
constant over the whole disk,  or $\sigma(r) = \sigmavint$.

% Necessity of intensity profile. Intro of Sersic profile.
The composite kinematic profile of our model, as would be observed with slit spectroscopy, consists of the 
combination of all the kinematic information of the portions of the galaxy lying within each slit pixel.  The relative 
weights of the individual kinematic components are determined by the associated intensities.  Thus we must 
consider the light distribution of our galaxy model. We assume that the light follows a modified S\'ersic intensity profile,
\begin{align}
\label{eq:I_sersic}
I(& r, n, \re, z_{\intrm}) = \\ 
& I_e \exp \left\{ -b_n \left[ \left( \frac{r}{\re}\right)^{1/n} - 1\right] \right\}
	\exp\left[-\frac{z_{\intrm}}{q_0 \re}\right], \notag
\end{align}
where $n$ and $\re$ are set to the \textsc{Galfit} best-fit parameters, $b_n \approx 2 n - 0.324$  \citep{Ciotti99}, 
and approximating the vertical scale height as $z_0 = q_0 \re$.  We show a simple example intensity profile, 
integrated over the line-of-sight, in the center panel of Figure~\ref{fig:v_los_example}.

% Construct composite 2D line model:
% I, V def
The composite 2D line model is constructed by combining the line-of-sight velocity information and the 
assumed intensity profile. We begin by calculating the intensity $I(x,y,z)$ and observed velocity
$V_{\mathrm{los}}(x,y,z)$  at every point in our slit coordinates. 
% Add sigmavint
To include the velocity dispersion, $\sigma_v(x,y,z)$, we assume that at every point $(x,y,z)$ the intensity has 
a gaussian distribution with wavelength, with center 
$\lambda_{\mathrm{los}}(x,y,z) = (1 + V_{\mathrm{los}}(x,y,z)/c) \lambda_0$ 
and standard deviation $\sigma_{\lambda} = (\sigmavint/c)\lambda_0$. 
We thus expand our intensity cube into wavelength space as
\begin{equation}
\label{eq:intensity_xyzlam}
I(x,y,z,\lambda) = \frac{I(x,y,z)}{\sigma_{\lambda} \sqrt{2\pi}}
\exp\left[-\frac{(\lambda-\lambda_{\mathrm{los}}(x,y,z))^2}{2\sigma_{\lambda}^2}\right],
\end{equation}
where we normalize the intensity distribution to ensure $\int_{\lambda} I(x,y,z,\lambda) d\lambda = I(x,y,z)$.

% Collapse z
We collapse the intensity over the $z$ (line-of-sight) direction, 
\begin{equation}
I(x,y,\lambda) = \int_{-\infty}^{+\infty} I(x,y,z,\lambda) dz,
\end{equation}
to estimate the observable spectral cube, which contains the combined line-of-sight velocity and velocity 
dispersion at every point $(x,y)$.

% Convolve with seeing 
The observed intensity is convolved with the atmospheric seeing.  We model the PSF as a 2D gaussian and take 
the blurred intensity cube to be $I(x,y,\lambda) \otimes \mathrm{PSF}(x,y)$.
% Collapse in slit direction: x
The MOSFIRE spectra are taken through a slit, so there is only one dimension of spatial information. 
Thus we collapse the intensity model in the slit width axis, $x$, over the width of the slit, $x_{\aper}$ by taking
\begin{equation} 
I(y,\lambda) =  \int_{-x_{\aper}/2}^{+x_{\aper}/2}
\left[I(x,y,\lambda) \otimes \mathrm{PSF}(x,y)\right] dx.
\end{equation}

% Convolve with instrument resolution
Finally, we include the effects of instrumental resolution by convolving the model with a Gaussian with width 
$\sigma_{\lambda, \, {\mathrm{inst}}}$ (measured from the width of sky lines).  
An example \Halpha emission line model is shown in the right panel of Figure~\ref{fig:v_los_example}.

% Practical approximations:
In practice, we generate a model by performing these calculation over a finite grid of values in $x, y, z$, and $\lambda$. 
We set $x_{\aper} = 0\farcs7 $ (MOSFIRE slit width), and $y_{\aper}$ equal to the spatial extent of the trimmed 2D 
spectrum to which we will compare the model.  We set $z_{\aper} = y_{\aper}$, to probe the same spatial extent both 
along the line-of-sight and along the slit.  We allow for sub-pixel sampling, and set the number of sub-pixels in 
$x, y, z$ so that the  sub-pixel width in each dimension is nearly equal, while preserving an integer number of 
whole pixels in the spectrum spatial direction, $y$.  Additionally, we pad the grid by 
$0. 5 \, \mathrm{FWHM}_{\mathrm{seeing}} \, \mathrm{arcsec}$ on both sides in the 
$x$ and $y$ directions, to accurately calculate the seeing-convolved intensity over the full $x_{\aper}, y_{\aper}$. 
We sample the model at the wavelengths of the associated 2D spectrum in the range around \Halpha.  
The array $I(y, \lambda)$ is oversampled in the $y$ direction.  Finally, we re-bin the data to match the observed 
spatial pixel size by adding the sub-pixels in $y$. The final model $f_{\mathrm{model}}(y,\lambda)$ now samples the intensity at 
the exact spatial positions $y$ and wavelengths $\lambda$ covered by the data.

% Model parameter dependence recap
The resulting model of the observed kinematic signature of a disk galaxy depends on fixed parameters 
$\deltPA$, $n$, $\re$, and $b/a$, all derived from the S\'ersic fits performed using \textsc{Galfit}. 
The seeing FWHM and instrument resolution are additional fixed parameters. 
Because we do not probe the flat part of the rotation curves for our galaxy sample, we fix $y_0$ and $\lambda_0$. 
We mask missing pixels and skylines for the 2D spectrum, collapse over $\lambda$ and
fix $y_0$ to the peak of a Gaussian fit. We similarly collapse over $y$ to fit $\lambda_0$. 
The free parameters are $V_a$, $r_t$, and \sigmavint.  $V_a$ and $r_t$ describe the arctan disk rotation model, 
while \sigmavint  introduces additional broadening in the wavelength direction.

%%%%%%%%%%%%%%%%%%%%%%%%%%%%%%%%%%%%%%%%%%%%%%%%%%
%%%%%%%%%%%%%%%%%%%%%%%%%%%%%%%%%%%%%%%%%%%%%%%%%%

\subsection{Procedure for measuring kinematics from 2D emission lines}
\label{sec:appendix2DsubB}

%%%%%%%%%%%%%%
% DATA EXTRACTION AND FITTING PROCEDURE:
In this appendix we describe how we measure rotation and dispersion velocities from \Halpha emission lines that 
exhibit resolved rotation. For each object, we start by subtracting out the continuum from the \Halpha 2D spectrum.  
First, we measure the continuum slope of the optimally extracted 1D spectrum using a noise-weighted linear fit in 
the wavelength range  $6454.6 \mathrm{\AA} \leq \lambda/(1+z_{\mathrm{MOS}}) \leq 6674.6 \mathrm{\AA}$, 
where we mask the \Halpha and \NII lines from 
$6533.6 \mathrm{\AA} \leq \lambda/(1+z_{\mathrm{MOS}})  \leq 6599.6 \mathrm{\AA}$.  
We then assume that the slope of the continuum in each spatial slice of the 2D spectrum is equal to the 1D 
continuum slope value, and perform a weighted linear fit in each spatial slice where only the intercept is allowed to 
vary. We subtract the best-fit continuum from each spatial slice to leave only the \Halpha line emission.  Next, we trim the 
continuum-subtracted emission line 2D spectrum to the wavelength range 
$6555.6 \mathrm{\AA} \leq \lambda/(1+z_{\mathrm{MOS}}) \leq 6573.6 \mathrm{\AA}$,  
to exclude the \NII emission lines from our resolved line fitting. As we exclude objects with outflows or AGN with very 
broad emission lines, \NII contamination within this trimmed range should be minimal.  We also trim the spectrum in the 
spatial direction so that only the positive emission line is retained.

% Mask definitions
To ensure that the model comparisons include only high S/N portions of each spectrum, we construct a mask $m(x,y)$, 
where $x$ is the wavelength dispersion direction of the spectrum and $y$ is the spatial direction. 
First, we mask pixels with missing data.  Second, we mask rows where $\mathrm{S/N}(y) < 2$, 
leaving only contiguous rows with high S/N unmasked. 
The row S/N is estimated to be the total row flux over the total row error, with the pixel errors added in quadrature.  
We mask pixels with missing data or columns affected by skyline contamination in this S/N measurement.  
The columns affected by skyline contamination are identified as those where the total column error, added in 
quadrature, is 2 and 3 times greater than the median error of all columns in the spectrum, for the K and H bands, 
respectively.\footnote{The  sky background in the K band is higher than in H, 
so we adopt a less conservative skyline identification  criterion for the K band.} 
We do not mask pixels affected by skyline contamination when fitting the models to the data, as these pixels are 
down-weighted in the fitting procedure because of their large errors relative to the  non-contaminated pixels.

% Example posterior: one column
\begin{figure}
	\centering
	\hglue -9.5pt
	\includegraphics[width=0.5\textwidth]{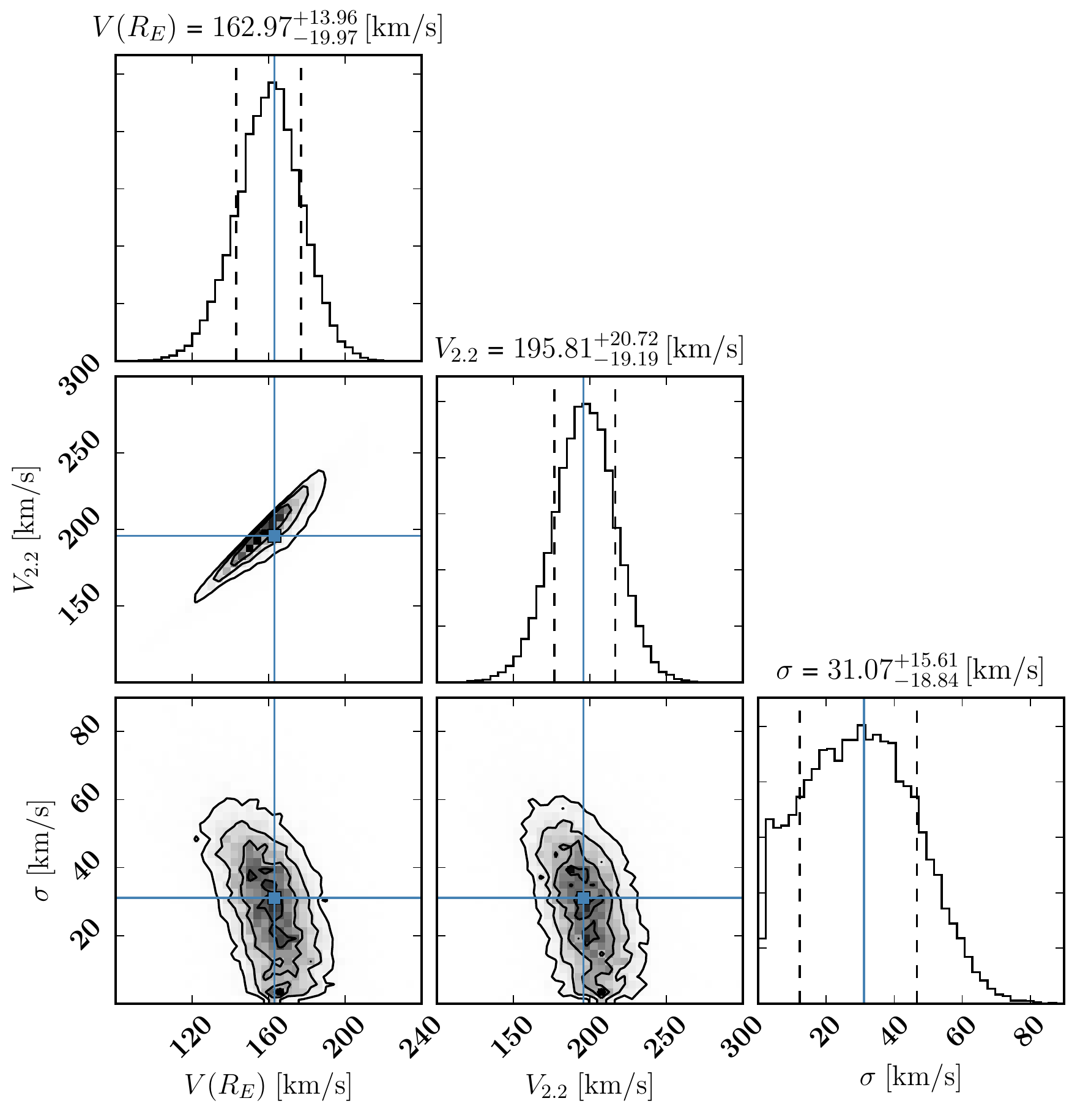}
	\caption{
		Example posterior distribution for $V(\re)$, $V_{2.2}$, and $\sigmavint$ for COSMOS-13701 (the 
		second galaxy shown in Figures \ref{fig:example_bestfits} and \ref{fig:light_profiles}). 
		The best-fit value of each parameter is taken to be the peak of the respective marginalized posterior distributions, 
		and are shown as the blue lines in the histograms. The lower and upper 68\% confidence intervals on each parameter 
		are shown as the dashed black lines. 
		The best-fit values are also shown as blue lines and blue squares in the various two-parameter posterior plots. 
		Figure made using \texttt{corner.py} \citep{corner_plot}.
	}
	\label{fig:example_mcmc_posterior}
\end{figure}

% Match intensity of model to data
To find the best-fit model to the data, we first match the model intensity to the data intensity profile.  
We perform a weighted least squares fit of the model $f_{\mathrm{model}}(x, y)$ to the data 
$f(x,y)$ (with errors $\sigma_{f}(x,y)$) at each $y$ and measure the appropriate scaling $S(y)$ 
between the model and data spatial rows
%\vspace{-4pt}
\begin{equation}
S(y) = \frac{\sum_{x} m(x,y) \left[ f(x,y) f_{\mathrm{model}}(x, y)/ \sigma_{f}(x, y)^2 \right] }
{\sum_{x}  m(x,y) \left[ f_{\mathrm{model}}(x, y)^2/ \sigma_{f}(x, y)^2 \right] }, 
\label{eq:scaling_y}
\end{equation}
where we mask missing data and columns contaminated by skyline emission with $m(x, y)$ 
(discussed in Section~\ref{sec:rot_meas}). We do not fix the scaling to the convolved and integrated (in the slit 
direction) \textsc{Galfit} profile, as the line emission may be distributed differently (see \citealt{Nelson12}).  
Nonetheless, in modeling the kinematics, we adopt a S\'ersic stellar light profile to determine the intensity-weighted velocities.  
In Section~\ref{sec:light_profiles}, we show that most objects have similar stellar light and \Halpha profiles, 
and also discuss the implications for the modeling results when the \Halpha profile differs from the stellar light profile.

% Goodness of fit metric definition
The goodness-of-fit of the model is determined using a weighted $\chi^2$ value. We choose the following weighting scheme to 
up-weight lower S/N rows, so the information in the fainter parts of the rotation curve is not lost: 
\begin{equation}
\label{eq:weights}
w_y =  
%\frac{1}{\mathrm{S/N}(y)} .
\left[\mathrm{S/N}(y)\right]^{-1}. 
\end{equation}
The weighted goodness-of-fit criterion is then 
\begin{align}
&\chi^2_{\mathrm{weighted}} = \notag \\  &\quad \sum\limits_{x,y} w_y
  \left[ m(x,y)  \frac{ f(x,y)-  S(y) f_{\mathrm{model}}(x,y)  }  {
      \sigma_{f}(x, y) } \right]^2 
\label{eq:chisq}
\end{align}
where we mask missing data and low S/N rows with $m(x,y)$, and the spectrum is dispersed in the $x$ direction.

% MCMC/emcee model fitting
We use the \texttt{python} MCMC package, \texttt{emcee} \citep{Foreman-Mackey13}  to find the best-fit models and 
confidence intervals.  We define flat priors $\log p(X)$ for each parameter $X$ ($V_a$, $r_t$, $\sigmavint$), 
with bounds calculated based on the spatial and wavelength coverage of the  trimmed 2D spectrum, yielding 
composite prior 
$\log p(V_a, r_t, \sigmavint) = \Sigma_{\{X=V_a, r_t, \sigmavint\}} \log p(X)$. 
The log posterior probability is taken to be 
\begin{align}
\label{eq:posterior}
\log &P(V_a, r_t, \sigmavint | \lambda, f) =  \\
& \log \mathcal{L} (f | \lambda, V_a, r_t, \sigmavint) + 
\log p(V_a, r_t, \sigmavint) + \mathrm{const}, \notag
\end{align} 
with log likelihood probability $\log \mathcal{L}=  - 0.5 \chi^2_{\mathrm{weighted}}$.

% #####################################
%% Figure for PV appendix:
% Halpha Position-Velocity Diagrams for galaxies with detected rotation
\begin{figure*}[ht!]
	\centering
	\hglue -4pt
	\includegraphics[width=1.01\textwidth]{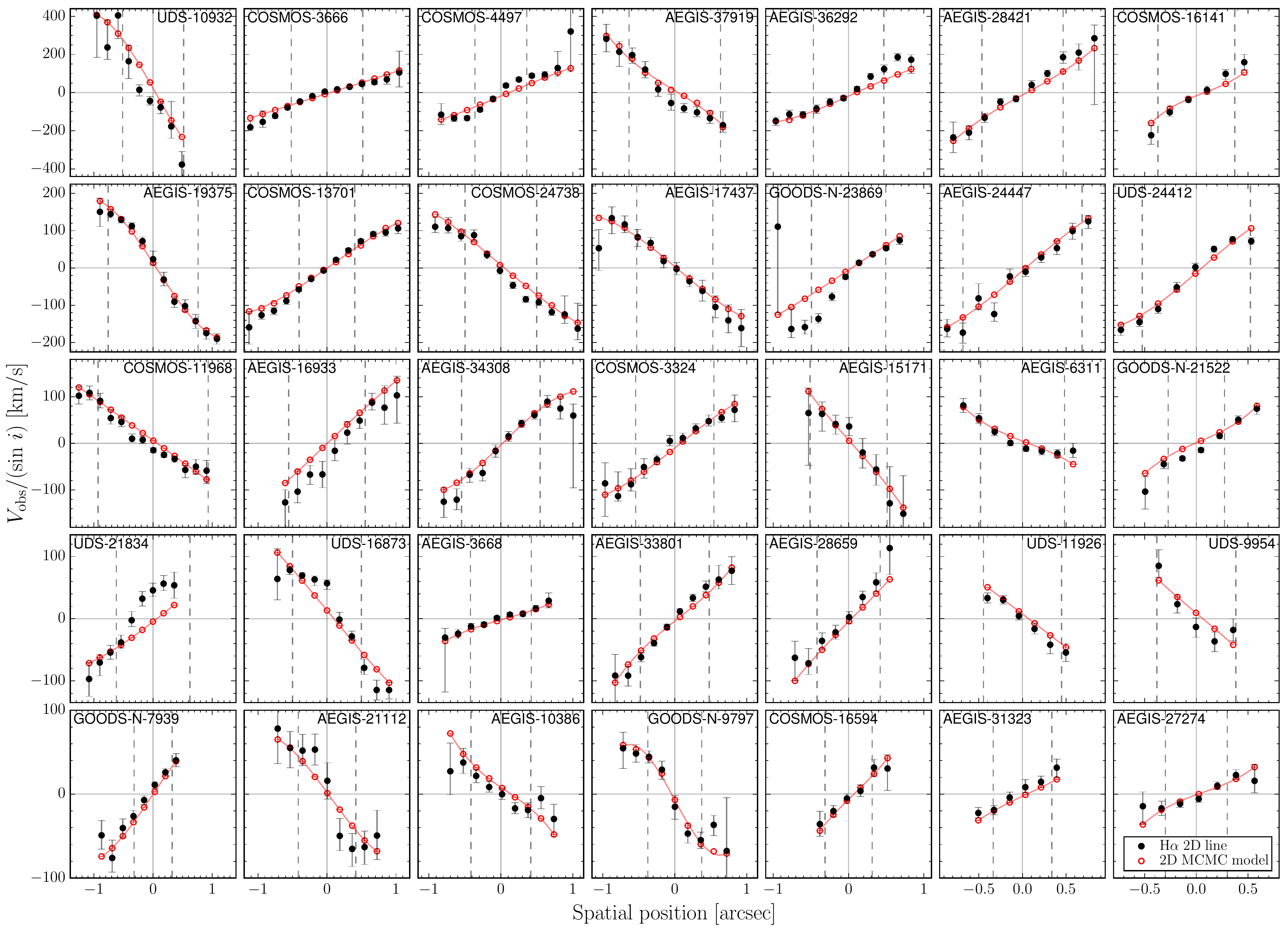}
	\caption{ 
		Position-velocity diagrams for the 35 MOSDEF galaxies with detected rotation. 
		The velocity profile corrected for inclination ($V_{\mathrm{obs}}(y)/(\sin i)$; $y$ axis) 
		is measured versus the spatial position ($x$ axis) from the 2D \Halpha emission line spectrum (black circles) 
		and from the best-fit 2D line model (open red circles) for each object. 
		The error bars do not include uncertainties from the inclination correction. 
		For reference, we fit a third order polynomial (red line) to the model velocity profile. 
		The vertical grey dashed lines show the projected effective radius convolved to match the MOSFIRE seeing.
		%Interp:
		The velocity profiles of the observed and model spectra are in good agreement.
	}
	\label{fig:PV_rot}
\end{figure*}

% #####################################

% Measuring best-fit values
The rotation curve turnover is not well constrained in our data, so there is a degeneracy in the values of $V_a$ and $r_t$. 
However, the values of $V(\re)$  and $V_{2.2} = V(2.2r_s)$ are much better constrained. 
Thus we calculate $V(\re)$ and $V_{2.2}$ for each pair of $(V_a, r_t)$ values in the posterior sampling, 
to determine the posterior distributions on $V(\re)$ and $V_{2.2}$.  We take the best-fit values of $V(\re)$, $V_{2.2}$, 
and $\sigmavint$ to be the peaks of the respective marginalized posterior distributions, and calculate the confidence 
intervals using the lower and upper 68-percentile bounds of the posterior distributions 
(e.g., see Figure~\ref{fig:example_mcmc_posterior}).

% ###############################################################################
% ###############################################################################
% ###############################################################################

\subsection{Position-velocity diagrams for galaxies with detected~rotation}
\label{sec:appendix_PV}

% Intro:
To demonstrate the agreement between the observed and modeled kinematics, 
we measure velocity as a function of position from both the observed and modeled 2D spectra 
for each of the 35 galaxies with detected rotation. 
% Method of measuring position-velocity diagram
For each unmasked, high S/N row (see Appendix~\ref{sec:appendix2DsubB}), we fit the flux $f(x, y)$ 
with a Gaussian and determine the central wavelength, $\lambda(y)$, constraining $\lambda$ to fall within 
$\pm 1.25 \, \mathrm{FWHM}_{\lambda, \, \mathrm{1D, \, obs}}$ of the fixed central wavelength $\lambda_0$ 
(see Appendix~\ref{sec:appendix2DsubA}). 
We then calculate $V_{\mathrm{obs}}(y)$ from $\lambda(y)$ and $\lambda_0$. 
We estimate the errors in $V_{\mathrm{obs}}(y)$ by creating 500 realizations in which we perturb the flux $f(x,y)$ 
by the errors $\sigma_{f}(x,y)$, and repeat the fitting procedure on each realization. 
Finally, we correct the observed velocities for inclination, yielding $V_{\mathrm{obs}}(y)/(\sin i)$.
% Introduce PV diagram plot
The velocity profiles are shown in Figure~\ref{fig:PV_rot}.  
The observed and model velocity profiles are in good agreement, suggesting that 
our modeling approach works well.

% ###############################################################################
% ###############################################################################
% ###############################################################################

\section{Appendix B: Inclination and aperture correction for unresolved disk galaxies}
\label{sec:appendix1D}

% Introduction: need for disk-specific "aperture" correction
If a disk galaxy is too small, it will be spatially unresolved and its rotation will not be detected. 
Additionally, some of the kinematic information may be missing 
because of slit losses. Furthermore, as our 1D spectra are optimally extracted, the observed velocity profile will depend on the 
inclination angle and the angle between the slit and the major axis of the galaxy.  In this appendix we estimate the correction 
between the intrinsic kinematics and the kinematics within the extracted aperture for galaxies without detected rotation, 
assuming that they are  rotationally supported disk galaxies. We make models of disk galaxies that account for variable
inclination angles and variable \deltPA and use these models to calculate
the integrated RMS velocity within the slit.  We follow the general
method presented in Appendix~B of \citet{vandeSande13} to calculate
the aperture correction  for a given kinematic and brightness
profile.

% Disk model details
% Coord system
The kinematics of the disk galaxy model are defined in the same way as in Appendix~\ref{sec:appendix2D}. 
The slit coordinate system relative to the intrinsic galaxy coordinates is defined following Equations \ref{eq:xyz_int}, 
\ref{eq:xyz_slit}. 
% Velocity profile assumptions: 
We assume that the rotation can be described with the arctan model (Equation~\ref{eq:v_rot}). As we have no 
spatial information, we must assume a radial profile for the rotation curve, that is determined entirely through 
turnover radius, $r_t$.  Based on the findings of \citet{Miller11}, we set $r_t = 0.4 \, r_s = 0.4 (\re/1.676)$. 
Following Equation~\ref{eq:v_los}, the relative line-of-sight radial profile of the model rotation curve is then 
$V_{\mathrm{los}} (x,y,z)/V(\re)$, in which we do not assume an absolute velocity scale. Since our galaxies may not 
be ideal disks, we assume a simple constant dispersion velocity $\sigmavint$ and a fixed value of 
$\vtosigre = V(\re)/\sigmavint$.

% Definition of RMS velocity
Following \citet{Cappellari08}, we assume the observed velocity dispersion is the square root of the second velocity moment, 
i.e. the RMS velocity,  $\Vrms ^2 = \sigma^2 + V^2$. To obtain relative quantities, we divide both sides by $V(R_E)$:
\begin{equation}
\label{eq:v_rms}
\!\!\!
\left( \frac{V_{\mathrm{RMS, \, los}} (x,y,z) }{V(\re)}  \right)^2
= \left( \frac{1}{ \vtosigre} \right)^2 + \left(
\frac{V_{\mathrm{los}} (x,y,z)}{V(\re)} \right)^2 
\end{equation}

% Intensity profile
The total observed velocity dispersion of a galaxy is the combination of the intensity-weighted velocity dispersions at every 
point of the galaxy, so we must also assume a light profile to include in our models. As in Appendix~\ref{sec:appendix2D}, 
we assume a modified S\'ersic intensity profile  $I(r,n,\re, z_{\intrm}, \sigma_z)$  (Equation~\ref{eq:I_sersic}).

% Aperture correction values for different delt PA, R_E, n, etc.
\begin{figure*}
	\centering
	\includegraphics[width=0.99\textwidth]{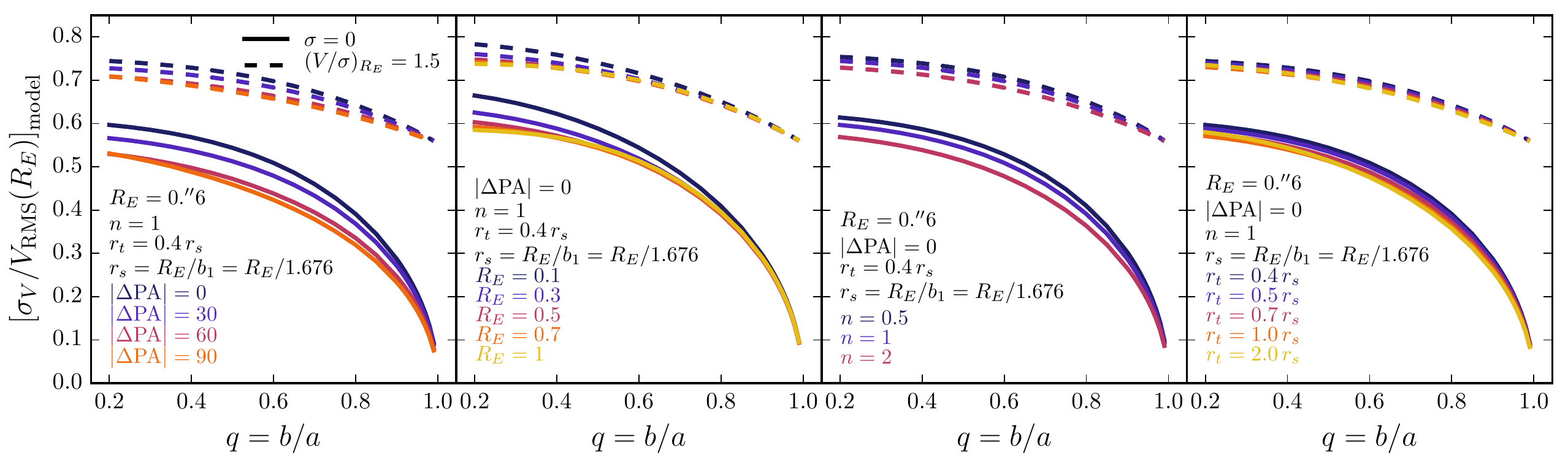}
	\caption{ 
		Aperture corrections, $\sigmavmodel/\Vrms(\re)_{\mathrm{model}}$, 
		for disk galaxies without resolved rotation, as a function of 
		(a) \deltPA, 
		(b) $\re$, 
		(c) $n$, and
		(d) $r_t$. 
		We show the aperture correction assuming no intrinsic velocity dispersion ($\sigmavint=0$, solid lines) 
		and partial rotational support ($\vtosigre=1.5$, dashed lines). 
		The non-variable parameters in each panel are set to $\deltPA=0$, $\re = 0\farcs6$, $n=1$, 
		and $r_t = 0.4 r_s = 0.4 (\re/1.676)$. 
		We assume a seeing FWHM of $0\farcs6$ for every model.
	}
	\label{fig:aper_corr}
\end{figure*}

% Intensity-weighted dispersion calculation
We calculate the intensity-weighted dispersion within the aperture from the RMS velocity relative to $V(\re)$,
following Equation B9 of \citet{vandeSande13}: 
\begin{align}
\label{eq:sigma_calculation}
&\left( \frac{\sigmavmodel } {V(\re)} \right)^2 = \\
& \mkern-28mu 
	\frac{\sum\limits_{-X}^{X} \sum\limits_{-Y}^{Y}  \left( \left[
    \sum\limits_{-Z}^{Z} \left( \frac{V_{\mathrm{RMS, los}} (x,y,z)
    }{V(\re)} \right)^2 I (x,y,z) \Delta z \right] \! \otimes
  \mathrm{PSF} \right) g(y) \Delta x \Delta y }  {\sum\limits_{-X}^{X}
  \sum\limits_{-Y}^{Y} \left(  \left[ \sum\limits_{-Z}^{Z} I (x,y,z)
    \Delta z \right] \! \otimes \mathrm{PSF}  \right) g(y) \Delta x
  \Delta y }  \notag 
\end{align}
Here we define $X = \frac{1}{2}  x_{\aper}$, $Y = \frac{1}{2}
y_{\aper}$, and $Z = \frac{1}{2} z_{\aper}$, and define $V_{\mathrm{RMS,los}}/V(\re)$ from Equation~\ref{eq:v_rms}.
We model the PSF as a 2D Gaussian with FWHM equal to that atmospheric seeing FWHM, 
and adopt the same spatial weighting function $g(y)$ as used in extracting the MOSDEF 1D spectra.

% Conversion to RMS velocity AT THE EFFECTIVE RADIUS, as used in our dynamical mass calculations
The dynamical masses of disk galaxies are calculated using the velocity at a specific radius, i.e. $V(\re)$,  
instead of an integrated velocity dispersion.  Thus, instead of calculating $\sigma_e$, the intrinsic intensity-weighted 
velocity dispersion within the effective radius $\re$,  we calculate the RMS velocity of the model at $r=\re$, 
$\Vrms(\re)_{\mathrm{model}} = \sqrt{\sigmavint^2 + V(\re)^2}$,  relative to $V(\re)$, which we write as: 
\begin{equation}
\label{eq:vrms_re}
\frac{ \Vrms(\re)_{\mathrm{model}} } { V(\re) }  = \sqrt{1 +  \frac{1}{\vtosigre^2} }. 
\end{equation}

The observed velocity dispersion corrected for both aperture and inclination effects, and converted to a RMS velocity, 
is the combination of Equations \ref{eq:sigma_calculation} and \ref{eq:vrms_re}: 
\begin{equation}
\label{eq:aper_corr}
\Vrms(\re)_{\mathrm{corr}} = \sigma_{\mathrm{obs}}  \left( \frac{
  \sigmavmodel }{ \Vrms(\re)_{\mathrm{model}} }
\right)^{-1}, 
\end{equation}
with 
\begin{equation}
\!\!\!
\frac{\sigmavmodel }{ \Vrms(\re)_{\mathrm{model}}} =
\left( \frac{\sigmavmodel }{ V(\re)} \right)
\frac{\vtosigre}{ \sqrt{ 1+\vtosigre^2} } . 
\end{equation}

% Method of applying aperture correction: Model input parameters and assumptions.
To calculate the correction $\sigmavmodel/\Vrms(\re)_{\mathrm{model}}$ for individual galaxies, we 
use the best-fit \textsc{Galfit} parameters for $n$, $\re=a$ (the semi-major axis), and $q=b/a$. 
We set $x_{\aper} = 0\farcs7$, the slit width for all observations, and set $y_{\aper} = y_{\mathrm{extract}}$, 
the actual width used to extract the 1D spectra.  We choose $z_{\aper} = y_{\aper}$, to probe the same spatial 
extent in the line-of-sight direction as we probe along the slit.

% Practical approximations/calculation methods:
In practice, we initially pad coordinate grids in the $x, y$ directions by 
$0. 5 \, \mathrm{FWHM}_{\mathrm{seeing}} \,\mathrm{arcsec}$, to accurately consider the convolution with the atmospheric 
seeing across the aperture edges. We include these pixels when calculating the collapse over  $z$ and the convolution 
with the seeing, then remove the padded pixels for the final sum within the aperture.  We sample the model over a large 
number of pixels, and choose the pixel sizes so they are nearly equal in all dimensions, $\Delta y = \Delta z \approx \Delta x$, 
with the constraint that there must be an integer number of pixels within  $x_{\aper}$ and $y_{\aper}$.

% Demonstration of aperture correction from the disk model: Figure
The effects of varying the model parameters $\axratio$, \deltPA, $\re$, $n$, and $r_t$ are demonstrated in Figure~\ref{fig:aper_corr}. 
In all cases, we assume a typical seeing of $0\farcs6$.  We adopt $x_{\aper} = 0\farcs7$, the slit width for all MOSDEF observations, 
and set $y_{\aper}= 4 R_{E, \,  \mathrm{proj+conv}}$, to approximate dependance of aperture size on the object size, 
misalignment, and seeing that is incorporated in the data extraction method.

% Discussion of trends of aperture correction with model parameters:
The inclination angle has the largest influence on the correction value.  At a fixed axis ratio, the inclusion of a finite
$\vtosigre$ value causes the largest difference in the aperture correction, as the intrinsic velocity dispersion increases 
the observed LOS velocity dispersion.  The position angle offset causes larger variations for more edge-on disks 
$\left( b/a \approx (b/a)_0 \right)$  than for disks closer to face-on ($b/a \approx 1$), as the more face-on disks are 
much closer to being round,  and the amount of the disk falling outside of the slit for any $\deltPA$ is similar. 
Variations with the S\'ersic index $n$ reflect how the different intensity profiles weight the velocity distribution. 
Changes of the assumed $r_t$ affect the rotational velocity profile, with larger $r_t$ moving the  velocity turnover to 
larger radii. When combined with the S\'ersic intensity weighting, this leads to smaller integrated velocity values.  
Finally, when the disk is aligned with the major axis, the aperture correction varies little with the effective radius $\re$. 
The aperture correction varies more with $\re$ when combined with larger $\deltPA$ offsets.

% ###############################################################################


\begin{thebibliography}{130}
\expandafter\ifx\csname natexlab\endcsname\relax\def\natexlab#1{#1}\fi

\bibitem[{Bell \& de~Jong(2001)}]{Bell01}
Bell, E.~F., \& de~Jong, R.~S. 2001, \apj, 550, 212

\bibitem[{Belli {et~al.}(2014)Belli, Newman, \& Ellis}]{Belli14a}
Belli, S., Newman, A.~B., \& Ellis, R.~S. 2014, \apj, 783, 117

\bibitem[{Bezanson {et~al.}(2015)Bezanson, Franx, \& van Dokkum}]{Bezanson15}
Bezanson, R., Franx, M., \& van Dokkum, P.~G. 2015, \apj, 799, 148

\bibitem[{Bezanson {et~al.}(2013)Bezanson, van Dokkum, van~de Sande, Franx, \&
  Kriek}]{Bezanson13}
Bezanson, R., van Dokkum, P., van~de Sande, J., Franx, M., \& Kriek, M. 2013,
  \apj, 764, L8

\bibitem[{Blanton \& Moustakas(2009)}]{Blanton09}
Blanton, M.~R., \& Moustakas, J. 2009, \araa, 47, 159

\bibitem[{Blumenthal {et~al.}(1984)Blumenthal, Faber, Primack, \&
  Rees}]{Blumenthal84}
Blumenthal, G.~R., Faber, S.~M., Primack, J.~R., \& Rees, M.~J. 1984, Nature,
  311, 517

\bibitem[{Bouch\'{e} {et~al.}(2007)Bouch\'{e}, Cresci, Davies, Eisenhauer,
  {Forster Schreiber}, Genzel, Gillessen, Lehnert, Lutz, Nesvadba, Shapiro,
  Sternberg, Tacconi, Verma, Cimatti, Daddi, Renzini, Erb, Shapley, \&
  Steidel}]{Bouche07}
Bouch\'{e}, N., Cresci, G., Davies, R., {et~al.} 2007, \apj, 671, 303

\bibitem[{Bournaud {et~al.}(2011)Bournaud, Chapon, Teyssier, Powell, Elmegreen,
  Elmegreen, Duc, Contini, Epinat, \& Shapiro}]{Bournaud11}
Bournaud, F., Chapon, D., Teyssier, R., {et~al.} 2011, \apj, 730, 4

\bibitem[{Brammer {et~al.}(2012)Brammer, van Dokkum, Franx, Fumagalli, Patel,
  Rix, Skelton, Kriek, Nelson, Schmidt, Bezanson, da~Cunha, Erb, Fan,
  {F\"{o}rster Schreiber}, Illingworth, Labb\'{e}, Leja, Lundgren, Magee,
  Marchesini, McCarthy, Momcheva, Muzzin, Quadri, Steidel, Tal, Wake, Whitaker,
  \& Williams}]{Brammer12}
Brammer, G.~B., van Dokkum, P.~G., Franx, M., {et~al.} 2012, \apjs, 200, 13

\bibitem[{Brewer {et~al.}(2012)Brewer, Dutton, Treu, Auger, Marshall,
  Barnab\`{e}, Bolton, Koo, \& Koopmans}]{Brewer12}
Brewer, B.~J., Dutton, A.~A., Treu, T., {et~al.} 2012, \mnras, 422, 3574

\bibitem[{Buitrago {et~al.}(2014)Buitrago, Conselice, Epinat, Bedregal,
  Gr{\"{u}}tzbauch, \& Weiner}]{Buitrago14}
Buitrago, F., Conselice, C.~J., Epinat, B., {et~al.} 2014, \mnras, 439, 1494

\bibitem[{Cacciato {et~al.}(2012)Cacciato, Dekel, \& Genel}]{Cacciato12}
Cacciato, M., Dekel, A., \& Genel, S. 2012, \mnras, 421, 818

\bibitem[{Calzetti {et~al.}(2000)Calzetti, Armus, Bohlin, Kinney, Koornneef, \&
  Storchi?Bergmann}]{Calzetti00}
Calzetti, D., Armus, L., Bohlin, R.~C., {et~al.} 2000, \apj, 533, 682

\bibitem[{Cappellari(2008)}]{Cappellari08}
Cappellari, M. 2008, \mnras, 390, 71

\bibitem[{Cappellari {et~al.}(2006)Cappellari, Bacon, Bureau, Damen, Davies,
  de~Zeeuw, Emsellem, Falcon-Barroso, Krajnovic, Kuntschner, McDermid,
  Peletier, Sarzi, van~den Bosch, \& van~de Ven}]{Cappellari06}
Cappellari, M., Bacon, R., Bureau, M., {et~al.} 2006, \mnras, 366, 1126

\bibitem[{{Cardelli} {et~al.}(1989){Cardelli}, {Clayton}, \&
  {Mathis}}]{Cardelli89}
{Cardelli}, J.~A., {Clayton}, G.~C., \& {Mathis}, J.~S. 1989, \apj, 345, 245

\bibitem[{Ceverino {et~al.}(2012)Ceverino, Dekel, Mandelker, Bournaud, Burkert,
  Genzel, \& Primack}]{Ceverino12}
Ceverino, D., Dekel, A., Mandelker, N., {et~al.} 2012, \mnras, 420, 3490

\bibitem[{Chabrier(2003)}]{Chabrier03}
Chabrier, G. 2003, \pasp, 115, 763

\bibitem[{Chapman {et~al.}(2005)Chapman, Blain, Smail, \& Ivison}]{Chapman05}
Chapman, S.~C., Blain, A.~W., Smail, I., \& Ivison, R.~J. 2005, \apj, 622, 772

\bibitem[{Ciotti \& Bertin(1999)}]{Ciotti99}
Ciotti, L., \& Bertin, G. 1999, A\&A, 352, 447

\bibitem[{Coil {et~al.}(2015)Coil, Aird, Reddy, Shapley, Kriek, Siana,
  Mobasher, Freeman, Price, \& Shivaei}]{Coil15}
Coil, A.~L., Aird, J., Reddy, N., {et~al.} 2015, \apj, 801, 35

\bibitem[{Conroy \& Gunn(2010)}]{Conroy10a}
Conroy, C., \& Gunn, J.~E. 2010, \apj, 712, 833

\bibitem[{Conroy {et~al.}(2009)Conroy, Gunn, \& White}]{Conroy09}
Conroy, C., Gunn, J.~E., \& White, M. 2009, \apj, 699, 486

\bibitem[{Contini {et~al.}(2012)Contini, Garilli, {Le F{\`{e}}vre},
  Kissler-Patig, Amram, Epinat, Moultaka, Paioro, Queyrel, Tasca, Tresse,
  Vergani, L{\'{o}}pez-Sanjuan, \& Perez-Montero}]{Contini12}
Contini, T., Garilli, B., {Le F{\`{e}}vre}, O., {et~al.} 2012, A{\&}A, 539, A91

\bibitem[{Courteau(1997)}]{Courteau97}
Courteau, S. 1997, \aj, 114, 2402

\bibitem[{Daddi {et~al.}(2008)Daddi, Dannerbauer, Elbaz, Dickinson, Morrison,
  Stern, \& Ravindranath}]{Daddi08}
Daddi, E., Dannerbauer, H., Elbaz, D., {et~al.} 2008, \apj, 673, L21

\bibitem[{Daddi {et~al.}(2010)Daddi, Bournaud, Walter, Dannerbauer, Carilli,
  Dickinson, Elbaz, Morrison, Riechers, Onodera, Salmi, Krips, \&
  Stern}]{Daddi10}
Daddi, E., Bournaud, F., Walter, F., {et~al.} 2010, \apj, 713, 686

\bibitem[{Dalcanton {et~al.}(1997)Dalcanton, Spergel, \& Summers}]{Dalcanton97}
Dalcanton, J.~J., Spergel, D.~N., \& Summers, F.~J. 1997, \apj, 482, 659

\bibitem[{Dav\'{e}(2008)}]{Dave08}
Dav\'{e}, R. 2008, \mnras, 385, 147

\bibitem[{Dekel \& Birnboim(2006)}]{Dekel06}
Dekel, A., \& Birnboim, Y. 2006, \mnras, 368, 2

\bibitem[{Dekel {et~al.}(2009)Dekel, Sari, \& Ceverino}]{Dekel09}
Dekel, A., Sari, R., \& Ceverino, D. 2009, \apj, 703, 785

\bibitem[{Dutton(2009)}]{Dutton09}
Dutton, A.~A. 2009, \mnras, 396, 121

\bibitem[{Dutton {et~al.}(2011{\natexlab{a}})Dutton, Conroy, van~den Bosch,
  Simard, Mendel, Courteau, Dekel, More, \& Prada}]{Dutton11a}
Dutton, A.~A., Conroy, C., van~den Bosch, F.~C., {et~al.} 2011{\natexlab{a}},
  \mnras, 345, 322

\bibitem[{Dutton {et~al.}(2011{\natexlab{b}})Dutton, van~den Bosch, Faber,
  Simard, Kassin, Koo, Bundy, Huang, Weiner, Cooper, Newman, Mozena, \&
  Koekemoer}]{Dutton11}
Dutton, A.~A., van~den Bosch, F.~C., Faber, S.~M., {et~al.} 2011{\natexlab{b}},
  \mnras, 410, 1660

\bibitem[{Elmegreen \& Elmegreen(2006)}]{Elmegreen06}
Elmegreen, B.~G., \& Elmegreen, D.~M. 2006, \apj, 650, 644

\bibitem[{Elmegreen {et~al.}(2009)Elmegreen, Elmegreen, Marcus, Shahinyan, Yau,
  \& Petersen}]{Elmegreen09}
Elmegreen, D.~M., Elmegreen, B.~G., Marcus, M.~T., {et~al.} 2009, \apj, 701,
  306

\bibitem[{Elmegreen {et~al.}(2007)Elmegreen, Elmegreen, Ravindranath, \&
  Coe}]{Elmegreen07}
Elmegreen, D.~M., Elmegreen, B.~G., Ravindranath, S., \& Coe, D.~a. 2007, \apj,
  658, 763

\bibitem[{Epinat {et~al.}(2010)Epinat, Amram, Balkowski, \&
  Marcelin}]{Epinat10}
Epinat, B., Amram, P., Balkowski, C., \& Marcelin, M. 2010, \mnras, 401, 2113

\bibitem[{Epinat {et~al.}(2008)Epinat, Amram, Marcelin, Balkowski, Daigle,
  Hernandez, Chemin, Carignan, Gach, \& Balard}]{Epinat08}
Epinat, B., Amram, P., Marcelin, M., {et~al.} 2008, \mnras, 388, 500

\bibitem[{Epinat {et~al.}(2009)Epinat, Contini, {Le F\`{e}vre}, Vergani,
  Garilli, Amram, Queyrel, Tasca, \& Tresse}]{Epinat09}
Epinat, B., Contini, T., {Le F\`{e}vre}, O., {et~al.} 2009, A\&A, 504, 789

\bibitem[{Epinat {et~al.}(2012)Epinat, Tasca, Amram, Contini, {Le F{\`{e}}vre},
  Queyrel, Vergani, Garilli, Kissler-Patig, Moultaka, Paioro, Tresse, Bournaud,
  L{\'{o}}pez-Sanjuan, \& Perret}]{Epinat12}
Epinat, B., Tasca, L., Amram, P., {et~al.} 2012, A{\&}A, 539, A92

\bibitem[{Erb {et~al.}(2006)Erb, Steidel, Shapley, Pettini, Reddy, \&
  Adelberger}]{Erb06c}
Erb, D.~K., Steidel, C.~C., Shapley, A.~E., {et~al.} 2006, \apj, 646, 107

\bibitem[{Fall \& Efstathiou(1980)}]{Fall80}
Fall, S.~M., \& Efstathiou, G. 1980, \mnras, 193, 189

\bibitem[{Fan {et~al.}(2001)Fan, Strauss, Schneider, Gunn, Lupton, Becker,
  Davis, Newman, Richards, White, {Anderson, Jr.}, Annis, Bahcall, Brunner,
  Csabai, Hennessy, Hindsley, Fukugita, Kunszt, Ivezi\'{c}, Knapp, McKay, Munn,
  Pier, Szalay, \& York}]{Fan01}
Fan, X., Strauss, M.~A., Schneider, D.~P., {et~al.} 2001, \aj, 121, 54

\bibitem[{Foreman-Mackey {et~al.}(2013)Foreman-Mackey, Hogg, Lang, \&
  Goodman}]{Foreman-Mackey13}
Foreman-Mackey, D., Hogg, D.~W., Lang, D., \& Goodman, J. 2013, \pasp, 125, 306

\bibitem[{Foreman-Mackey {et~al.}(2014)Foreman-Mackey, Price-Whelan, Ryan,
  Emily, Smith, Barbary, Hogg, \& Brewer}]{corner_plot}
Foreman-Mackey, D., Price-Whelan, A., Ryan, G., {et~al.} 2014, corner.py
  v0.1.1, Zenodo, doi:10.5281/zenodo.11020

\bibitem[{{F\"{o}rster Schreiber} {et~al.}(2006){F\"{o}rster Schreiber},
  Genzel, Lehnert, Bouche, Verma, Erb, Shapley, Steidel, Davies, Lutz,
  Nesvadba, Tacconi, Eisenhauer, Abuter, Gilbert, Gillessen, \&
  Sternberg}]{ForsterSchreiber06}
{F\"{o}rster Schreiber}, N.~M., Genzel, R., Lehnert, M.~D., {et~al.} 2006,
  \apj, 645, 17

\bibitem[{{F\"{o}rster Schreiber} {et~al.}(2009){F\"{o}rster Schreiber},
  Genzel, Bouch\'{e}, Cresci, Davies, Buschkamp, Shapiro, Tacconi, Hicks,
  Genel, Shapley, Erb, Steidel, Lutz, Eisenhauer, Gillessen, Sternberg,
  Renzini, Cimatti, Daddi, Kurk, Lilly, Kong, Lehnert, Nesvadba, Verma,
  McCracken, Arimoto, Mignoli, \& Onodera}]{ForsterSchreiber09}
{F\"{o}rster Schreiber}, N.~M., Genzel, R., Bouch\'{e}, N., {et~al.} 2009,
  \apj, 706, 1364

\bibitem[{{F\"{o}rster Schreiber} {et~al.}(2014){F\"{o}rster Schreiber},
  Genzel, Newman, Kurk, Lutz, Tacconi, Wuyts, Bandara, Burkert, Buschkamp,
  Carollo, Cresci, Daddi, Davies, Eisenhauer, Hicks, Lang, Lilly, Mainieri,
  Mancini, Naab, Peng, Renzini, Rosario, {Shapiro Griffin}, Shapley, Sternberg,
  Tacchella, Vergani, Wisnioski, Wuyts, \& Zamorani}]{ForsterSchreiber14}
{F\"{o}rster Schreiber}, N.~M., Genzel, R., Newman, S.~F., {et~al.} 2014, \apj,
  787, 38

\bibitem[{Freeman(1970)}]{Freeman70}
Freeman, K.~C. 1970, \apj, 160, 811

\bibitem[{Gabor \& Bournaud(2014)}]{Gabor14}
Gabor, J.~M., \& Bournaud, F. 2014, \mnras, 441, 1615

\bibitem[{Genel {et~al.}(2012)Genel, Dekel, \& Cacciato}]{Genel12}
Genel, S., Dekel, A., \& Cacciato, M. 2012, \mnras, 425, 788

\bibitem[{Genzel {et~al.}(2008)Genzel, Burkert, Bouch\'{e}, Cresci,
  {F\"{o}rster Schreiber}, Shapley, Shapiro, Tacconi, Buschkamp, Cimatti,
  Daddi, Davies, Eisenhauer, Erb, Genel, Gerhard, Hicks, Lutz, Naab, Ott,
  Rabien, Renzini, Steidel, Sternberg, \& Lilly}]{Genzel08}
Genzel, R., Burkert, A., Bouch\'{e}, N., {et~al.} 2008, \apj, 687, 59

\bibitem[{Genzel {et~al.}(2010)Genzel, Tacconi, Gracia-Carpio, Sternberg,
  Cooper, Shapiro, Bolatto, Bouch\'{e}, Bournaud, Burkert, Combes, Comerford,
  Cox, Davis, Schreiber, Garcia-Burillo, Lutz, Naab, Neri, Omont, Shapley, \&
  Weiner}]{Genzel10}
Genzel, R., Tacconi, L.~J., Gracia-Carpio, J., {et~al.} 2010, \mnras, 407, 2091

\bibitem[{Genzel {et~al.}(2011)Genzel, Newman, Jones, {F\"{o}rster Schreiber},
  Shapiro, Genel, Lilly, Renzini, Tacconi, Bouch\'{e}, Burkert, Cresci,
  Buschkamp, Carollo, Ceverino, Davies, Dekel, Eisenhauer, Hicks, Kurk, Lutz,
  Mancini, Naab, Peng, Sternberg, Vergani, \& Zamorani}]{Genzel11}
Genzel, R., Newman, S., Jones, T., {et~al.} 2011, \apj, 733, 101

\bibitem[{Genzel {et~al.}(2014)Genzel, {F\"{o}rster Schreiber}, Lang,
  Tacchella, Tacconi, Wuyts, Bandara, Burkert, Buschkamp, Carollo, Cresci,
  Davies, Eisenhauer, Hicks, Kurk, Lilly, Lutz, Mancini, Naab, Newman, Peng,
  Renzini, {Shapiro Griffin}, Sternberg, Vergani, Wisnioski, Wuyts, \&
  Zamorani}]{Genzel14}
Genzel, R., {F\"{o}rster Schreiber}, N.~M., Lang, P., {et~al.} 2014, \apj, 785,
  75

\bibitem[{Genzel {et~al.}(2015)Genzel, Tacconi, Lutz, Saintonge, Berta,
  Magnelli, Combes, Garc\'{\i}a-Burillo, Neri, Bolatto, Contini, Lilly,
  Boissier, Boone, Bouch\'{e}, Bournaud, Burkert, Carollo, Colina, Cooper, Cox,
  Feruglio, {F\"{o}rster Schreiber}, Freundlich, Gracia-Carpio, Juneau, Kovac,
  Lippa, Naab, Salome, Renzini, Sternberg, Walter, Weiner, Weiss, \&
  Wuyts}]{Genzel15}
Genzel, R., Tacconi, L.~J., Lutz, D., {et~al.} 2015, \apj, 800, 20

\bibitem[{Gnerucci {et~al.}(2011)Gnerucci, Marconi, Cresci, Maiolino, Mannucci,
  Calura, Cimatti, Cocchia, Grazian, Matteucci, Nagao, Pozzetti, \&
  Troncoso}]{Gnerucci11}
Gnerucci, A., Marconi, A., Cresci, G., {et~al.} 2011, A\&A, 528, A88

\bibitem[{Governato {et~al.}(2007)Governato, Willman, Mayer, Brooks, Stinson,
  Valenzuela, Wadsley, \& Quinn}]{Governato07}
Governato, F., Willman, B., Mayer, L., {et~al.} 2007, \mnras, 374, 1479

\bibitem[{Green {et~al.}(2014)Green, Glazebrook, McGregor, Damjanov, Wisnioski,
  Abraham, Colless, Sharp, Crain, Poole, \& McCarthy}]{Green14}
Green, A.~W., Glazebrook, K., McGregor, P.~J., {et~al.} 2014, \mnras, 437, 1070

\bibitem[{Grogin {et~al.}(2011)Grogin, Kocevski, Faber, Ferguson, Koekemoer,
  Riess, Acquaviva, Alexander, Almaini, Ashby, Barden, Bell, Bournaud, Brown,
  Caputi, Casertano, Cassata, Castellano, Challis, Chary, Cheung, Cirasuolo,
  Conselice, Cooray, Croton, Daddi, Dahlen, Dav\'{e}, de~Mello, Dekel,
  Dickinson, Dolch, Donley, Dunlop, Dutton, Elbaz, Fazio, Filippenko,
  Finkelstein, Fontana, Gardner, Garnavich, Gawiser, Giavalisco, Grazian, Guo,
  Hathi, H\"{a}ussler, Hopkins, Huang, Huang, Jha, Kartaltepe, Kirshner, Koo,
  Lai, Lee, Li, Lotz, Lucas, Madau, McCarthy, McGrath, McIntosh, McLure,
  Mobasher, Moustakas, Mozena, Nandra, Newman, Niemi, Noeske, Papovich,
  Pentericci, Pope, Primack, Rajan, Ravindranath, Reddy, Renzini, Rix, Robaina,
  Rodney, Rosario, Rosati, Salimbeni, Scarlata, Siana, Simard, Smidt,
  Somerville, Spinrad, Straughn, Strolger, Telford, Teplitz, Trump, van~der
  Wel, Villforth, Wechsler, Weiner, Wiklind, Wild, Wilson, Wuyts, Yan, \&
  Yun}]{Grogin11}
Grogin, N.~A., Kocevski, D.~D., Faber, S.~M., {et~al.} 2011, \apjs, 197, 35

\bibitem[{Hopkins \& Beacom(2006)}]{Hopkins06}
Hopkins, A.~M., \& Beacom, J.~F. 2006, \apj, 651, 142

\bibitem[{Kassin {et~al.}(2007)Kassin, Weiner, Faber, Koo, Lotz, Diemand,
  Harker, Bundy, Metevier, Phillips, Cooper, Croton, Konidaris, Noeske, \&
  Willmer}]{Kassin07}
Kassin, S.~A., Weiner, B.~J., Faber, S.~M., {et~al.} 2007, \apj, 660, L35

\bibitem[{Kassin {et~al.}(2012)Kassin, Weiner, Faber, Gardner, Willmer, Coil,
  Cooper, Devriendt, Dutton, Guhathakurta, Koo, Metevier, Noeske, \&
  Primack}]{Kassin12}
---. 2012, \apj, 758, 106

\bibitem[{Kennicutt(1998)}]{Kennicutt98}
Kennicutt, R.~C. 1998, \apj, 498, 541

\bibitem[{Kere\v{s} {et~al.}(2009)Kere\v{s}, Katz, Fardal, Dav\'{e}, \&
  Weinberg}]{Keres09}
Kere\v{s}, D., Katz, N., Fardal, M., Dav\'{e}, R., \& Weinberg, D.~H. 2009,
  \mnras, 395, 160

\bibitem[{Kere\v{s} {et~al.}(2005)Kere\v{s}, Katz, Weinberg, \& Dave}]{Keres05}
Kere\v{s}, D., Katz, N., Weinberg, D.~H., \& Dave, R. 2005, \mnras, 363, 2

\bibitem[{Koekemoer {et~al.}(2011)Koekemoer, Faber, Ferguson, Grogin, Kocevski,
  Koo, Lai, Lotz, Lucas, McGrath, Ogaz, Rajan, Riess, Rodney, Strolger,
  Casertano, Castellano, Dahlen, Dickinson, Dolch, Fontana, Giavalisco,
  Grazian, Guo, Hathi, Huang, van~der Wel, Yan, Acquaviva, Alexander, Almaini,
  Ashby, Barden, Bell, Bournaud, Brown, Caputi, Cassata, Challis, Chary,
  Cheung, Cirasuolo, Conselice, Cooray, Croton, Daddi, Dav\'{e}, de~Mello,
  de~Ravel, Dekel, Donley, Dunlop, Dutton, Elbaz, Fazio, Filippenko,
  Finkelstein, Frazer, Gardner, Garnavich, Gawiser, Gruetzbauch, Hartley,
  H\"{a}ussler, Herrington, Hopkins, Huang, Jha, Johnson, Kartaltepe,
  Khostovan, Kirshner, Lani, Lee, Li, Madau, McCarthy, McIntosh, McLure,
  McPartland, Mobasher, Moreira, Mortlock, Moustakas, Mozena, Nandra, Newman,
  Nielsen, Niemi, Noeske, Papovich, Pentericci, Pope, Primack, Ravindranath,
  Reddy, Renzini, Rix, Robaina, Rosario, Rosati, Salimbeni, Scarlata, Siana,
  Simard, Smidt, Snyder, Somerville, Spinrad, Straughn, Telford, Teplitz,
  Trump, Vargas, Villforth, Wagner, Wandro, Wechsler, Weiner, Wiklind, Wild,
  Wilson, Wuyts, \& Yun}]{Koekemoer11}
Koekemoer, A.~M., Faber, S.~M., Ferguson, H.~C., {et~al.} 2011, \apjs, 197, 36

\bibitem[{Kriek {et~al.}(2009)Kriek, van Dokkum, Labb\'{e}, Franx, Illingworth,
  Marchesini, \& Quadri}]{Kriek09}
Kriek, M., van Dokkum, P.~G., Labb\'{e}, I., {et~al.} 2009, \apj, 700, 221

\bibitem[{{Kriek} {et~al.}(2015){Kriek}, {Shapley}, {Reddy}, {Siana}, {Coil},
  {Mobasher}, {Freeman}, {de Groot}, {Price}, {Sanders}, {Shivaei}, {Brammer},
  {Momcheva}, {Skelton}, {van Dokkum}, {Whitaker}, {Aird}, {Azadi}, {Kassis},
  {Bullock}, {Conroy}, {Dav{\'e}}, {Kere{\v s}}, \& {Krumholz}}]{Kriek15}
{Kriek}, M., {Shapley}, A.~E., {Reddy}, N.~A., {et~al.} 2015, \apjs, 218, 15

\bibitem[{Law {et~al.}(2007{\natexlab{a}})Law, Steidel, Erb, Larkin, Pettini,
  Shapley, \& Wright}]{Law07a}
Law, D.~R., Steidel, C.~C., Erb, D.~K., {et~al.} 2007{\natexlab{a}}, \apj, 669,
  929

\bibitem[{Law {et~al.}(2009)Law, Steidel, Erb, Larkin, Pettini, Shapley, \&
  Wright}]{Law09}
---. 2009, \apj, 697, 2057

\bibitem[{Law {et~al.}(2007{\natexlab{b}})Law, Steidel, Erb, Pettini, Reddy,
  Shapley, Adelberger, \& Simenc}]{Law07}
---. 2007{\natexlab{b}}, \apj, 656, 1

\bibitem[{Law {et~al.}(2012{\natexlab{a}})Law, Steidel, Shapley, Nagy, Reddy,
  \& Erb}]{Law12}
Law, D.~R., Steidel, C.~C., Shapley, A.~E., {et~al.} 2012{\natexlab{a}}, \apj,
  759, 29

\bibitem[{Law {et~al.}(2012{\natexlab{b}})Law, Steidel, Shapley, Nagy, Reddy,
  \& Erb}]{Law12a}
---. 2012{\natexlab{b}}, \apj, 745, 85

\bibitem[{McLean {et~al.}(2010)McLean, Steidel, Epps, Matthews, Adkins,
  Konidaris, Weber, Aliado, Brims, Canfield, Cromer, Fucik, Kulas, Mace,
  Magnone, Rodriguez, Wang, \& Weiss}]{McLean10}
McLean, I.~S., Steidel, C.~C., Epps, H., {et~al.} 2010, in Proceedings of SPIE,
  ed. I.~S. McLean, S.~K. Ramsay, \& H.~Takami, Vol. 7735, 77351E--77351E--12

\bibitem[{McLean {et~al.}(2012)McLean, Steidel, Epps, Konidaris, Matthews,
  Adkins, Aliado, Brims, Canfield, Cromer, Fucik, Kulas, Mace, Magnone,
  Rodriguez, Rudie, Trainor, Wang, Weber, \& Weiss}]{McLean12}
McLean, I.~S., Steidel, C.~C., Epps, H.~W., {et~al.} 2012, in Proceedings of
  SPIE, ed. I.~S. McLean, S.~K. Ramsay, \& H.~Takami, Vol. 8446, 84460J

\bibitem[{Meurer {et~al.}(1996)Meurer, Carignan, Beaulieu, \&
  Freeman}]{Meurer96}
Meurer, G.~R., Carignan, C., Beaulieu, S.~F., \& Freeman, K.~C. 1996, \aj, 111,
  1551

\bibitem[{Miller {et~al.}(2011)Miller, Bundy, Sullivan, Ellis, \&
  Treu}]{Miller11}
Miller, S.~H., Bundy, K., Sullivan, M., Ellis, R.~S., \& Treu, T. 2011, \apj,
  741, 115

\bibitem[{Miller {et~al.}(2012)Miller, Ellis, Sullivan, Bundy, Newman, \&
  Treu}]{Miller12}
Miller, S.~H., Ellis, R.~S., Sullivan, M., {et~al.} 2012, \apj, 753, 74

\bibitem[{Miller {et~al.}(2013)Miller, Sullivan, \& Ellis}]{Miller13}
Miller, S.~H., Sullivan, M., \& Ellis, R.~S. 2013, \apj, 762, L11

\bibitem[{Mo {et~al.}(1998)Mo, Mao, \& White}]{Mo98}
Mo, H.~J., Mao, S., \& White, S. D.~M. 1998, \mnras, 295, 319

\bibitem[{Momcheva {et~al.}(2015)Momcheva, Brammer, van Dokkum, Skelton,
  Whitaker, Nelson, Fumagalli, Maseda, Leja, Franx, Rix, Bezanson, {Da Cunha},
  Dickey, Schreiber, Illingworth, Kriek, Labb{\'{e}}, Lange, Lundgren, Magee,
  Marchesini, Oesch, Pacifici, Patel, Price, Tal, Wake, van~der Wel, \&
  Wuyts}]{Momcheva15}
Momcheva, I.~G., Brammer, G.~B., van Dokkum, P.~G., {et~al.} 2015, submitted to
  ApJSS, arXiv: 1510.02106

\bibitem[{Nelson {et~al.}(2012)Nelson, van Dokkum, Brammer, {F\"{o}rster
  Schreiber}, Franx, Fumagalli, Patel, Rix, Skelton, Bezanson, {Da Cunha},
  Kriek, Labbe, Lundgren, Quadri, \& Schmidt}]{Nelson12}
Nelson, E.~J., van Dokkum, P.~G., Brammer, G., {et~al.} 2012, \apj, 747, L28

\bibitem[{Nelson {et~al.}(2013)Nelson, van Dokkum, Momcheva, Brammer, Lundgren,
  Skelton, Whitaker, {Da Cunha}, {F\"{o}rster Schreiber}, Franx, Fumagalli,
  Kriek, Labbe, Leja, Patel, Rix, Schmidt, van~der Wel, \& Wuyts}]{Nelson13}
Nelson, E.~J., van Dokkum, P.~G., Momcheva, I., {et~al.} 2013, \apj, 763, L16

\bibitem[{Nelson {et~al.}(2015)Nelson, van Dokkum, Schreiber, Franx, Brammer,
  Momcheva, Wuyts, Whitaker, Skelton, Fumagalli, Kriek, Labb\'{e}, Leja, Rix,
  Tacconi, van~der Wel, van~den Bosch, Oesch, Dickey, \& Lange}]{Nelson15}
Nelson, E.~J., van Dokkum, P.~G., Schreiber, N. M.~F., {et~al.} 2015, submitted
  to ApJ, arXiv: 1507.03999

\bibitem[{Newman {et~al.}(2010)Newman, Ellis, Treu, \& Bundy}]{ANewman10}
Newman, A.~B., Ellis, R.~S., Treu, T., \& Bundy, K. 2010, \apj, 717, L103

\bibitem[{Newman {et~al.}(2013)Newman, Genzel, {F\"{o}rster Schreiber},
  {Shapiro Griffin}, Mancini, Lilly, Renzini, Bouch\'{e}, Burkert, Buschkamp,
  Carollo, Cresci, Davies, Eisenhauer, Genel, Hicks, Kurk, Lutz, Naab, Peng,
  Sternberg, Tacconi, Wuyts, Zamorani, \& Vergani}]{Newman13}
Newman, S.~F., Genzel, R., {F\"{o}rster Schreiber}, N.~M., {et~al.} 2013, \apj,
  767, 104

\bibitem[{Noeske {et~al.}(2007)Noeske, Weiner, Faber, Papovich, Koo,
  Somerville, Bundy, Conselice, Newman, Schiminovich, {Le Floc'h}, Coil, Rieke,
  Lotz, Primack, Barmby, Cooper, Davis, Ellis, Fazio, Guhathakurta, Huang,
  Kassin, Martin, Phillips, Rich, Small, Willmer, \& Wilson}]{Noeske07}
Noeske, K.~G., Weiner, B.~J., Faber, S.~M., {et~al.} 2007, \apj, 660, L43

\bibitem[{Oser {et~al.}(2010)Oser, Ostriker, Naab, Johansson, \&
  Burkert}]{Oser10}
Oser, L., Ostriker, J.~P., Naab, T., Johansson, P.~H., \& Burkert, A. 2010,
  \apj, 725, 2312

\bibitem[{Osterbrock \& Ferland(2006)}]{Osterbrock06}
Osterbrock, D.~E., \& Ferland, G.~J. 2006, {Astrophysics of gaseous nebulae and
  active galactic nuclei}, 2nd edn. (Sausalito, CA: University Science Books)

\bibitem[{Peng {et~al.}(2010)Peng, Ho, Impey, \& Rix}]{Peng10a}
Peng, C.~Y., Ho, L.~C., Impey, C.~D., \& Rix, H.-W. 2010, \aj, 139, 2097

\bibitem[{Pettini {et~al.}(2001)Pettini, Shapley, Steidel, Cuby, Dickinson,
  Moorwood, Adelberger, \& Giavalisco}]{Pettini01}
Pettini, M., Shapley, A.~E., Steidel, C.~C., {et~al.} 2001, \apj, 554, 981

\bibitem[{Pizagno {et~al.}(2005)Pizagno, Prada, Weinberg, Rix, Harbeck, Grebel,
  Bell, Brinkmann, Holtzman, \& West}]{Pizagno05}
Pizagno, J., Prada, F., Weinberg, D.~H., {et~al.} 2005, \apj, 633, 844

\bibitem[{Price {et~al.}(2014)Price, Kriek, Brammer, Conroy, Schreiber, Franx,
  Fumagalli, Lundgren, Momcheva, Nelson, Skelton, van Dokkum, Whitaker, \&
  Wuyts}]{Price14}
Price, S.~H., Kriek, M., Brammer, G.~B., {et~al.} 2014, \apj, 788, 86

\bibitem[{Reddy \& Steidel(2009)}]{Reddy09}
Reddy, N.~A., \& Steidel, C.~C. 2009, \apj, 692, 778

\bibitem[{Reddy {et~al.}(2015)Reddy, Kriek, Shapley, Freeman, Siana, Coil,
  Mobasher, Price, Sanders, \& Shivaei}]{Reddy15}
Reddy, N.~A., Kriek, M., Shapley, A.~E., {et~al.} 2015, \apj, 806, 259

\bibitem[{Salpeter(1955)}]{Salpeter55}
Salpeter, E.~E. 1955, \apj, 121, 161

\bibitem[{{S\'{e}rsic}(1968)}]{Sersic68}
{S\'{e}rsic}, J.~L. 1968, {Atlas de galaxias australes} (Cordoba, Argentina:
  Observatorio Astronomico)

\bibitem[{Sharples {et~al.}(2013)Sharples, Bender, {Agudo Berbel}, Bezawada,
  Castillo, Cirasuolo, Davidson, Davies, Dubbeldam, Fairley, Finger,
  {F\"{o}rster Schreiber}, Gonte, Hess, Jung, Lewis, Lizon, Muschielok,
  Pasquini, Pirard, Popovic, Ramsay, Rees, Richter, Riquelme, Rodrigues,
  Saviane, Schlichter, Schmidtobreick, Segovia, Smette, Szeifert, van Kesteren,
  Wegner, \& Wiezorrek}]{Sharples13}
Sharples, R., Bender, R., {Agudo Berbel}, A., {et~al.} 2013, The Messenger,
  151, 21

\bibitem[{Sharples {et~al.}(2004)Sharples, Bender, Lehnert, {Ramsay Howat},
  Bremer, Davies, Genzel, Hofmann, Ivison, Saglia, \& Thatte}]{Sharples04}
Sharples, R.~M., Bender, R., Lehnert, M.~D., {et~al.} 2004, in Proceedings of
  SPIE, ed. A.~F.~M. Moorwood \& M.~Iye, Vol. 5492, 1179--1186

\bibitem[{Shivaei {et~al.}(2015)Shivaei, Reddy, Shapley, Kriek, Siana,
  Mobasher, Coil, Freeman, Sanders, Price, Groot, \& Azadi}]{Shivaei15}
Shivaei, I., Reddy, N.~A., Shapley, A.~E., {et~al.} 2015, accepted to ApJ,
  arXiv: 1507.03017

\bibitem[{Simard \& Pritchet(1999)}]{Simard99}
Simard, L., \& Pritchet, C.~J. 1999, \pasp, 111, 453

\bibitem[{Skelton {et~al.}(2014)Skelton, Whitaker, Momcheva, Brammer, van
  Dokkum, Labb\'{e}, Franx, van~der Wel, Bezanson, {Da Cunha}, Fumagalli,
  {F\"{o}rster Schreiber}, Kriek, Leja, Lundgren, Magee, Marchesini, Maseda,
  Nelson, Oesch, Pacifici, Patel, Price, Rix, Tal, Wake, \& Wuyts}]{Skelton14}
Skelton, R.~E., Whitaker, K.~E., Momcheva, I.~G., {et~al.} 2014, \apjs, 214, 24

\bibitem[{Swinbank {et~al.}(2011)Swinbank, Papadopoulos, Cox, Krips, Ivison,
  Smail, Thomson, Neri, Richard, \& Ebeling}]{Swinbank11}
Swinbank, A.~M., Papadopoulos, P.~P., Cox, P., {et~al.} 2011, \apj, 742, 11

\bibitem[{Szomoru {et~al.}(2011)Szomoru, Franx, Bouwens, van Dokkum, Labb\'{e},
  Illingworth, \& Trenti}]{Szomoru11}
Szomoru, D., Franx, M., Bouwens, R.~J., {et~al.} 2011, \apj, 735, L22

\bibitem[{Szomoru {et~al.}(2013)Szomoru, Franx, van Dokkum, Trenti,
  Illingworth, Labbe, \& Oesch}]{Szomoru13}
Szomoru, D., Franx, M., van Dokkum, P.~G., {et~al.} 2013, \apj, 763, 73

\bibitem[{Tacconi {et~al.}(2008)Tacconi, Genzel, Smail, Neri, Chapman, Ivison,
  Blain, Cox, Omont, Bertoldi, Greve, {F\"{o}rster Schreiber}, Genel, Lutz,
  Swinbank, Shapley, Erb, Cimatti, Daddi, \& Baker}]{Tacconi08}
Tacconi, L.~J., Genzel, R., Smail, I., {et~al.} 2008, \apj, 680, 246

\bibitem[{Tacconi {et~al.}(2010)Tacconi, Genzel, Neri, Cox, Cooper, Shapiro,
  Bolatto, Bouch\'{e}, Bournaud, Burkert, Combes, Comerford, Davis, Schreiber,
  Garcia-Burillo, Gracia-Carpio, Lutz, Naab, Omont, Shapley, Sternberg, \&
  Weiner}]{Tacconi10}
Tacconi, L.~J., Genzel, R., Neri, R., {et~al.} 2010, Nature, 463, 781

\bibitem[{Tacconi {et~al.}(2013)Tacconi, Neri, Genzel, Combes, Bolatto, Cooper,
  Wuyts, Bournaud, Burkert, Comerford, Cox, Davis, {F\"{o}rster Schreiber},
  Garc\'{\i}a-Burillo, Gracia-Carpio, Lutz, Naab, Newman, Omont, Saintonge,
  {Shapiro Griffin}, Shapley, Sternberg, \& Weiner}]{Tacconi13}
Tacconi, L.~J., Neri, R., Genzel, R., {et~al.} 2013, \apj, 768, 74

\bibitem[{Taylor {et~al.}(2010)Taylor, Franx, Brinchmann, van~der Wel, \& van
  Dokkum}]{Taylor10a}
Taylor, E.~N., Franx, M., Brinchmann, J., van~der Wel, A., \& van Dokkum, P.~G.
  2010, \apj, 722, 1

\bibitem[{Tully \& Fisher(1977)}]{Tully77}
Tully, R.~B., \& Fisher, J.~R. 1977, A\&A, 54, 661

\bibitem[{van~de Sande {et~al.}(2013)van~de Sande, Kriek, Franx, van Dokkum,
  Bezanson, Bouwens, Quadri, Rix, \& Skelton}]{vandeSande13}
van~de Sande, J., Kriek, M., Franx, M., {et~al.} 2013, \apj, 771, 85

\bibitem[{van~den Bosch(2001)}]{vandenBosch01}
van~den Bosch, F.~C. 2001, \mnras, 327, 1334

\bibitem[{van~der Wel {et~al.}(2014{\natexlab{a}})van~der Wel, Franx, van
  Dokkum, Skelton, Momcheva, Whitaker, Brammer, Bell, Rix, Wuyts, Ferguson,
  Holden, Barro, Koekemoer, Chang, McGrath, H\"{a}ussler, Dekel, Behroozi,
  Fumagalli, Leja, Lundgren, Maseda, Nelson, Wake, Patel, Labb\'{e}, Faber,
  Grogin, \& Kocevski}]{vanderWel14a}
van~der Wel, A., Franx, M., van Dokkum, P.~G., {et~al.} 2014{\natexlab{a}},
  \apj, 788, 28

\bibitem[{van~der Wel {et~al.}(2014{\natexlab{b}})van~der Wel, Chang, Bell,
  Holden, Ferguson, Giavalisco, Rix, Skelton, Whitaker, Momcheva, Brammer,
  Kassin, Martig, Dekel, Ceverino, Koo, Mozena, van Dokkum, Franx, Faber, \&
  Primack}]{vanderWel14}
van~der Wel, A., Chang, Y.-Y., Bell, E.~F., {et~al.} 2014{\natexlab{b}}, \apj,
  792, L6

\bibitem[{van Dokkum {et~al.}(2013)van Dokkum, Leja, Nelson, Patel, Skelton,
  Momcheva, Brammer, Whitaker, Lundgren, Fumagalli, Conroy, {F\"{o}rster
  Schreiber}, Franx, Kriek, Labb\'{e}, Marchesini, Rix, van~der Wel, \&
  Wuyts}]{vanDokkum13}
van Dokkum, P.~G., Leja, J., Nelson, E.~J., {et~al.} 2013, \apj, 771, L35

\bibitem[{van Dokkum {et~al.}(2015)van Dokkum, Nelson, Franx, Oesch, Momcheva,
  Brammer, Schreiber, Skelton, Whitaker, van~der Wel, Bezanson, Fumagalli,
  Illingworth, Kriek, Leja, \& Wuyts}]{vanDokkum15}
van Dokkum, P.~G., Nelson, E.~J., Franx, M., {et~al.} 2015, \apj, 813, 23

\bibitem[{Vergani {et~al.}(2012)Vergani, Epinat, Contini, Tasca, Tresse, Amram,
  Garilli, Kissler-Patig, {Le F\`{e}vre}, Moultaka, Paioro, Queyrel, \&
  L\'{o}pez-Sanjuan}]{Vergani12}
Vergani, D., Epinat, B., Contini, T., {et~al.} 2012, A\&A, 546, A118

\bibitem[{Vogt {et~al.}(1996)Vogt, Forbes, Phillips, Gronwall, Faber,
  Illingworth, \& Koo}]{Vogt96}
Vogt, N.~P., Forbes, D.~A., Phillips, A.~C., {et~al.} 1996, \apj, 465, L15

\bibitem[{Weiner {et~al.}(2006)Weiner, Willmer, Faber, Harker, Kassin,
  Phillips, Melbourne, Metevier, Vogt, \& Koo}]{Weiner06a}
Weiner, B.~J., Willmer, C. N.~A., Faber, S.~M., {et~al.} 2006, \apj, 653, 1049

\bibitem[{White \& Frenk(1991)}]{White91}
White, S. D.~M., \& Frenk, C.~S. 1991, \apj, 379, 52

\bibitem[{White \& Rees(1978)}]{White78}
White, S. D.~M., \& Rees, M.~J. 1978, \mnras, 183, 341

\bibitem[{Williams {et~al.}(2009)Williams, Quadri, Franx, van Dokkum, \&
  Labb\'{e}}]{Williams09}
Williams, R.~J., Quadri, R.~F., Franx, M., van Dokkum, P., \& Labb\'{e}, I.
  2009, \apj, 691, 1879

\bibitem[{Williams {et~al.}(2010)Williams, Quadri, Franx, van Dokkum, Toft,
  Kriek, \& Labb\'{e}}]{Williams10}
Williams, R.~J., Quadri, R.~F., Franx, M., {et~al.} 2010, \apj, 713, 738

\bibitem[{Wisnioski {et~al.}(2012)Wisnioski, Glazebrook, Blake, Poole, Green,
  Wyder, \& Martin}]{Wisnioski12}
Wisnioski, E., Glazebrook, K., Blake, C., {et~al.} 2012, \mnras, 422, 3339

\bibitem[{Wisnioski {et~al.}(2015)Wisnioski, {F\"{o}rster Schreiber}, Wuyts,
  Wuyts, Bandara, Wilman, Genzel, Bender, Davies, Fossati, Lang, Mendel,
  Beifiori, Brammer, Chan, Fabricius, Fudamoto, Kulkarni, Kurk, Lutz, Nelson,
  Momcheva, Rosario, Saglia, Seitz, Tacconi, \& van Dokkum}]{Wisnioski15}
Wisnioski, E., {F\"{o}rster Schreiber}, N.~M., Wuyts, S., {et~al.} 2015, \apj,
  799, 209

\bibitem[{Wright {et~al.}(2009)Wright, Larkin, Law, Steidel, Shapley, \&
  Erb}]{Wright09}
Wright, S.~A., Larkin, J.~E., Law, D.~R., {et~al.} 2009, \apj, 699, 421

\bibitem[{Wright {et~al.}(2007)Wright, Larkin, Barczys, Erb, Iserlohe, Krabbe,
  Law, McElwain, Quirrenbach, Steidel, \& Weiss}]{Wright07}
Wright, S.~A., Larkin, J.~E., Barczys, M., {et~al.} 2007, \apj, 658, 78

\bibitem[{Wuyts {et~al.}(2007)Wuyts, Labbe, Franx, Rudnick, van Dokkum, Fazio,
  {Forster Schreiber}, Huang, Moorwood, Rix, Rottgering, \& van~der
  Werf}]{Wuyts07}
Wuyts, S., Labbe, I., Franx, M., {et~al.} 2007, \apj, 655, 51

\end{thebibliography}
\end{document}